\documentclass[12pt]{article}
\pdfoutput=1

\usepackage{tabu}
\usepackage{geometry}

\usepackage{putex}
\usepackage{float}
\usepackage{caption}
\usepackage{psfrag}
\usepackage{amssymb}
\usepackage{amsmath}
\usepackage{empheq}
\usepackage{amsthm}
\usepackage{epsf}
\usepackage{graphicx}
\usepackage{epstopdf}

\usepackage{enumerate}
\usepackage{cite}
\usepackage[font=small]{caption}
\usepackage{subcaption}

\usepackage[utf8]{inputenc}
\usepackage{braket}
\usepackage{titlesec}

\usepackage{psfrag}
\usepackage{color}
\definecolor{darkblue}{rgb}{0.1,0.1,.7}
\usepackage[colorlinks, linkcolor=darkblue, citecolor=darkblue, urlcolor=darkblue, linktocpage]{hyperref}

\hypersetup{breaklinks}
\usepackage{comment}

\newcommand{\dDisc}{\text{dDisc}}

\usepackage{braket}

\def\<{\langle}
\def\>{\rangle}

\newcommand   \f  {\phi}
\newcommand   \p  {\psi}
\newcommand{\bea}{\begin{eqnarray}}
\newcommand{\eea}{\end{eqnarray}}

\DeclareMathOperator*{\res}{Res}

\usepackage{stackengine}

\usepackage{accents}

\def\O{{\cal O}}

\newcommand {\be} {\begin {equation}}
\newcommand {\ee} {\end {equation}}

\newcommand {\bes} {\begin {equation*}}
\newcommand {\ees} {\end {equation*}}

\newcommand{\beq}{\begin{equation}}
\newcommand{\eeq}{\end{equation}}

\def\be{ \begin{equation} }
\def\ee{ \end{equation} }

 \newmuskip\pFqmuskip

\newcommand*\pFq[6][8]{%
  \begingroup 
  \pFqmuskip=#1mu\relax
  \mathcode`\,=\string"8000
  \begingroup\lccode`\~=`\,
  \lowercase{\endgroup\let~}\pFqcomma
  {}_{#2}F_{#3}{\left[\genfrac..{0pt}{}{#4}{#5};#6\right]}%
  \endgroup
}
\newcommand{\pFqcomma}{\mskip\pFqmuskip}

\makeatletter
\renewcommand{\@maketitle}{
\newpage
 \begin{center}%
  {\large\bfseries \@title \par}%
 \end{center}%
 \par} \makeatother

\numberwithin{equation}{section}

\institution{CT}{Walter Burke Institute for Theoretical Physics, California Institute of Technology, \cr Pasadena, CA, 91125}

\begin{document}

\preprint{CALT-TH-2019-053}

\title{AdS/CFT Unitarity at Higher Loops: 
\\
High-Energy String Scattering}
 
\authors{David Meltzer}
\abstract{What is the space of weakly-coupled, gravitational theories which contain massive, higher-spin particles? This class of theories is highly constrained and it is conjectured their ultraviolet completion must be string theory. We provide more evidence for this conjecture by studying the Regge limit in large $N$, $4d$ CFTs with single-trace operators of unbounded spin. We show that in the Regge limit, these theories have bulk scattering amplitudes which are consistent with the string theory prediction to all orders in $1/N$ for large, but finite, coupling. In the language of Regge theory, we show Pomeron exchange naturally exponentiates in the $1/N$ expansion. To do this, we solve the bootstrap equations at tree-level and then use the Lorentzian inversion formula to find the one-loop correlator in the Regge limit. This is a unitarity method for AdS/CFT which can be repeated iteratively to make all orders statements. We also explain under what conditions the tree-level result exponentiates in the $1/N$ expansion at arbitrary coupling.  Finally, we comment on further inelastic effects and show they give subleading contributions at large coupling. As a consistency check, we recover results from bulk Einstein gravity in the limit where all higher-spin particles decouple.}

\date{ }

\maketitle
\setcounter{tocdepth}{3}
\tableofcontents

\newpage
\section{Introduction}
A fundamental problem in the conformal bootstrap \cite{Polyakov:1974gs,Rattazzi:2008pe,Poland:2018epd} is to determine how the basic principles of causality and unitarity constrain the space of large $N$ conformal field theories (CFTs). Using the AdS/CFT correspondence \cite{Maldacena:1997re,Witten:1998qj,Gubser:1998bc}, this is equivalent to studying the structure of quantum gravity in an AdS spacetime. In order to understand universal features of AdS/CFT, we will use ideas from the analytic bootstrap \cite{Fitzpatrick:2012yx,Komargodski:2012ek,Caron-Huot:2017vep} to study high-energy scattering in AdS. 

The motivating question we would like to answer is: what is the space of consistent, weakly-coupled, gravitational theories in AdS, and is it possible to show every such theory necessarily has an ultraviolet completion given by string/M-theory \cite{Heemskerk:2009pn,Maldacena:2012sf,Camanho:2014apa,Caron-Huot:2016icg}? Specifically, we consider this question for gravitational theories which also contain a massive, higher spin ($J>2$) particle. We will show that in the Regge limit, theories satisfying this criteria have scattering amplitudes that agree with the string theory prediction \cite{Brower:2006ea,Brower:2007xg} to all orders in $1/N$. The bulk scattering amplitudes have been calculated using semi-classical worldsheet methods, or when the higher-spin particles are heavy, and we will find agreement in this limit. In other words, although we do not impose our bulk theory has strings, we find evidence they emerge naturally.

The physical motivation for studying the CFT Regge limit is manifest in AdS, where it is dual to high-energy scattering at fixed impact parameter \cite{Cornalba:2006xk,Cornalba:2006xm,Cornalba:2007zb}. In this limit, the dominant contribution at tree-level in $1/N$ comes from the leading Regge trajectory, or the set of operators with the lowest dimension for each even-spin \cite{Cornalba:2007fs,Costa:2012cb}. The Regge limit can change qualitatively depending on the relative size of the bulk, center of mass energy, $\sqrt{S}$, and the dimension of the lightest, spin-four, single-trace operator, $\Delta_{gap}$. In the limit $\Delta_{gap}^{2}\gg S\gg1$, the Regge limit is controlled by graviton exchange in the bulk. By studying this limit and imposing causality and unitarity of the boundary CFT, it has been shown that tree-level, cubic interactions involving gravitons are completely fixed to be given by Einstein gravity minimally coupled to matter \cite{Afkhami-Jeddi:2016ntf,Afkhami-Jeddi:2017rmx,Afkhami-Jeddi:2018own,KPZ2017,Costa:2017twz,Meltzer:2017rtf}.

In this paper, we extend this program by studying the effects of higher-spin particles in the bulk theory, both at tree and loop-level. That is, we will study the limit $S\gg\Delta_{gap}^{2}\gg1$, where we are more sensitive to the higher-spin spectrum, in the $1/N$ expansion. As is well-known, it is not possible to consistently add a single, higher-spin particle to Einstein gravity and maintain causality. Instead, as in string theory, we must always add an infinite tower of particles with unbounded spin \cite{Camanho:2014apa,Maldacena:2015waa,Hartman:2015lfa,DAppollonio:2015fly}. After resumming the leading trajectory, we find the Regge limit is controlled by the exchange of an operator with non-integer spin, which is known as the ``Pomeron" or reggeized graviton\cite{Brower:2006ea,Brower:2007xg,Cornalba:2007fs,Costa:2012cb,Kravchuk:2018htv}. In general weakly-coupled theories of higher-spin particles \cite{Caron-Huot:2016icg,KPZ2017,Li:2017lmh}, the resummation of the leading trajectory reproduces features of string theory, such as the transverse spreading of strings at high energies.

A natural question to consider is: does this universality also extend to loop-level in $1/N$? Our expectations from high-energy, fixed impact parameter scattering in the bulk \cite{Brower:2006ea,Brower:2007xg,Cornalba:2006xk,Cornalba:2006xm,Cornalba:2007zb} is that at higher-loops the tree-level amplitude should exponentiate, a notion which we will make more precise shortly. In Einstein gravity, this exponentiation is found by resumming a set of graviton ladder diagrams, and the final result matches the prediction from scattering in a shockwave background \cite{Aichelburg:1970dh,Dray:1984ha,tHooft:1987vrq}. There is a similar story in string theory \cite{Amati:1987wq,Amati:1987uf}, once we include new tree-level stringy effects. Therefore, the goal of this paper is to prove this exponentiation is universal in the presence of higher-spin particles using the boundary CFT.\footnote{In the limit where higher-spin particles decouple, it has been demonstrated in \cite{Cornalba:2007zb} that the phase shift due to graviton exchange exponentiates.}

In order to understand when tree-level, Regge behavior exponentiates, we will study a four-point function of scalars in the CFT. At each order in $1/N$, and using the minimal ansatz for the spectrum, we will find results consistent with exponentiation at leading order in the $1/\Delta_{gap}$ expansion. Since we also take $S\gg \Delta_{gap}^{2}$, the higher-spin operators will not decouple when we also take $\Delta_{gap}$ large. By relaxing our assumption on the spectrum and allowing for inelastic effects, we will give conditions for the tree-level phase shift to exponentiate at arbitrary $\Delta_{gap}$. In a string theory dual, these inelastic effects correspond to one-loop corrections to long string creation. We also demonstrate how tidal excitations, an inelastic effect dual to new double-trace exchanges at one-loop, are suppressed when we take $\Delta_{gap}$ large.

The results presented here rely crucially on the Lorentzian inversion formula \cite{Caron-Huot:2017vep}. The inversion formula implies that the operator product expansion (OPE) data for general, non-perturbative CFTs can be analytically continued in spin, and therefore that local operators can be organized on Regge trajectories. We will use the inversion formula to study the Regge limit at higher loops and also to determine one-loop corrections to the anomalous dimensions and OPE coefficients of double-trace operators.

\subsection*{Summary and Outline}
We can now spell out our main assumptions and conclusions. First, we will be studying a four-point function of pairwise identical scalars, $\<\f\f\p\p\>$, in a $4d$ CFT. We assume the central charge of the CFT is large, $N^{2}\sim C_{T}\gg1$,\footnote{The central charge, $C_{T}$, is defined to be the normalization of $T$, i.e. $\<TT\>\sim C_{T}$. Although we do not assume the CFT has a Lagrangian description, we will use $N$ to count bulk loops.} so the bulk theory is weakly-coupled. We also use the dimension of the lightest, spin-four single-trace operator, $\Delta_{gap}$, as a proxy for the coupling.\footnote{As an example, in planar $\mathcal{N}=4$ SYM the relation between the `t Hooft coupling and the gap is $\sqrt{\lambda}\sim \Delta_{gap}^{2}$.} Throughout this work, our focus will be on the leading Regge behavior at each order in $1/N$.

The AdS dual of the Regge limit is high-energy, fixed impact parameter scattering, so it is natural to study this limit using bulk impact parameter space. That is, we study the bulk scattering amplitude, $\mathcal{B}(S,L)$, as a function of the center of mass energy, $\sqrt{S}$, and the impact parameter, $L$, on the transverse space $H_{d-1}$. We then want to understand if it exponentiates in the Regge limit:
\begin{align}
\mathcal{B}(S,L)\propto e^{i \delta(S,L)}, \label{eq:eikV0}
\end{align}
where $\delta(S,L)$ is the tree-level phase shift. We can also expand the phase shift at large $\Delta_{gap}$,
\begin{align}
\delta(S,L)=\sum\limits_{i=0}^{\infty} \delta^{(i)}(S,L)\Delta_{gap}^{-2i}.
\end{align}
In a theory of gravity plus matter of bounded spin, or when $\Delta_{gap}\rightarrow\infty$, the tree-level phase shift $\delta^{(0)}(S,L)$ is purely real. On the other hand, a tower of higher-spin particles can generate an imaginary piece for $\delta^{(1)}(S,L)$ at tree-level in $1/N$. 

Our final assumption is that the CFT contains a tower of higher-spin, single-trace operators with the following relation between their dimension and spin:
\begin{align}
j(\Delta)=2-2\frac{\Delta(4-\Delta)}{\Delta_{gap}^{2}} +O(\Delta_{gap}^{-4}).
\end{align}
This is the minimal ansatz consistent with locality, causality, and the existence of a flat-space limit \cite{Cornalba:2007fs,Maldacena:2015waa}. We will need this relation when making statements about the large $\Delta_{gap}$ expansion. 

Given these assumptions, our main result will be to show \eqref{eq:eikV0} is a good approximation in the limit $S\gg\Delta_{gap}^{2}\gg 1$ at fixed $L$ and to all orders in $1/N$. This agrees with the prediction from string theory in AdS \cite{Brower:2006ea,Brower:2007xg}, where exponentiation was also found to hold in this regime.

Let us now give a brief outline of the paper. In Section \ref{sec:AdSUnitarityRegge}, we review conformal Regge theory, the Lorentzian inversion formula, and the impact parameter transform \cite{Cornalba:2006xk,Cornalba:2006xm,Cornalba:2007zb,Cornalba:2007fs,Cornalba:2008qf,Cornalba:2009ax}. In Section \ref{sec:Regge_One_Lp}, we calculate the Regge limit at higher orders in $1/N$ using the inversion formula and demonstrate exponentiation for large $\Delta_{gap}$. In Section \ref{sec:MoreStringyCorrections}, we study how including new single and double-trace operators affects the Regge limit and their relation to inelastic effects in string theory \cite{Amati:1987wq,Amati:1987uf,Amati:1988tn,Giddings:2006vu,Shenker:2014cwa}. In Section \ref{sec:Discussion}, we conclude with a summary and possible future directions. Appendix \ref{app:Conv} contains technical details and definitions used in the paper. Appendix \ref{sec:Regge2d} contains results for $2d$ CFTs in the Regge limit, when we restrict to the global conformal group.

\section{Conformal Regge Theory and AdS/CFT}
\label{sec:AdSUnitarityRegge}
\subsection{Regge Limit and Analyticity in Spin}
\label{sec:RLimit_Trajectories}
We will start by reviewing conformal Regge theory \cite{Cornalba:2007fs,Costa:2012cb}. To make the connection to the Lorentzian inversion formula manifest \cite{Caron-Huot:2017vep,ssw}, we will follow the presentation given in \cite{Kravchuk:2018htv}. 

\begin{figure}[H]
\centering
\includegraphics[scale=.35]{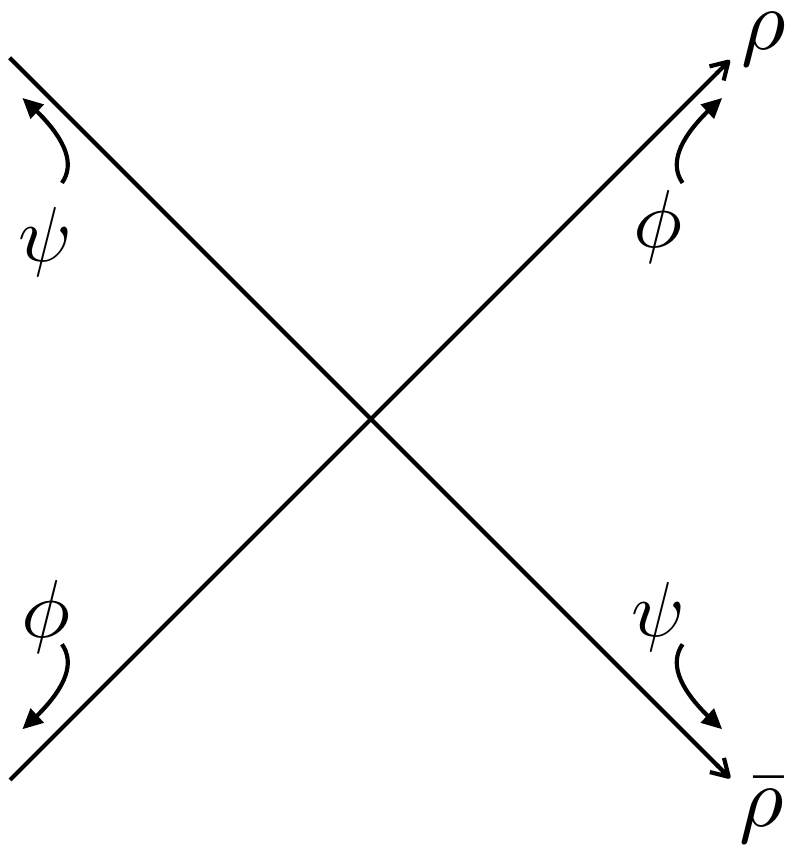}
\caption{The Regge limit is defined by sending the four operators to null infinity while keeping the spacelike distance between the pairs of identical operators fixed.}
\label{fig:CFTReggeKinematics}
\end{figure}

Our main object of study will be a correlation function of pairwise identical scalars:
\begin{align}
G(x_i)=\<\f(x_1)\f(x_2)\p(x_3)\p(x_4)\>.
\end{align}
To reach the Regge limit we need to work in Lorentzian signature. The physical picture is clearest in the lightcone coordinates $\rho, \bar{\rho}=x^{1}\pm x^{0}$. The Regge limit is defined by setting:
\begin{align}
x_{1}&= -x_{2}=(\sqrt{\rho},\sqrt{\bar{\rho}}),
\nonumber \\
x_{4}&=-x_{3}=(1/\sqrt{\rho},1/\sqrt{\bar{\rho}}),
\end{align}
and then taking $\rho,\bar{\rho}^{-1}\rightarrow \infty$ with $\rho\bar{\rho}$ held fixed\footnote{To keep track of the $i\epsilon$ prescriptions we should write $\rho=r e^{-t+i\epsilon}, \bar{\rho}=re^{t-i\epsilon}$ and then take $t\rightarrow \infty$.}, as shown in figure \ref{fig:CFTReggeKinematics}.

It is useful to phrase this in terms of the cross ratios $z$ and $\bar{z}$,
\begin{align}
z\bar{z}=\frac{x_{12}^{2}x_{34}^{2}}{x_{13}^{2}x_{24}^{2}}, \qquad (1-z)(1-\bar{z})=\frac{x_{14}^{2}x_{23}^{2}}{x_{13}^{2}x_{24}^{2}},
\end{align}
where the relation between the two variables is
\begin{align}
z=\frac{4\rho}{(1+\rho)^{2}}, \qquad \bar{z}=\frac{4\bar{\rho}}{(1+\bar{\rho})^{2}}.
\end{align}
Taking the Regge limit then corresponds to rotating $\bar{z}$ counterclockwise around a branch point at $\bar{z}=1$, keeping $z$ on the first sheet, and then sending $z,\bar{z}\rightarrow 0$ with $z/\bar{z}$ fixed. To see Regge growth, we need to look at the $s$-channel, or $\f\f \rightarrow \p\p$, expansion. For a generic four-point function of scalars the $s$-channel conformal block expansion takes the form: 
\begin{align}
&\<\f_1\f_2\f_3\f_4\>=T_{s}(x_i)\mathcal{G}(z,\bar{z}), \label{eq:CGDef}
\\
&\mathcal{G}(z,\bar{z})=\sum\limits_{\O}\left(-\frac{1}{2}\right)^{J_{\O}}c_{\f_1\f_2\O}c_{\f_3\f_4\O}g^{a,b}_{\O}(z,\bar{z}), \qquad a=-\frac{1}{2}\Delta_{12}, \ b=\frac{1}{2}\Delta_{34}, \label{eq:CBDecomp}
\end{align}
where $g^{a,b}_{\O}(z,\bar{z})$ are the conformal blocks, $\Delta_{ij}=\Delta_i-\Delta_j$, and $T_{s}(x_i)$ is a kinematic prefactor
\begin{align}
T_{s}(x_i)=\frac{1}{x_{12}^{\Delta_1+\Delta_2}x_{34}^{\Delta_3+\Delta_4}}\left(\frac{x_{14}}{x_{24}}\right)^{\Delta_{21}}\left(\frac{x_{14}}{x_{13}}\right)^{\Delta_{34}}.
\end{align}
The factors $c_{\f_1\f_2\O}c_{\f_3\f_4\O}$ are the OPE coefficients, see Appendix \ref{app:Conv} for our conventions. We will also need the $t$-channel expansion, which corresponds to $1\leftrightarrow 3$. We take this to act on the cross ratios as $(z,\bar{z})\rightarrow (1-\bar{z},1-z)$. Finally, we drop the superscripts for the blocks when $a=b=0$.

Now we find the Regge limit gives:
\begin{align}
g^{\circlearrowleft a,b}_{\Delta,J}(z,\bar{z})\sim (z\bar{z})^{\frac{1}{2}(1-J)}\left(\frac{z}{\bar{z}}\right)^{\frac{\Delta-J}{2}},
\end{align} 
where the arrow denotes the analytic continuation of $\bar{z}$ around 1. We see that in the Regge limit, it is the operators with the largest spin which dominate. Generically the sum over spin is unbounded, so we must first resum this expansion via a Sommerfeld-Watson transform. The first step is to define the $s$-channel conformal partial wave expansion for $\<\f\f\p\p\>$:
\begin{align}
\mathcal{G}(z,\bar{z})&=\sum\limits_{J=0}^{\infty}\int\limits_{-\infty}^{\infty}\frac{d\nu}{2\pi}c\left(\text{\large$\tfrac{d}{2}$}+i\nu,J\right)F_{\frac{d}{2}+i\nu,J}(z,\bar{z}), \label{eq:CPW_Decomp}
\\
F_{\Delta,J}(z,\bar{z})&\equiv\frac{1}{2}\left(g_{\Delta,J}(z,\bar{z})+\frac{S^{\f\f}_{\O}}{S^{\p\p}_{\widetilde{\O}}}g_{d-\Delta,J}(z,\bar{z})\right), \label{eq:CPW_Def}
\end{align}
where $F_{\Delta,J}(z,\bar{z})$ are the partial waves and $\widetilde{\O}$ is an operator with dimension $d-\Delta$ and spin $J_{\O}$. The shadow coefficients, $S^{\f_1\f_2}_{\O}$, are defined in \eqref{eq:ShadowDef}. The relation to the conformal block decomposition (\ref{eq:CBDecomp}) can be found by closing the $\nu$ contour in the lower half-plane for the first term in (\ref{eq:CPW_Def}) and in the upper half-plane for the second. Shadow symmetry of the OPE function guarantees they give the same result \cite{Costa:2012cb,Caron-Huot:2017vep} and we find:
\begin{align}
\res_{\text{\hspace{.08in}$\Delta=\Delta_{\O}$}}c(\Delta,J_{\O})=-\left(-\frac{1}{2}\right)^{J_{\O}}c_{\f\f\O}c_{\p\p\O}.
\end{align}
The Lorentzian inversion formula \cite{Caron-Huot:2017vep} gives a natural split for the OPE function into two terms,
\begin{align}
c(\Delta,J)=c^{t}(\Delta,J)+(-1)^{J}c^{u}(\Delta,J),
\end{align}
each of which can be analytically continued in spin.\footnote{The Lorentzian inversion formula is guaranteed to reproduce the Euclidean OPE data for $J>j_{0}$, where $j_0$ is the effective ``spin" controlling the Regge limit.} The functions $c^{t,u}(\Delta,J)$ come from the $t$ and $u$-channel contributions to the Lorentzian inversion formula \cite{Caron-Huot:2017vep},
\begin{align}
c^{t}(\Delta,J)&=\frac{\kappa_{\Delta+J}}{4}\int\limits_{0}^{1}\frac{dzd\bar{z}}{(z\bar{z})^{2}}\bigg|\frac{z-\bar{z}}{z\bar{z}}\bigg|^{d-2}g_{J+d-1,\Delta+1-d}(z,\bar{z})\dDisc_{t}[\mathcal{G}(z,\bar{z})],
\\
\kappa_{\beta}&=\frac{\Gamma\left(\frac{\beta}{2}-a\right)\Gamma\left(\frac{\beta}{2}+a\right)\Gamma\left(\frac{\beta}{2}-b\right)\Gamma\left(\frac{\beta}{2}+b\right)}{2\pi^{2}\Gamma(\beta-1)\Gamma(\beta)}.
\end{align}
When $a=b=0$, the $t$-channel double-discontinuity is defined by:
\begin{align}
\dDisc_{t}[\mathcal{G}(z,\bar{z})]=\mathcal{G}(z,\bar{z})-\frac{1}{2}\left(\mathcal{G}^{\circlearrowleft}(z,\bar{z})+\mathcal{G}^{\circlearrowright}(z,\bar{z})\right),
\end{align}
where the arrows indicate the continuation around $\bar{z}=1$ in a (counter-)clockwise fashion. The $u$-channel term is identical, $c^{u}(\Delta,J)=c^{t}(\Delta,J)$, since $\<\f\f\p\p\>$ is symmetric under $x_3\leftrightarrow x_4$.

\begin{figure}
\centering
\includegraphics[scale=.35]{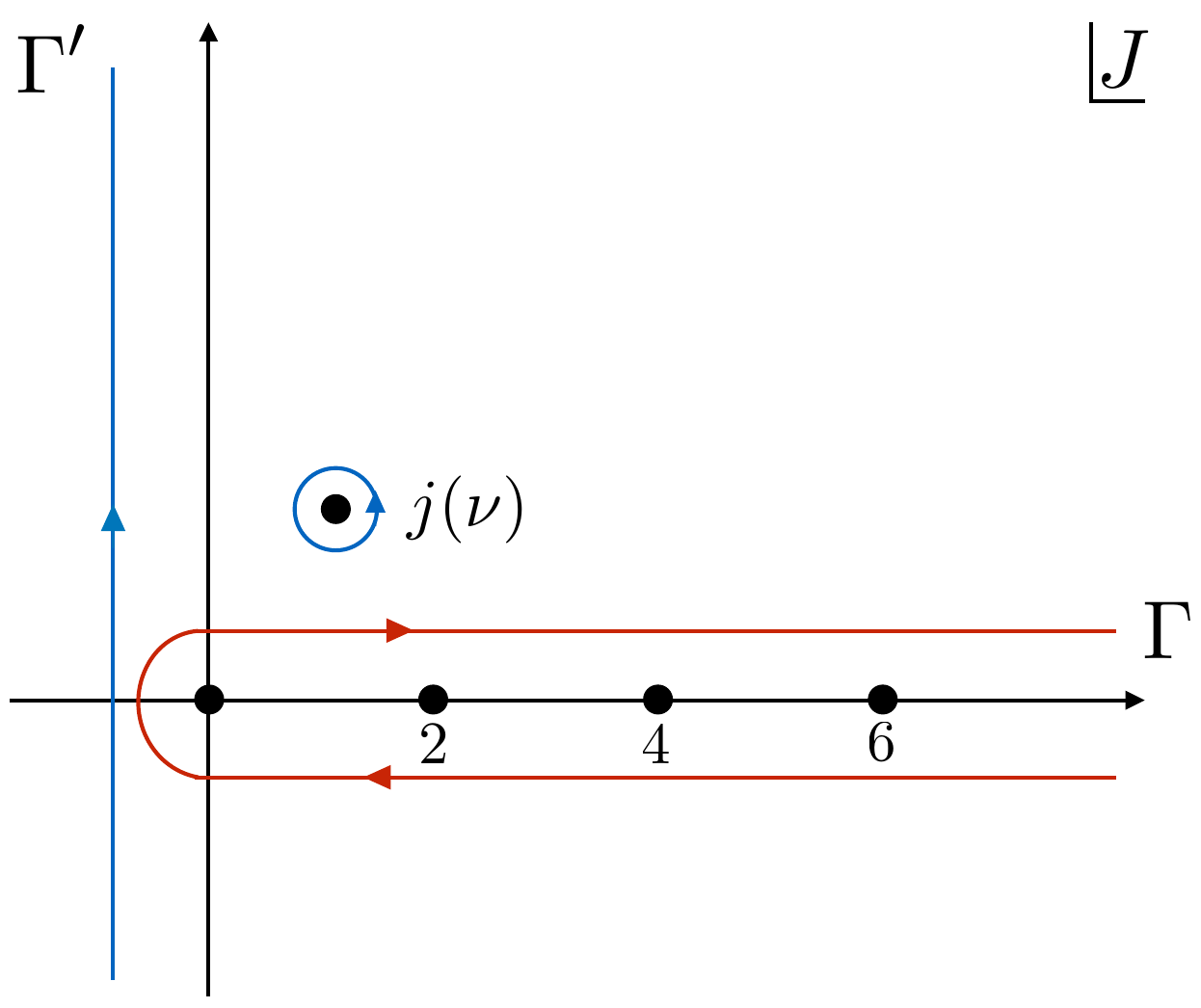}
\caption{The red contour $\Gamma$ comes from the original, partial wave decomposition. We deform it to the blue contour $\Gamma'$ and in the process pick up a pole, $j(\nu)$, which gives the leading Regge growth. We have suppressed other possible non-analyticities, such as branch cuts.}
\label{fig:Sommerfeld_watson}
\end{figure}

To perform the Sommerfeld-Watson transform we replace the sum over spin in (\ref{eq:CPW_Decomp}) with a contour integral: 
\begin{align}
\hspace{-.4in}\mathcal{G}(z,\bar{z})&=\sum\limits_{J=0}^{\infty}\int\limits_{-\infty}^{\infty} \frac{d\nu}{2\pi}c\left(\text{\large$\tfrac{d}{2}$}+i\nu,J\right)F_{\frac{d}{2}+i\nu,J}(z,\bar{z})
\\ &=-\int_{\Gamma}dJ\int\limits_{-\infty}^{\infty}  \frac{d\nu}{2\pi}\frac{1+e^{-i \pi J}}{1-e^{-2\pi i J}}c^{t}\left(\text{\large$\tfrac{d}{2}$}+i\nu,J\right)F_{\frac{d}{2}+i\nu,J}(z,\bar{z}),
\end{align}
where the contour $\Gamma$ is shown in figure \ref{fig:Sommerfeld_watson}. Finally, we push the contour $\Gamma$ to the left such that it runs over the contour $\Gamma'$. In the process we can pick up poles and branch cuts in the complex $J$ plane which determine the leading Regge behavior \cite{Cornalba:2007fs,Costa:2012cb,Kravchuk:2018htv}. 

In general, we do not know how $c^{t}(\Delta,J)$ behaves for complex $J$. However, for large $N$ CFTs we expect the leading, tree-level, contribution to the Regge limit comes from a pole at $J=j(\nu)$. This pole resums the exchange of the entire leading, single-trace Regge trajectory \cite{Cornalba:2007fs,Brower:2006ea,Brower:2007xg,Costa:2012cb,Costa:2013zra}.\footnote{In large $N$ CFTs we are interested in the single-trace trajectory, but the leading trajectory at finite $N$ is generically composed of double-twist operators \cite{Fitzpatrick:2012yx,Komargodski:2012ek}.} This trajectory is defined to be the set of even-spin, single-trace operators with the smallest dimension, $\Delta(J)$, for each spin. Therefore, in all CFTs it includes the stress-tensor $T$. The function $j(\nu)$ is the inverse of $\Delta(J)$:
\begin{align}
\nu^{2}+\left(\Delta(j(\nu))-d/2\right)^{2}=0.
\end{align}
With this assumption, the correlator in the Regge limit is:
\begin{align}
\mathcal{G}^{\circlearrowleft}(z,\bar{z})\approx -\int\limits_{-\infty}^{\infty}\frac{d\nu}{2\pi}\res_{J=j(\nu)}\frac{1+e^{-i\pi J}}{1-e^{-2\pi i J}}\frac{1}{\kappa_{\frac{d}{2}+i\nu+J}}c^{t}\left(\text{\large$\tfrac{d}{2}$}+i\nu,J\right)g_{1-J,1-\frac{d}{2}-i\nu}(z,\bar{z}), \label{eq:LeadingReggeTerm}
\end{align}
where the block $g_{1-J,1-\Delta}(z,\bar{z})$ appears after analytic continuation and taking the small $z,\bar{z}$ limit. The Regge behavior in (\ref{eq:LeadingReggeTerm}) comes from the exchange of the Pomeron, or a family of operators with dimension $\Delta=\frac{d}{2}+i\nu$ and spin $j(\nu)$, where $\nu$ is real.\footnote{Sometimes the Pomeron is identified with just the intercept point $\nu=0$, i.e. it has dimension $\Delta=\frac{d}{2}$ and spin $j(0)$. Here we will identify it with the entire $\nu$ integral.} For CFTs with a string theory dual the Pomeron is dual to a worldsheet vertex operator \cite{Brower:2006ea}. For general CFTs it can be identified with a generalized light-ray operator \cite{Kravchuk:2018htv}. For this work, we will simply use the term Pomeron as a shorthand for the contribution of the leading, single-trace trajectory to the Regge limit.

To compare with \cite{Costa:2012cb}, we use the small $z,\bar{z}$ behavior of the conformal block in (\ref{eq:LeadingReggeTerm}) \cite{Caron-Huot:2017vep}:
\begin{align}
g_{1-J,1-\Delta}(z,\bar{z})\approx \frac{4\pi^{\frac{d}{2}}\Gamma(\Delta-\frac{d}{2})}{\Gamma(\Delta-1)}(z\bar{z})^{\frac{1}{2}(1-J)}\Omega_{i\nu}\left(\frac{1}{2}\log(z/\bar{z})\right),\label{eq:gLRtoHarm}
\end{align}
where $\Delta=\frac{d}{2}+i\nu$ and $\Omega_{i\nu}(L)$ is the harmonic function on $H_{d-1}$, see \eqref{eq:HarmDefd}. In \cite{Costa:2012cb} they showed that the correlator in the Regge limit is
\begin{align}
\mathcal{G}^{\circlearrowleft}(z,\bar{z})\approx 2\pi i \int\limits_{-\infty}^{\infty}d\nu \alpha(\nu)(z\bar{z})^{\frac{1}{2}(1-j(\nu))}\Omega_{i\nu}\left(\frac{1}{2}\log(z/\bar{z})\right), \label{eqn:ConfReg}
\end{align}
so the conversion between the two is given by
\begin{align}
\alpha(\nu)=-\res_{J=j(\nu)}i\frac{ e^{i \pi  J}}{\left(1-e^{i \pi J}\right) }\frac{ \pi ^{\frac{d}{2}-2} \Gamma \left(i\nu\right)}{\Gamma (\frac{d}{2}+i\nu -1) \kappa_{\frac{d}{2}+i\nu +J}}c^{t}\left(\text{\large$\tfrac{d}{2}$}+i\nu,J\right). 
\end{align}
\subsection{Crossing Symmetry at Tree-Level}
\label{ssec:Tree_Crossing}
Next, let us review how to solve crossing in the Regge limit at tree-level in $1/N$ \cite{Li:2017lmh}. We will set $d=4$, since then we know the blocks in closed form  \cite{DO1,DO2,DO3}, see \eqref{eq:4dBlocks}. We will use the results of this section as the input for the loop-level computations in Section \ref{sec:Regge_One_Lp}.

Crossing symmetry on the first, or Euclidean, sheet equates the $s$ and $t$-channel OPEs:
\begin{align}
\left(z\bar{z}\right)^{-\Delta_{\p}}((1-z)(1-\bar{z}))^{\frac{1}{2}(\Delta_{\f}+\Delta_{\p})}&\sum\limits_{\O}\left(-\frac{1}{2}\right)^{J_{\O}}c_{\f\f\O}c_{\p\p\O}g_{\O}(z,\bar{z})
\nonumber \\
=&\sum\limits_{\O'}\left(-\frac{1}{2}\right)^{J_{\O'}}c_{\f\p\O'}c_{\p\f\O'}g^{a,a}_{\O'}(1-z,1-\bar{z}), \label{eq:crossing_Euclidean}
\end{align}
where $a=\frac{1}{2}(\Delta_{\f}-\Delta_{\p})$. In the $t$-channel, or $\f\p\rightarrow \p\f$ expansion, we will use $c_{\f\p\O}=(-1)^{J_\O}c_{\p\f\O}$ and adopt the notation:
\begin{align}
P_{\O}=\frac{1}{2^{J_{\O}}}c_{\f\p\O}^{2}.
\end{align}

If we analytically continue around $\bar{z}=1$ and then take $z,\bar{z}$ small, we need to use conformal Regge theory for the first line of (\ref{eq:crossing_Euclidean}). For the second line, we can do the analytic continuation term by term, as each $t$-channel block just picks up a phase.\footnote{Recall the blocks have the expansion $g_{\Delta,J}(z,\bar{z})=\sum\limits_{n,m}c_{n,m}z^{\frac{\Delta-J}{2}+n}\bar{z}^{\frac{\Delta+J}{2}+m}+z\leftrightarrow\bar{z}$ where $n,m$ are integers so sending $z\rightarrow ze^{2\pi i}$ gives the overall factor $e^{i\pi(\Delta-J)}$.} Furthermore, at leading order in $1/N$ the $t$-channel sum contains both double-trace operators, $[\f\p]_{n,J}$, and generic single-trace operators, which we label as $\mathcal{X}_{\Delta,J}$. The double-trace operators have the form:
\begin{align}
[\f\p]_{n,J}\approx \f\partial^{\mu_1}...\partial^{\mu_J}\partial^{2n}\p-traces,
\end{align}
and classical dimension $\Delta^{(0)}_{n,J}=\Delta_\f+\Delta_\p+2n+J$. For the $t$-channel double-trace operators it will be convenient to introduce the variables
\begin{align}
h=\frac{1}{2}(\Delta-J), \qquad \bar{h}=\frac{1}{2}(\Delta+J).
\end{align}

Next, we will take (\ref{eq:crossing_Euclidean}), expand to tree-level in $1/N$, and then go to the Regge limit. We write the $1/N$ expansion for the four-point function and double-trace OPE data as,
\begin{align}
\mathcal{G}(z,\bar{z})&=\mathcal{G}^{(0)}(z,\bar{z})+\frac{1}{N^{2}}\mathcal{G}^{(1)}(z,\bar{z})+...
\\
\Delta_{h,\bar{h}}&= \Delta^{(0)}_{h,\bar{h}} +\frac{1}{N^{2}}\gamma^{(1)}_{h,\bar{h}}+...
\\ 
P_{h,\bar{h}}&= P^{MFT}_{h,\bar{h}}\left(1+\frac{1}{N^{2}}\delta P^{(1)}_{h,\bar{h}}+...\right),
\end{align}
where the superscript denotes the order in $1/N$. Here $P^{MFT}_{h,\bar{h}}$ are the mean field theory (MFT) OPE coefficients which reproduce $\mathcal{G}^{(0)}(z,\bar{z})$, or identity exchange in the $s$-channel, from the $t$-channel expansion \cite{Heemskerk:2009pn,Fitzpatrick:2011dm}. For later convenience, we have labeled the double-trace OPE data using $h$ and $\bar{h}$. 

Expanding in $1/N$ and taking the Regge limit yields the tree-level, Regge, crossing equation:
\begin{align}
\hspace{-.5cm}
e^{i \pi (\Delta_{\f}+\Delta_{\p})}&(z\bar{z})^{-\Delta_{\f}}\mathcal{G}^{(1)\circlearrowleft}(z,\bar{z}) \nonumber\\
\approx & \, e^{i\pi(\Delta_{\f}+\Delta_{\p})}\sum_{h,\bar{h}}P^{MFT}_{h,\bar{h}}\left[ \gamma^{(1)}_{h,\bar{h}}\left(i\pi+\frac{1}{2}(\partial_{h}+\partial_{\bar{h}})\right)+ \delta P^{(1)}_{h,\bar{h}}\right]g^{a,a}_{h,\bar{h}}(1-z,1-\bar{z}) \nonumber \\ &+\sum_{\mathcal{X}}e^{2i\pi h_{\mathcal{X}}}P^{(1)}_{\mathcal{X}}g^{a,a}_{\mathcal{X}}(1-z,1-\bar{z}).
\label{eq:treeCrossing}
\end{align}
The leading behavior for the first line will come from Pomeron exchange, see (\ref{eqn:ConfReg}). We will give the tree-level results for $\alpha(\nu)$ and $j(\nu)$ in \eqref{eq:alpha_def_Regge} and \eqref{eq:jlargeN}. The second line includes corrections to the double-trace operators and the third line runs over all single-trace operators in the $\f\times\p$ OPE.

We see from (\ref{eqn:ConfReg}) that $\mathcal{G}^{(1)\circlearrowleft}(z,\bar{z})$ grows like $(z\bar{z})^{\frac{1}{2}(1-j(0))}$ in the Regge limit, where we have approximated the integral by its behavior around $\nu=0$. If $j(0)>1$, then this is an enhanced divergence with respect to identity exchange. The single-trace blocks generically sum out of phase, so they cannot reproduce this Regge behavior. On the other hand, the double-trace blocks add in phase and have a chance of producing this divergence. In fact the double-trace sum is dominated by the regime \cite{Cornalba:2006xm,Cornalba:2007zb}:
\begin{align}
z^{-\frac{1}{2}}\sim\bar{z}^{-\frac{1}{2}}\sim h\sim\bar{h}\gg1. \label{eq:regionhhb}
\end{align}
Therefore, the divergence is reproduced by the large $h$ and $\bar{h}$ asymptotics of the double-trace data, where we keep the ratio $h/\bar{h}$ arbitrary. We can then ignore terms with $(\partial_{h}+\partial_{\bar{h}})$, since they will give a subleading effect. 

In the limit \eqref{eq:regionhhb}, the $4d$, $t$-channel blocks and MFT OPE coefficients are \cite{Fitzpatrick:2012yx,Komargodski:2012ek,Li:2017lmh}:
\begin{align}
g^{a,b}_{h,\bar{h}}(1-z,1-\bar{z}) \ \approx \ & \frac{\sqrt{h\bar{h}}}{\pi}2^{2(h+\bar{h}-1)}\frac{1}{\bar{z}-z}K_{a+b}(2h\sqrt{z})K_{a+b}(2\bar{h}\sqrt{\bar{z}})\left(z\bar{z}\right)^{-\frac{1}{2}(a+b)}+(z\leftrightarrow \bar{z}), \label{eq:g4dBessel}
\\
P^{MFT}_{h,\bar{h}}\ \approx \ &\frac{\pi  2^{6-2 (h+\bar{h})} \left(\bar{h}^2-h^2\right) h^{\Delta_{\f}+\Delta_{\p}-\frac{7}{2}} \bar{h}^{\Delta_{\f}+\Delta_{\p}-\frac{7}{2}}}{\Gamma (\Delta_{\f}-1) \Gamma (\Delta_{\f}) \Gamma (\Delta_{\p}-1) \Gamma (\Delta_{\p})}.
\label{eq:PMFT4d}
\end{align}
To solve crossing we will use the explicit form of the $4d$ harmonic function\footnote{We will label the harmonic functions by the dimension of the boundary spacetime.}
\begin{align}
\Omega_{i\nu}(L)=\frac{\nu \sin(\nu L)}{4\pi^{2}\sinh(L)}.
\end{align} 
Then, as shown in \cite{Li:2017lmh}, crossing symmetry implies the double-trace OPE data can be written as a spectral integral over $\Omega_{i\nu}(L)$: 
\begin{align}
\frac{1}{N^{2}}\gamma^{(1)}_{h,\bar{h}}\ \approx \ &2\Gamma(\Delta_\f-1)\Gamma(\Delta_\f)\Gamma(\Delta_\p-1)\Gamma(\Delta_\p)
\nonumber \\ & Re \int\limits_{-\infty}^{\infty} d\nu \frac{\alpha(\nu)}{\chi_{j(\nu)}(\nu)\chi_{j(\nu)}(-\nu)}(h\bar{h})^{j(\nu)-1}\Omega_{i\nu}(\log(h/\bar{h})), \label{eq:treegamma}
\\[7.5pt]
\frac{1}{N^{2}}\delta P^{(1)}_{h,\bar{h}} \ \approx  &-2\pi\Gamma(\Delta_\f-1)\Gamma(\Delta_\f)\Gamma(\Delta_\p-1)\Gamma(\Delta_\p)
\nonumber \\ & Im\int\limits_{-\infty}^{\infty} d\nu \frac{\alpha(\nu)}{\chi_{j(\nu)}(\nu)\chi_{j(\nu)}(-\nu)}(h\bar{h})^{j(\nu)-1}\Omega_{i\nu}(\log(h/\bar{h})) \label{eq:treedP}.
\end{align}
The function $\chi_{j}(\nu)$ in general $d$ is defined as
\begin{align}
\chi_{j}(\nu)=\Gamma\left(\frac{1}{2}(2\Delta_{\f}+j+i\nu-\frac{d}{2})\right)\Gamma\left(\frac{1}{2}(2\Delta_{\p}+j+i\nu-\frac{d}{2})\right).\label{eq:chiDef}
\end{align}
To check this result, we first approximate the sum over $h$ and $\bar{h}$ in the second line of (\ref{eq:treeCrossing}) by an integral
\begin{align}
\sum\limits_{h,\bar{h}} \ \Rightarrow \ \int\limits_{0}^{\infty} d\bar{h} \int\limits_{0}^{\bar{h}}dh=\frac{1}{2}\int\limits_{0}^{\infty} d\bar{h} \int\limits_{0}^{\infty}dh.
\end{align}
We used the symmetry of the integrand under $h\leftrightarrow \bar{h}$ to extend both integrals over $(0,\infty)$. Finally to perform the integrals, given the approximation (\ref{eq:g4dBessel}), we use:
\begin{align}
\int\limits_{0}^{\infty}dh \hspace{.1cm} h^{a}K_{b}(2h\sqrt{z})=\frac{1}{4} z^{-\frac{a}{2}-\frac{1}{2}} \Gamma \left(\frac{1}{2} (a-b+1)\right) \Gamma \left(\frac{1}{2} (a+b+1)\right).\label{eq:KInt}
\end{align}
It is then straightforward to check that our results for the double-trace OPE data, (\ref{eq:treegamma}) and (\ref{eq:treedP}), when plugged into the crossing equation (\ref{eq:treeCrossing}), reproduce the tree-level Regge behavior given by (\ref{eqn:ConfReg}) \cite{Li:2017lmh}.

To interpret this result we need more information about the tree-level Regge limit in large $N$ theories. This problem was studied in \cite{Costa:2012cb} where they found
\begin{align}
\alpha(\nu)=i\frac{e^{i\pi j(\nu)}}{1-e^{i\pi j(\nu)}}\pi^{\frac{d}{2}-1}2^{j(\nu)-1}\chi_{j(\nu)}(\nu)\chi_{j(\nu)}(-\nu)\frac{\pi}{2\nu}j'(\nu)K_{\frac{d}{2}+i\nu,j(\nu)}c_{\f\f j(\nu)}c_{\p\p j(\nu)}. \label{eq:alpha_def_Regge}
\end{align}
Here $c_{\f\f j(\nu)}c_{\p\p j(\nu)}$ is the coupling of the external scalars to the leading trajectory, parameterized by $j(\nu)$. The product $\chi_{j(\nu)}(\nu)\chi_{j(\nu)}(-\nu)$ has poles corresponding to the double-trace operators, $[\f\f]_{n,j(\nu)}$ and $[\p\p]_{n,j(\nu)}$, which have also been analytically continued in spin. These poles have been projected out in (\ref{eq:treegamma}) and (\ref{eq:treedP}). The function $K$ is defined in \eqref{eq:KDef}, but we will not need its explicit form. 

In large $N$ theories, we also know the form of the function $j(\nu)$ for $\Delta_{gap}\gg1$ \cite{Cornalba:2007fs,Costa:2012cb}:
\begin{align}
j(\nu)=2-2\frac{d^{2}/4+\nu^{2}}{\Delta_{gap}^{2}}+O(\Delta_{gap}^{-4}).\label{eq:jlargeN}
\end{align}
This is the minimal answer consistent with having $j(\nu)\rightarrow 2$ as $\Delta_{gap}\rightarrow \infty$, conservation of the stress-tensor, $j(id/2)=2$, and the existence of the flat space limit, e.g. $j(\nu)$ is finite when $\nu\sim\Delta_{gap}\gg1$. Requiring that we have linear trajectories in the flat space limit \cite{Caron-Huot:2016icg} also prevents us from multiplying by additional powers of $\nu/\Delta_{gap}$. Shadow symmetry also implies $j(\nu)$ is an even function of $\nu$. Finally, causality tells us that in a large $N$ CFT we must have $j(0)\leq2$ \cite{Camanho:2014apa,Maldacena:2015waa,Caron-Huot:2017vep}, and this fixes the sign of the leading correction. This answer for $j(\nu)$ was also found by studying string theory in AdS \cite{Brower:2006ea}.

We can also observe that $j(\nu)$ has a maximum at $\nu=0$. This implies we can approximate the $\nu$ integral in (\ref{eqn:ConfReg}) by a saddle point approximation in the limit $z,\bar{z}\rightarrow0$ and see explicitly the Regge growth is controlled by the $\nu=0$ point. Finally, in the limit $\Delta_{gap}\rightarrow\infty$, where $j(\nu)\rightarrow 2$, our results for the anomalous dimensions and OPE coefficients, \eqref{eq:treegamma} and \eqref{eq:treedP}, reproduce the pure gravity result \cite{Cornalba:2006xm,Cornalba:2007zb,Li:2017lmh}. 
\subsection{Impact Parameter Transform}
\label{sec:Impact_Transform}
In this section we will review the impact parameter transform \cite{Cornalba:2006xm,Cornalba:2007zb,Cornalba:2007fs,Cornalba:2008qf,Cornalba:2009ax}. This transform makes manifest the bulk, physical picture of high-energy scattering. Moreover, the fact we observe exponentiation after performing the impact parameter transform makes it clear that this is the most natural space to study loop corrections at high energies.

\begin{figure}[H]
\centering
\includegraphics[scale=.35]{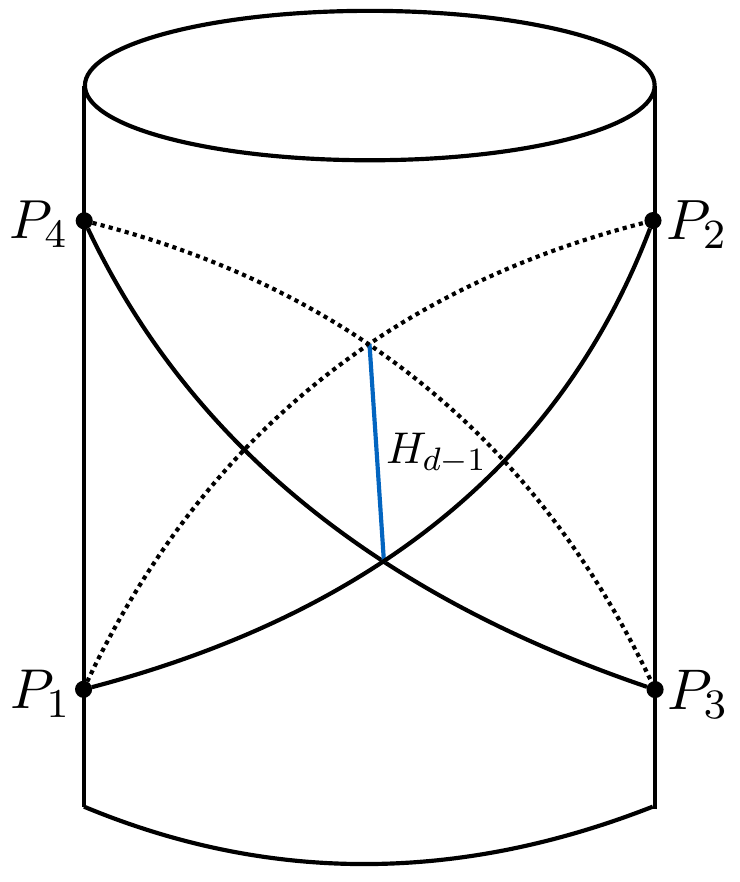}
\caption{High-energy, fixed impact parameter scattering in AdS. The boundary operators are centered around $P_i$, the centers of different Poincar\'e patches. The impact parameter space is the spatial $H_{d-1}$ slice, drawn in blue.}
\label{fig:Regge_Kinematics_AdS}
\end{figure}

For the four-point function $\<\f\f\p\p\>=T_{s}(x_i)\mathcal{G}(z,\bar{z})$, the impact parameter correlator, $\mathcal{B}(S,L)$, is defined by:
\begin{align}
(z\bar{z})^{-\Delta_{\f}}\mathcal{G}^{\circlearrowleft}(z,\bar{z})=(-1)^{-\Delta_{\f}-\Delta_{\p}}\int\limits_{V^{+}} dp d\bar{p}e^{-2i(p\cdot x +\bar{p}\cdot \bar{x})}\frac{\mathcal{B}(S,L)}{(-p^{2})^{\frac{d}{2}-\Delta_{\f}}(-\bar{p}^{2})^{\frac{d}{2}-\Delta_{\p}}}.
\end{align}
The vectors $x,p$, and their barred versions are real vectors in Minkowski space. The integration for $p$ and $\bar{p}$ is over the future light-cone $V^{+}$, where we take the metric to be mostly plus. The relation to the cross ratios and AdS impact parameter variables is
\begin{align}
z\bar{z}=x^{2}\bar{x}^{2}, \qquad z+\bar{z}=-2x\cdot\bar{x},
\nonumber \\
S=|p||\bar{p}|, \qquad \cosh L=-\frac{p\cdot \bar{p}}{|p||\bar{p}|}.
\end{align}
Here $\sqrt{S}$ is the energy of the bulk scattering process and $L$ is the distance on the transverse, impact parameter space, $H_{d-1}$, see figure \ref{fig:Regge_Kinematics_AdS}. The Regge limit is now $S\rightarrow\infty$ with $L$ held fixed.
\noindent If the position space correlator takes the form (\ref{eqn:ConfReg}), then the impact parameter transform yields:
\begin{align}
\mathcal{B}(S,L)=2\pi i \int\limits_{-\infty}^{\infty} d\nu \beta(\nu)S^{j(\nu)-1}\Omega_{i\nu}(L), \label{eqn:BImpactRegge}
\\
\beta(\nu)=\frac{4\pi^{2-d}}{\chi_{j(\nu)}(\nu)\chi_{j(\nu)}(-\nu)}\alpha(\nu). \label{eqn:BetaDef}
\end{align}
The bulk eikonal approximation says that in the $1/N$ expansion, the correlator in impact parameter space should exponentiate:
\begin{align}
\mathcal{B}_{eik}(S,L)=\mathcal{N}e^{i\delta(S,L)}, \label{eq:EikForm}
\end{align}
where the constant is given by
\begin{align}
\mathcal{N}=\frac{4\pi^{2-d}}{\Gamma(\Delta_{\f})\Gamma(\Delta_{\f}+1-\frac{d}{2})\Gamma(\Delta_{\p})\Gamma(\Delta_{\p}+1-\frac{d}{2})}.
\label{eq:NConst}
\end{align}
If we expand (\ref{eq:EikForm}) to tree level in $1/N$ and match to (\ref{eqn:ConfReg}), (\ref{eq:treegamma}), and (\ref{eq:treedP}) we can make the identifications:
\begin{align}
&Re \ \delta(S,L)  \ = \ \frac{1}{N^2} \pi \gamma^{(1)}_{h,\bar{h}}, \label{eq:redeltaanom}
\\
&Im \ \delta(S,L) \ = \ -\frac{1}{N^2} \delta P^{(1)}_{h,\bar{h}}, \label{eq:imdeltadp}
\end{align}
where $S=h\bar{h}$ and $L= \log(\bar{h}/h)$. This identification says studying $\gamma^{(1)}_{h,\bar{h}}$ and $\delta P^{(1)}_{h,\bar{h}}$ in the limit $h\bar{h}\gg1$, with $h/\bar{h}$ held fixed, probes the Regge limit in the bulk. 

The main goal of this paper is to understand: to what extent does (\ref{eq:EikForm}), with the identifications (\ref{eq:redeltaanom}) and (\ref{eq:imdeltadp}), provide a good approximation for the full correlator? In other words, when does the tree-level result exponentiate in AdS at large $S$? We will show exponentiation is universal in the limit $S \gg \Delta_{gap}^{2}\gg1$ and to all orders in $1/N$. 

This matches the results of \cite{Brower:2006ea,Brower:2007xg} where they proved exponentiation in the same limit by studying string theory in AdS. To see this, we first note that the function $j(\nu)$ at large $\Delta_{gap}$, \eqref{eq:jlargeN}, matches the result from string theory \cite{Brower:2006ea,Brower:2007xg}. Then we expand $\mathcal{B}(S,L)$ to tree-level in $1/N$ and evaluate the $\nu$ integral at large $S$ by a saddle-point approximation. The final answer is determined by an overall constant given by the OPE coefficients. Therefore, the tree-level result already has the expected stringy behavior and the non-trivial problem is proving exponentiation from the boundary CFT.
\subsection{Unitarity Method for AdS/CFT}
\label{ssec:Unitarity}
To determine the correlator at one-loop and higher, we will use an AdS/CFT unitarity method. In particular, we will use crossing symmetry of the boundary CFT \cite{Aharony:2016dwx} to bootstrap the correlator at all loops from tree-level data.\footnote{For other work on computing loops in AdS see \cite{Giombi:2017hpr,Cardona:2017tsw,Yuan:2017vgp,Yuan:2018qva,Liu:2018jhs,Bertan:2018afl,Bertan:2018khc,Ponomarev:2019ofr,Carmi:2019ocp,Meltzer:2019nbs}.}

\begin{figure}
\centering
\includegraphics[scale=.35]{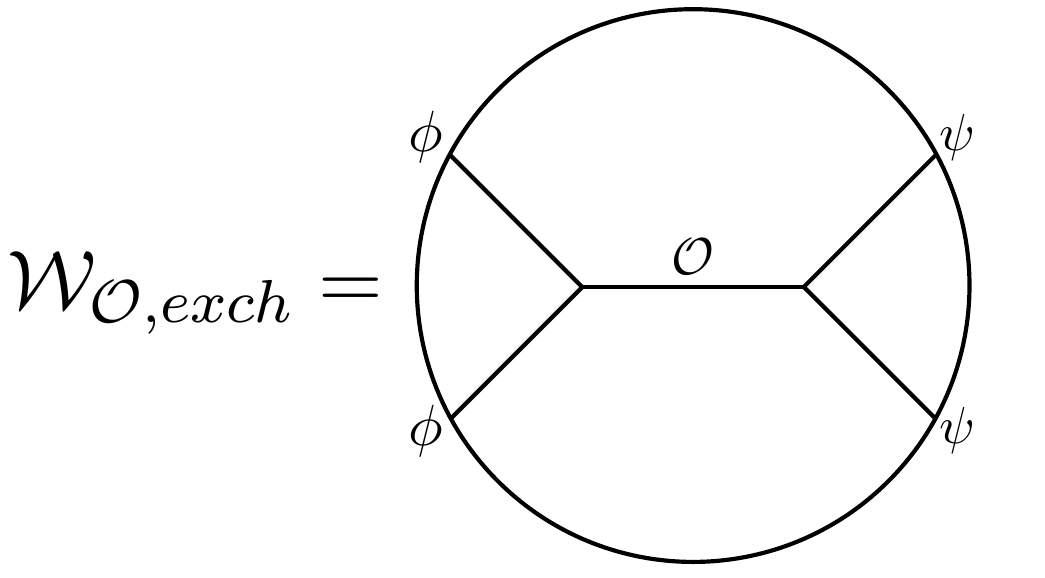}
\caption{Exchange Witten diagram where we label the bulk field by its dual CFT operator.}
\label{fig:Exchange_Diagram}
\end{figure}

To explain this procedure, let us consider the exchange Witten diagram, $\mathcal{W}_{\O,exch}(x_i)$, with $\O$ a scalar, as shown in figure \ref{fig:Exchange_Diagram}. Expanding this diagram in the $\f\p \rightarrow \p\f$ OPE channel, we can determine tree-level corrections to the spectrum and couplings of the double-trace operators $[\f\p]_{n,J}$, that is $\gamma^{(1)}_{h,\bar{h}}$ and $\delta P^{(1)}_{h,\bar{h}}$. Both are non-zero for unbounded spin, $J=\bar{h}-h$ \cite{Fitzpatrick:2012yx,Komargodski:2012ek,Alday:2017gde}.

To study loop corrections, we can plug these results into the inversion formula, which we reproduce here for $\<\f\f\p\p\>$,
\begin{align}
c(\Delta,J)&=\left(1+(-1)^{J}\right)c^{t}(\Delta,J),
\\
c^{t}(\Delta,J)&=\frac{\kappa_{\Delta+J}}{4}\int\limits_{0}^{1}\frac{dzd\bar{z}}{(z\bar{z})^{2}}\bigg|\frac{z-\bar{z}}{z\bar{z}}\bigg|^{d-2}g_{J+d-1,\Delta+1-d}(z,\bar{z})\dDisc_{t}[\mathcal{G}(z,\bar{z})].\label{eq:InvFormPt2}
\end{align}
The function $\dDisc_{t}[\mathcal{G}(z,\bar{z})]$ has a simple expansion in $t$-channel blocks,
\begin{align}
\dDisc_{t}[\mathcal{G}(z,\bar{z})]=\frac{(z\bar{z})^{\Delta_{\f}}}{\left[(1-z)(1-\bar{z})\right]^{\frac{\Delta_{\f}+\Delta_{\p}}{2}}}\sum\limits_{\O}2P_{\O}\sin^{2}\left(\frac{\pi}{2}(2h_{\O}-\Delta_\f-\Delta_\p)\right)g^{a,a}_{\O}(1-z,1-\bar{z}), \label{eq:dDiscExp}
\end{align}
where the prefactor comes from our definition of $\mathcal{G}(z,\bar{z})$, see (\ref{eq:CGDef}). The $\dDisc_{t}$ produces the $\sin^{2}$ factors and these ensure the double-trace operators contribute starting at order $N^{-4}$ with their MFT OPE coefficients,
\begin{align}
\dDisc_{t}[\mathcal{G}^{(2)}(z,\bar{z})]\supset \sum\limits_{h,\bar{h}}\frac{\pi^{2}}{2}\left(\gamma^{(1)}_{h,\bar{h}}\right)^{2}P^{MFT}_{h,\bar{h}}g^{a,a}_{h,\bar{h}}(1-z, 1-\bar{z}). \label{eq:dDiscSqAnom}
\end{align}
Once we have this double-discontinuity, we can perform the inversion integral in \eqref{eq:InvFormPt2} to determine one-loop corrections to the OPE data. These can then be used to determine the full correlator at one-loop \cite{Bissi:2019kkx,Carmi:2019cub}. 

\begin{figure}
\centering
\includegraphics[scale=.35]{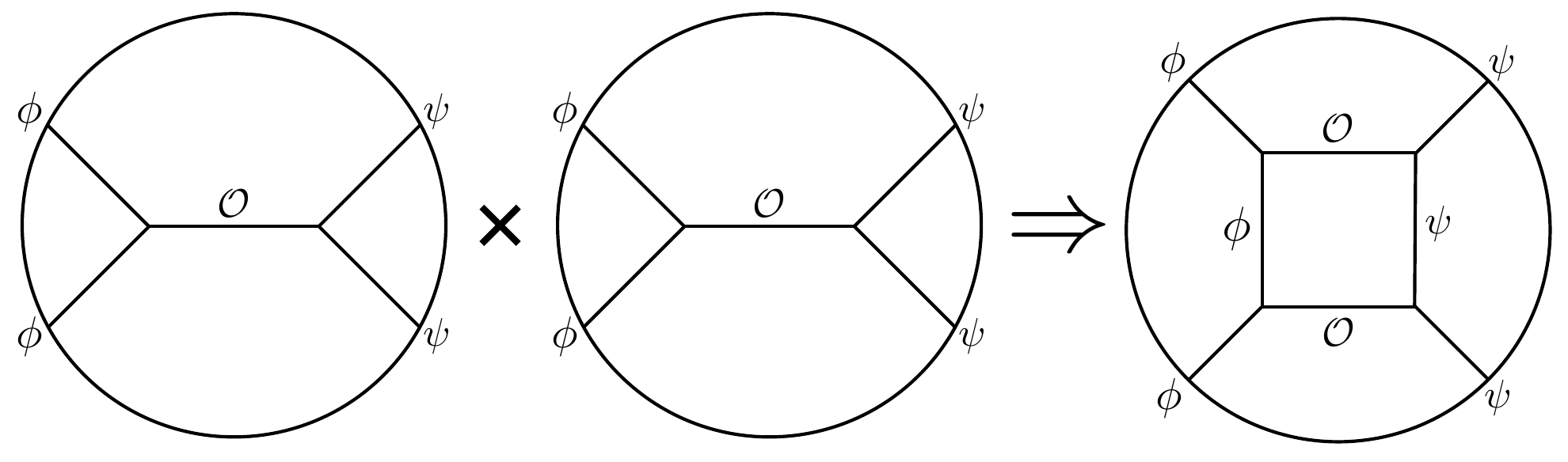}
\caption{We glue two, tree-level Witten diagrams along the $\f$ and $\p$ lines to create a box diagram.}
\label{fig:Gluing_Trees}
\end{figure}

It is also useful to understand how this approach maps to the standard loop expansion for Witten diagrams. If we use the anomalous dimensions $\gamma^{(1)}_{h,\bar{h}}$ due to the exchange Witten diagram $\mathcal{W}_{\O,exch}(x_i)$ in the expansion (\ref{eq:dDiscSqAnom}), then the resulting sum is equal to the double-discontinuity of a one-loop box diagram. The full box diagram and its OPE expansion is then determined via the inversion formula. Diagrammatically, we are gluing two tree-level diagrams in the bulk by leveraging crossing symmetry on the boundary, see figure \ref{fig:Gluing_Trees}. In the bulk, the action of $\dDisc_{t}$ can be visualized as projecting onto internal, horizontal line cuts, or the $[\f\p]_{n,J}$ double-trace operators \cite{Meltzer:2019nbs}.

In this work however, we are studying the Pomeron, which is not an integer spin, local operator. We can follow the same procedure, using the anomalous dimensions due to Pomeron exchange in (\ref{eq:dDiscSqAnom}), and ask: does this process compute the Pomeron box diagram studied in \cite{Brower:2007xg}?\footnote{In string theory, the Pomeron box is one term which contributes to the Regge limit of the torus diagram.} We will find agreement only to leading order at large $\Delta_{gap}$. At higher orders in $1/\Delta_{gap}$ there are corrections to the full correlator due to tidal excitations which are not included in the bulk or boundary analysis. Therefore, we are only justified in comparing the leading order terms of each method. Nevertheless, it is interesting to ask: how do we reproduce the Pomeron box calculated in \cite{Brower:2007xg} from the AdS/CFT unitarity method? As we will show in Section \ref{sec:MoreStringyCorrections}, this will require imposing a one-loop sum rule on the single-trace operators.\footnote{We thank Eric Perlmutter for discussions on this topic.}

\section{Regge Limit at One Loop and Higher}
\label{sec:Regge_One_Lp}
In this section we will prove exponentiation holds in the limit $S \gg \Delta_{gap}^{2}\gg1$. We will first give a brief summary of our method. In Section \ref{ssec:InvReggeLimit} we show how to determine the Regge limit at one-loop by solving a simplified inversion problem. Our main assumption is that the double-discontinuity at one-loop is determined by the exchange of the $[\f\p]_{n,J}$ operators. In Section \ref{ssec:Double_Inv} we solve the crossing symmetry equations at one-loop. We show that to solve the one-loop crossing at large $\Delta_{gap}$, we only need to include tree-level corrections to the double-trace operators $[\f\p]_{n,J}$. In Section \ref{ssec:All_Loop} we prove analogous statements carry over to all orders in $1/N$. Finally, in Section \ref{ssec:Eikonalization} we use these results to prove exponentiation in impact parameter space. The fact we only need tree-level corrections to solve crossing at large $\Delta_{gap}$ maps exactly to the statement that the correlator in impact parameter space exponentiates at high energies.

\subsection{Inversion Formula in the Regge Limit}
\label{ssec:InvReggeLimit}
To study the one-loop correlator, we will follow a procedure similar to the one outlined in the previous section. That is, we will use tree-level data in the inversion formula to determine $1/N^{4}$ corrections.

We claim that to determine the one-loop correlator in the Regge limit, it is sufficient to know $\dDisc_{t}[\mathcal{G}(z,\bar{z})]$ in the limit $z,\bar{z}\rightarrow 0$. We will call this the Regge limit for the double-discontinuity.\footnote{Strictly speaking, the Regge limit is defined on the second-sheet of the full correlator. However, this limit of the double-discontinuity has qualitatively similar features, so we will also call it a Regge limit. It has also been studied in \cite{Kologlu:2019bco,Kologlu:2019mfz}.} Determining the Regge limit at higher-loops is then a simpler problem than determining the full correlator as we only have to know $\dDisc_{t}[\mathcal{G}(z,\bar{z})]$ in a certain limit.

To start, the inversion formula in $d=4$ for $\<\f\f\p\p\>$ is,
\begin{align}
c^{t}(\Delta,J)&=\frac{\kappa_{\Delta+J}}{4}\int\limits_{0}^{1}\frac{dzd\bar{z}}{(z\bar{z})^{2}}\bigg(\frac{z-\bar{z}}{z\bar{z}}\bigg)^{2}g_{J+3,\Delta-3}(z,\bar{z})\dDisc_{t}[\mathcal{G}(z,\bar{z})].\label{eq:invd4}
\end{align}
To illustrate the logic, let us consider some simple scalings for $\dDisc_{t}[\mathcal{G}(z,\bar{z})]$:
\begin{align}
z\ll 1 &\rightarrow \dDisc_{t}[\mathcal{G}(z,\bar{z})]\sim z^{h_{*}},
\\
z\sim\bar{z}\ll 1&\rightarrow \dDisc_{t}[\mathcal{G}(z,\bar{z})]\sim(z\bar{z})^{\frac{1}{2}(1-j_{*})},
\end{align}
for some $h_{*}$ and $j_{*}$. These two limits are the lightcone and Regge limits for the double-discontinuity, respectively. In the same limits, the block in the integrand scales like:
\begin{align}
z\ll 1 &\rightarrow g_{J+3,\Delta-3}(z,\bar{z})\sim z^{\frac{1}{2}(J-\Delta)+3},
\\
z\sim\bar{z}\ll 1 &\rightarrow g_{J+3,\Delta-3}(z,\bar{z})\sim (\sqrt{z\bar{z}})^{J+3}.
\end{align}
If we perform the inversion formula in the limit $z\ll1$, we see a pole at fixed  $h=\frac{1}{2}(\Delta-J)$:
\begin{align}
c(\Delta,J)\sim \frac{1}{\Delta-J-2h_{*}}. \label{eq:poleintwist}
\end{align}
These are the familiar double-twist trajectories which are studied in the lightcone bootstrap \cite{Fitzpatrick:2012yx,Komargodski:2012ek,Caron-Huot:2017vep}. If we perform the inversion integral in the limit $z\sim\bar{z}\ll1$, we find a pole at fixed spin:
\begin{align}
c(\Delta,J)\sim\frac{1}{J-j_{*}}.
\end{align}
When performing a Sommerfeld-Watson transform, both poles can be picked up and contribute to the Regge limit. If we set $\Delta=\frac{d}{2}+i\nu$, then \eqref{eq:poleintwist} gives a pole at $J=\frac{d}{2}-2h_{*}\pm i\nu$. By unitarity $2h_{*}\geq \frac{d-2}{2}$, where the inequality is saturated by a free scalar. Therefore, in interacting theories this pole gives a term which vanishes in the Regge limit. On the other hand, we expect poles which come from the region $z\sim\bar{z}\ll1$ to give a growing contribution. We will show this explicitly by studying the Regge growth of $\dDisc_{t}[\mathcal{G}(z,\bar{z})]$ at order $N^{-4}$. 

Next, we assume the double-discontinuity at one-loop is determined by the double-trace operators $[\f\p]_{n,J}$. As reviewed in Section \ref{ssec:Unitarity}, this implies $\dDisc_{t}[\mathcal{G}^{(2)}(z,\bar{z})]$ is determined by the squared anomalous dimensions due to Pomeron exchange,
\begin{align}
\dDisc_{t}[\mathcal{G}^{(2)}(z,\bar{z})]=\frac{(z\bar{z})^{\Delta_{\f}}}{\left[(1-z)(1-\bar{z})\right]^{\frac{\Delta_{\f}+\Delta_{\p}}{2}}}\sum\limits_{h,\bar{h}} \frac{\pi^{2}}{2}P^{MFT}_{h,\bar{h}}\left(\gamma^{(1)}_{h,\bar{h}}\right)^{2}g^{a,a}_{h,\bar{h}}(1-z,1-\bar{z}). \label{eq:ddiscPt1}
\end{align}
Comparing with our results in Section \ref{ssec:Tree_Crossing} and using the Bessel function approximation (\ref{eq:g4dBessel}), we can recognize that (\ref{eq:ddiscPt1}) grows like $(z\bar{z})^{1-j(0)}$ in the Regge limit. Therefore, we expect $c^{t}(\Delta,J)$ has a non-analyticity at $J=2j(0)-1$. In a theory of gravity plus low-spin matter\footnote{By low-spin matter we mean only fields of spin $J\leq 2$ are allowed.} we have $j(\nu)=2$, so at one-loop we find a pole at $J=3$ due to two-graviton exchange \cite{Fitzpatrick:2019efk}.\footnote{We cannot apply the saddle point approximation for the $\nu$ integral in a theory dual to gravity since $j(\nu)$ is constant, but we can instead just close the contour in the lower half-plane.} Following this terminology, we will say that for generic $\Delta_{gap}$ the non-analyticity at $J=2j(0)-1$ is due to two-Pomeron exchange.

To calculate (\ref{eq:ddiscPt1}), we will use (\ref{eq:treegamma}) and (\ref{eq:treedP}), which we will write for compactness as:
\begin{align}
\gamma^{(1)}_{h,\bar{h}}&\approx \int\limits_{-\infty}^{\infty}d\nu \widehat{\gamma}^{(1)}(\nu)(h\bar{h})^{j(\nu)-1}\Omega_{i\nu}\left(\log(h/\bar{h})\right),
\\
\delta P^{(1)}_{h,\bar{h}}&\approx \int\limits_{-\infty}^{\infty}d\nu \widehat{\delta P}^{(1)}(\nu)(h\bar{h})^{j(\nu)-1}\Omega_{i\nu}\left(\log(h/\bar{h})\right).
\end{align}
Then to determine the double-discontinuity at one-loop, we have to calculate:
\begin{align}
\dDisc_{t}[\mathcal{G}^{(2)}(z,\bar{z})]\approx\sum\limits_{h,\bar{h}}\int\limits_{-\infty}^{\infty}&d\nu_1d\nu_2\frac{\pi^{2}}{2}P^{MFT}_{h,\bar{h}}\widehat{\gamma}^{(1)}(\nu_1)\widehat{\gamma}^{(1)}(\nu_2)(h\bar{h})^{j(\nu_1)+j(\nu_2)-2}
\nonumber \\ &\Omega_{i\nu_1}\left(\log(h/\bar{h})\right)\Omega_{i\nu_2}\left(\log(h/\bar{h})\right)g^{a,a}_{h,\bar{h}}(1-z,1-\bar{z}).
\end{align}
Unfortunately, we can no longer use (\ref{eq:KInt}) to do the $h$ and $\bar{h}$ integrals directly. The reason is simple, in $d=4$ each $\Omega_{i\nu}\left(\log(h/\bar{h})\right)$ contains a factor of $(h^{2}-\bar{h}^{2})^{-1}$,
\begin{align}
\Omega_{i\nu}\left(\log(h/\bar{h})\right)=-\frac{i \nu  h^{1-i \nu } \bar{h}^{1-i \nu } \left(h^{2 i \nu }-\bar{h}^{2 i \nu }\right)}{4 \pi ^2 \left(h^2-\bar{h}^2\right)}.
\end{align}
At tree-level this is cancelled by a factor of $(h^{2}-\bar{h}^{2})$ in $P^{MFT}_{h,\bar{h}}$, see \eqref{eq:PMFT4d}, but when we square the anomalous dimensions we have an extra factor left over in the denominator. Instead of performing this integral directly, we instead use completeness of the harmonic function, $\Omega_{i\nu}(L)$, on $H_{d-1}$ \cite{Penedones:2007ns} to write\footnote{Similar identities have been used to study one-loop bubble diagrams in AdS \cite{Penedones:2010ue,Fitzpatrick:2011dm,Fitzpatrick:2011hu,Yuan:2018qva,Carmi:2018qzm}.}:
\begin{align}
\Omega_{i\nu_1}(L)\Omega_{i\nu_2}(L)=\int\limits_{-\infty}^{\infty} d\nu B(\nu_1,\nu_2;\nu)\Omega_{i\nu}(L). \label{eq:HBubble}
\end{align}
With this identity, we can now use the integral (\ref{eq:KInt}) at the price of introducing an extra spectral integral. To prove exponentiation we do not need the explicit form of the bubble function $B(\nu_1,\nu_2;\nu)$, although it could be useful for other applications.\footnote{In $d=2$ the bubble function is very simple:
\begin{align}
B^{(d=2)}(\nu_1,\nu_2;\nu)=\frac{1}{4\pi}\left(\delta(\nu-\nu_1-\nu_2)+\delta(\nu-\nu_1+\nu_2)\right).
\end{align}
Since the analysis is identical in form to the $d=4$ case we will present the $d=2$ results in Appendix \ref{sec:Regge2d}.}

Now using the identity (\ref{eq:HBubble}) and approximating the $h$ and $\bar{h}$ sum by an integral, we can calculate the double-discontinuity in the Regge limit:
\begin{align}
\hspace{-.2in}\dDisc_{t}[\mathcal{G}^{(2)}(z,\bar{z})]&\approx\frac{\pi^2}{4}\frac{(z\bar{z})^{\Delta_{\f}}}{[(1-z)(1-\bar{z})]^{\frac{\Delta_\p+\Delta_\f}{2}}}\int\limits_{0}^{\infty} dh d\bar{h}\int\limits_{-\infty}^{\infty}d\nu_1d\nu_2d\nu\widehat{\gamma}(\nu_1)\widehat{\gamma}(\nu_2)B(\nu_1,\nu_2;\nu)
\nonumber \\ &\hspace{1.75in}(h\bar{h})^{j(\nu_1)+j(\nu_2)-2}\Omega_{i\nu}(\log(h/\bar{h}))g^{a,a}_{h,\bar{h}}(1-z,1-\bar{z})
\nonumber \\[10pt] &\approx \frac{\pi^{2}}{2}\int\limits_{-\infty}^{\infty}d\nu_1d\nu_2d\nu\frac{\chi_{j(\nu_1)+j(\nu_2)-1}(\nu)\chi_{j(\nu_1)+j(\nu_2)-1}(-\nu)  }{\Gamma (\Delta_{\f}-1) \Gamma (\Delta_{\f}) \Gamma (\Delta_{\p}-1) \Gamma (\Delta_{\p})}\widehat{\gamma}^{(1)}(\nu_1) \widehat{\gamma}^{(1)}(\nu_2) B(\nu_1,\nu_2;\nu )
\nonumber \\ &\hspace{1.34in} (z\bar{z})^{\frac{1}{2}(2-j(\nu_1)-j(\nu_2))} \Omega_{i\nu}\left(\frac{1}{2}\log(z/\bar{z})\right).
\end{align}

Next, we use this expression in the inversion formula, (\ref{eq:invd4}), and evaluate the integrals in the small $z,\bar{z}$ limit,
\begin{align}
\hspace{-.3in}c^{t,(2)}(\Delta,J)\approx -(\Delta-2) \kappa_{\Delta +J}\int\limits_{-\infty}^{\infty}&d\nu_1d\nu_2d\nu\frac{\chi_{j(\nu_1)+j(\nu_2)-1}(\nu)\chi_{j(\nu_1)+j(\nu_2)-1}(-\nu)}{\Gamma (\Delta_{\f}-1) \Gamma (\Delta_{\f}) \Gamma (\Delta_{\p}-1) \Gamma (\Delta_{\p})}\nu ^2 \widehat{\gamma}^{(1)}(\nu_1) \widehat{\gamma}^{(1)}(\nu_2)B(\nu_1,\nu_2;\nu )
\nonumber \\ & \frac{1}{(\Delta -j(\nu_1)-j(\nu_2)+J+i \nu -1) (\Delta -j(\nu_1)-j(\nu_2)+J-i \nu -1)}
\nonumber \\ &\frac{1}{  (\Delta +j(\nu_1)+j(\nu_2)-J-i \nu -3) (\Delta +j(\nu_1)+j(\nu_2)-J+i \nu -3)},
\label{eq:invHarm}
\end{align}
where $c^{t,(2)}(\Delta,J)$ is the one-loop OPE function. We claim that \eqref{eq:invHarm} will determine the leading Regge growth at one-loop.

We can see the integrand of (\ref{eq:invHarm}) does not have a pole at fixed $J$. To see how the expected Regge growth emerges, without directly computing the $\nu_i$ integrals in (\ref{eq:invHarm}), we need to plug this into the Euclidean partial wave decomposition and then perform a Sommerfeld-Watson transform. Recall that the full OPE function for $\<\f\f\p\p\>$ is given by
\begin{align}
c(\Delta,J)=(1+(-1)^J)c^{t}(\Delta,J).
\end{align}
The partial wave decomposition and its analytic continuation in spin take the usual form,
\begin{align}
\mathcal{G}(z,\bar{z})&=\sum\limits_{J}\int\limits_{-\infty}^{\infty}d\nu' c\left(2+i\nu',J\right)F_{2+i\nu',J}(z,\bar{z})
\nonumber \\ &=-\oint_{\Gamma}dJ\int\limits_{-\infty}^{\infty}\frac{d\nu'}{2\pi}\frac{1+e^{-i\pi J}}{1-e^{-2\pi i J}}c^{t}\left(2+i\nu',J\right)F_{2+i\nu',J}(z,\bar{z}),
\end{align}
where we set $d=4$. The function $c^{t,(2)}(2+i\nu',J)$ has poles in the lower half-plane of $\nu'$ when:
\begin{align}
i\nu'=1+J -j(\nu_1)-j(\nu_2)\pm i\nu.
\end{align}
Therefore, when we close the $\nu'$ contour we find
\begin{align}
\mathcal{G}^{(2)}(z,\bar{z})\approx-\oint_{\Gamma}dJ\frac{1+e^{-i\pi J}}{1-e^{-2\pi i J}}\int\limits_{-\infty}^{\infty} &d\nu_1d\nu_2d\nu\frac{i \nu   \chi_{j(\nu_1)+j(\nu_2)-1}(\nu)\chi_{j(\nu_1)+j(\nu_2)-1}(-\nu)   }{8 \Gamma (\Delta_{\f}-1) \Gamma (\Delta_{\f}) \Gamma (\Delta_{\p}-1) \Gamma (\Delta_{\p}) }\frac{\kappa_{2J+3-j(\nu_1)-j(\nu_2)-i \nu}}{(J-j(\nu_1)-j(\nu_2)+1)}
\nonumber \\ &    \widehat{\gamma}^{(1)}(\nu_1) \widehat{\gamma}^{(1)}(\nu_2) B(\nu_1,\nu_2;\nu ) g_{J+3-j(\nu_1)-j(\nu_2)-i \nu,J}(z,\bar{z}) +(\nu\leftrightarrow -\nu). \label{eq:ReggePostNuCont}
\end{align}

Now we can deform the $\Gamma$ contour to the $\Gamma'$ contour, see figure \ref{fig:Sommerfeld_watson}, and in doing so we pick up a pole at $J=j(\nu_1)+j(\nu_2)-1$. To go to the Regge limit we also need to analytically continue the blocks around $\bar{z}=1$. To perform the analytic continuation, we will introduce the pure blocks \cite{Caron-Huot:2017vep}. These are solutions to the conformal Casimir equation with a pure power-law behavior in the limit $z\ll\bar{z}\ll1$:
\begin{align}
g^{pure}_{\Delta,J}(z,\bar{z})\approx z^{\frac{\Delta-J}{2}}\bar{z}^{\frac{\Delta+J}{2}}.
\end{align}
The standard conformal blocks in general $d$ have the decomposition:
\begin{align}
g_{\Delta,J}(z,\bar{z})=g^{pure}_{\Delta,J}(z,\bar{z})+\frac{\Gamma(J+d-2)\Gamma(-J-\frac{d-2}{2})}{\Gamma(J+\frac{d-2}{2})\Gamma(-J)}g^{pure}_{\Delta,-J-d+2}. \label{eq:gDecompPure}
\end{align}
If we send $\bar{z}$ counter-clockwise around the branch point $\bar{z}=1$ and then take $z,\bar{z}\ll1$ we have\footnote{In general, this monodromy contains more terms but they are subleading in the limit $z,\bar{z}\rightarrow 0$ \cite{Caron-Huot:2017vep}.}:
\begin{align}
g^{\circlearrowleft,pure}_{\Delta,J}(z,\bar{z})\approx-\frac{i}{\pi}\frac{1}{\kappa_{\Delta+J}}g^{pure}_{1-J,1-\Delta}(z,\bar{z}).
\end{align}
When we go to the Regge limit the first term in (\ref{eq:gDecompPure}) is dominant and the second term can be dropped. Finally, this gives us the Regge limit of the one-loop correlator:
\begin{align}
\mathcal{G}^{(2)\circlearrowleft}(z,\bar{z})\approx& \ -\pi^{2}\int\limits_{-\infty}^{\infty} d\nu_1d\nu_2d\nu\frac{1+e^{-i\pi (j(\nu_1)+j(\nu_2)-1)}}{1-e^{-2\pi i (j(\nu_1)+j(\nu_2)-1)}}\frac{ \widehat{\gamma}^{(1)}(\nu_1) \widehat{\gamma}^{(1)}(\nu_2) B(\nu_1,\nu_2;\nu ) }{ \Gamma (\Delta_{\f}-1) \Gamma (\Delta_{\f}) \Gamma (\Delta_{\p}-1) \Gamma (\Delta_{\p}) }
\nonumber \\ &\hspace{.5in}\chi_{j(\nu_1)+j(\nu_2)-1}(\nu)\chi_{j(\nu_1)+j(\nu_2)-1}(-\nu) (z\bar{z})^{\frac{1}{2}(2-j(\nu_1)-j(\nu_2))}\Omega_{i\nu}\left(\frac{1}{2}\log(z/\bar{z})\right).
\label{eq:1lpReggeFinal}
\end{align}
To obtain this expression, we used that $B(\nu_1,\nu_2;\nu)$ is symmetric in $\nu$ and recognized that the two terms in \eqref{eq:ReggePostNuCont} combine to give a single block $g_{2-j(\nu_1)-j(\nu_2),i\nu -1}(z,\bar{z})$. Finally we use \eqref{eq:gLRtoHarm} to write the final answer in terms of the harmonic function. The expression (\ref{eq:1lpReggeFinal}) tells us that the leading Regge singularity at one-loop is located at $J=j(\nu_1)+j(\nu_2)-1$. If we assume at tree-level that,
\begin{align}
j(\nu)=2-2\frac{4+\nu^{2}}{\Delta_{gap}^{2}} + O(\Delta_{gap}^{-4}), \label{eq:j4LargeN}
\end{align}
and keep $\Delta_{gap}$ finite, but large, we can perform the $\nu_{1,2}$ integrals using a saddle-point approximation. This reveals a non-analyticity at $J=2j(0)-1$. The integrand of (\ref{eq:ReggePostNuCont}) has a continuous family of poles, parameterized by $\nu_{1,2}$ and starting at $J=2j(0)-1$, so this will be a branch point of the full OPE function. 

\subsection{Inversion of the One-Loop Correlator}
\label{ssec:Double_Inv}
Now that we have the one-loop correlator in the Regge limit, \eqref{eq:1lpReggeFinal}, we can redo the analysis of Section \ref{ssec:Tree_Crossing} and study corrections to the double-trace data by imposing crossing symmetry. 

At one-loop, keeping only the double-traces $[\f\p]_{n,J}$ in the $t$-channel, crossing symmetry in the Regge limit says:
\begin{align}
(z\bar{z})^{-\Delta_{\f}}\mathcal{G}^{(2)\circlearrowleft}(z,\bar{z})\approx \sum_{h,\bar{h}}&
P^{MFT}_{h,\bar{h}}\bigg[\gamma^{(2)}_{h,\bar{h}}\left(i\pi+\frac{1}{2}(\partial_{h}+\partial_{\bar{h}})\right)+\delta P^{(2)}_{h,\bar{h}}+\gamma^{(1)}_{h,\bar{h}}\delta P^{(1)}_{h,\bar{h}}\left(i\pi +\frac{1}{2}(\partial_{h}+\partial_{\bar{h}})\right)
\nonumber \\ &+\left(\gamma^{(1)}_{h,\bar{h}}\right)^{2}\left(-\frac{\pi^{2}}{2}+\frac{i\pi}{2}(\partial_{h}+\partial_{\bar{h}})+\frac{1}{8}\left(\partial_{h}+\partial_{\bar{h}}\right)^{2}\right)\bigg]g^{a,a}_{h,\bar{h}}(1-z,1-\bar{z}),
\label{eq:loopCrossingV1}
\end{align}
where $h\approx n$ and $\bar{h}\approx n+J$. We can simplify the expression by dropping all powers of $(\partial_{h}+\partial_{\bar{h}})$, since they reduce the Regge growth of the double-trace sum by a factor of $(z\bar{z})^{\frac{1}{4}}$. The crossing equation then becomes:
\begin{align}
(z\bar{z})^{-\Delta_{\f}}\mathcal{G}^{(2)\circlearrowleft}(z,\bar{z})\approx \sum_{h,\bar{h}}
P^{MFT}_{h,\bar{h}}\bigg[&i\pi \gamma^{(2)}_{h,\bar{h}}+\delta P^{(2)}_{h,\bar{h}}+i\pi \gamma^{(1)}_{h,\bar{h}}\delta P^{(1)}_{h,\bar{h}}-\frac{\pi^{2}}{2}\left(\gamma^{(1)}_{h,\bar{h}}\right)^{2}\bigg]g^{a,a}_{h,\bar{h}}(1-z,1-\bar{z}).
\label{eq:loopCrossingV2}
\end{align}
We have two unknowns, $\gamma^{(2)}_{h,\bar{h}}$ and $\delta P^{(2)}_{h,\bar{h}}$, and two equations, corresponding to the real and imaginary pieces of $\mathcal{G}^{(2)\circlearrowleft}(z,\bar{z})$, so it straightforward to solve for both. In practice, to solve crossing we will write the one-loop corrections as spectral integrals over the harmonic functions $\Omega_{i\nu}(L)$ and use (\ref{eq:KInt}) to evaluate the integrals over $h$ and $\bar{h}$. Furthermore, when performing the $h$ and $\bar{h}$ integrals for the $\gamma^{(1)}_{h,\bar{h}}\delta P^{(1)}_{h,\bar{h}}$ term, it will be convenient to symmetrize their spectral integral representation:
\begin{align}
\gamma^{(1)}_{h,\bar{h}}\delta P^{(1)}_{h,\bar{h}}=\frac{1}{2}\int\limits_{-\infty}^{\infty} d\nu_1 d\nu_2 \widehat{\gamma}^{(1)}(\nu_1)\widehat{\delta P}^{(1)}(\nu_2)(h\bar{h})^{j(\nu_1)+j(\nu_2)-2}\Omega_{i\nu_1}(\log(h/\bar{h}))\Omega_{i\nu_2}(\log(h/\bar{h})) + \left(\nu_1\leftrightarrow \nu_2\right).
\end{align}
Crossing symmetry then implies the one-loop correction to $P_{h,\bar{h}}$ vanishes,
\begin{align}
\delta P^{(2)}_{h,\bar{h}}\approx 0, \label{eq:dpOneLoop}
\end{align}
to leading order at large $h$ and $\bar{h}$.\footnote{More precisely, the statement is $\delta P^{(2)}_{h,\bar{h}}$ does not grow like $(h\bar{h})^{2j(0)-2}$ at large $h,\bar{h}$.} For the one-loop anomalous dimensions we find:
\begin{align}
\gamma^{(2)}_{h,\bar{h}}&\approx\int\limits_{-\infty}^{\infty} d\nu_{1}d\nu_2d\nu \ \widehat{\gamma}^{(2)}(\nu_1,\nu_2;\nu) (h\bar{h})^{j(\nu_1)+j(\nu_2)-2}\Omega_{i\nu}\left(\log(h/\bar{h})\right), 
\\
\widehat{\gamma}^{(2)}(\nu_1,\nu_2;\nu)&=-\frac{1}{2} \pi \tan \left(\frac{1}{2} \pi  j(\nu_1)\right) \tan \left(\frac{1}{2} \pi  j(\nu_2)\right) \tan \left(\frac{1}{2} \pi (j(\nu_1)+j(\nu_2))\right) 
\nonumber \\ & \hspace{2.45in}\times\widehat{\gamma}^{(1)}(\nu_1) B(\nu_1,\nu_2;\nu )  \widehat{\gamma}^{(1)}(\nu_2). \label{eq:gammaOneLoop}
\end{align}

It is useful to rephrase (\ref{eq:dpOneLoop}) in terms of the underlying OPE coefficients. The dictionary is:
\begin{align}
2^{h-\bar{h}}c_{h,\bar{h}}^{2}=P_{h,\bar{h}}=P^{MFT}_{h,\bar{h}}\left(1+\frac{1}{N^{2}}\delta P^{(1)}_{h,\bar{h}}+\frac{1}{N^{4}}\delta P^{(2)}_{h,\bar{h}}+...\right).
\end{align}
We can also expand the individual OPE coefficients at large $N$:
\begin{align}
c_{h,\bar{h}}=\sum\limits_{i=0}^{\infty} \frac{1}{N^{2i}}c^{(i)}_{h,\bar{h}},
\end{align}
and compare the two sides. For example at tree-level we find:
\begin{align}
\frac{c^{(1)}_{h,\bar{h}}}{c^{(0)}_{h,\bar{h}}}=\frac{1}{2}\delta P^{(1)}_{h,\bar{h}}.
\end{align}
Imposing $\delta P^{(2)}_{h,\bar{h}}=0$ gives us the condition
\begin{align}
c^{(2)}_{h,\bar{h}}\approx -\frac{\left(c^{(1)}_{h,\bar{h}}\right)^{2}}{2c^{(0)}_{h,\bar{h}}},
\end{align}
to leading order at large $h$ and $\bar{h}$. So (\ref{eq:dpOneLoop}) determines the one-loop OPE coefficients at large $h$ and $\bar{h}$ in terms of lower loop data.

The one-loop anomalous dimension (\ref{eq:gammaOneLoop}) is exact in $\Delta_{gap}$, but to compare with the eikonal ansatz \eqref{eq:EikForm} we should use the explicit form of $j(\nu)$, as given in \eqref{eq:j4LargeN}. Then if we expand at large $\Delta_{gap}$ we find 
\begin{align}
\gamma^{(2)}_{h,\bar{h}}\sim \Delta_{gap}^{-6}.
\end{align}
Therefore, the contribution of $\gamma^{(2)}_{h,\bar{h}}$ to the imaginary piece of $\mathcal{G}^{(2)\circlearrowleft}(z,\bar{z})$ is suppressed by $\Delta_{gap}^{-4}$ in comparison to the contribution of the $\gamma^{(1)}_{h,\bar{h}}\delta P^{(1)}_{h,\bar{h}}$ term. This implies that in (\ref{eq:loopCrossingV2}), the leading Regge behavior at large $\Delta_{gap}$ comes from the $\left(\gamma^{(1)}_{h,\bar{h}}\right)^{2}$ and $ \gamma^{(1)}_{h,\bar{h}}\delta P^{(1)}_{h,\bar{h}}$ terms. In other words, the one-loop Regge limit at large, but finite, coupling is determined by the tree-level, double-trace anomalous dimensions and OPE coefficients. This suppression is the first hint that the bulk scattering will exponentiate, as we discuss in more detail in Section \ref{ssec:Eikonalization}.

The fact $\gamma^{(2)}_{h,\bar{h}}\rightarrow 0$ as $\Delta_{gap}\rightarrow \infty$ is also consistent with the pure gravity result. It was shown in \cite{Cornalba:2007zb} that the graviton ladder diagrams exponentiate in the Regge limit, and to reproduce them in the CFT we only keep powers of $\gamma^{(1)}_{h,\bar{h}}$ in the $t$-channel, double-trace sum. In other words, when $\Delta_{gap}\rightarrow \infty$ the one and higher-loop corrections to the anomalous dimensions do not vanish exactly, but they will give a softer Regge growth in comparison to the tree-level anomalous dimensions.


\subsection{All Order Results}
\label{ssec:All_Loop}
We can now ask what happens at higher orders in $1/N$. From Section \ref{ssec:Double_Inv}, we learned the leading contribution to the one-loop correlator, at large $\Delta_{gap}$, comes from the tree-level corrections to the OPE data. We will now show this result carries over to all orders in $1/N$. The proof is simple and follows from an inductive argument. The one-loop case considered in the previous section was our base case, so in this section we need to prove the inductive part.

We assume that at $L-1$ loops and for large $\Delta_{gap}$, we can solve crossing for $\mathcal{G}^{\circlearrowleft}(z,\bar{z})$ by only including tree-level corrections to the double-trace operators, $[\f\p]_{n,J}$, in the $t$-channel. This implies that the Regge limit of $\dDisc_{t}[\mathcal{G}(z,\bar{z})]$ at $L$-loops and large $\Delta_{gap}$ is also determined by the same tree-level corrections, since the double-discontinuity always factorizes onto lower-loop data. Following the same procedure as in the previous sections, we can then reconstruct the $L$-loop correlator in the Regge limit from the inversion formula. Finally, we solve crossing at $L$-loops and large $\Delta_{gap}$ using only tree-level corrections to the double-trace data in the $t$-channel. This will complete the inductive part of the proof.

For technical reasons, we will need to separately consider this problem at $2M-1$ loops and $2M$ loops. This comes from the $t$-channel expansion of the double-discontinuity:
\begin{align}
\dDisc_{t}[\mathcal{G}(z,\bar{z})]=\sum\limits_{h,\bar{h}}2\sin^{2}\left(\frac{\pi}{2}\gamma_{h,\bar{h}}\right)P_{h,\bar{h}}g^{a,a}_{h,\bar{h}}(1-z,1-\bar{z}).
\end{align} 
At $2M-1$ loops, the leading contribution to the Regge limit of $\dDisc_{t}[\mathcal{G}(z,\bar{z})]$ comes from the $\left(\gamma^{(1)}_{h,\bar{h}}\right)^{2M}$ term while at $2M$ loops it comes from the $\left(\gamma^{(1)}_{h,\bar{h}}\right)^{2M}\delta P^{(1)}_{h,\bar{h}}$ term.

For convenience, let us also define the function $\Theta$ as:
\begin{align}
\prod_{j=1}^{m}\Omega_{i\nu_{j}}(L)=\int\limits_{-\infty}^{\infty} d\nu \ \Theta(\nu_1,...,\nu_{m};\nu)\Omega_{i\nu}(L),
\end{align}
where the relation to the bubble factor $B(\nu_1,\nu_2;\nu)$ is
\begin{align}
\Theta(\nu_1,...,\nu_{m};\omega_{m})=\int\limits_{-\infty}^{\infty} d\omega_{2}...d\omega_{m-1} B(\nu_1,\nu_2;\omega_2)\prod_{j=2}^{m-1}B(\omega_{j},\nu_{j+1};\omega_{j+1}).
\end{align}

Then at $2M-1$ loops the double-discontinuity in the Regge limit is,
\begin{align}
\dDisc_{t}[\mathcal{G}^{(2M)}(z,\bar{z})]\approx& -\frac{(z\bar{z})^{\Delta_{\f}}}{\left[(1-z)(1-\bar{z})\right]^{\frac{\Delta_\f+\Delta_\p}{2}}}\sum_{h,\bar{h}}P^{MFT}_{h,\bar{h}}\left(\gamma^{(1)}_{h,\bar{h}}\right)^{2M}\frac{(i\pi)^{2M}}{(2M)!}g^{a,a}_{h,\bar{h}}(1-z,1-\bar{z})
 \nonumber \\[10pt]
\approx& - \int\limits_{-\infty}^{\infty} d\nu_1...d\nu_{2M} d\nu\frac{(i \pi )^{2 M}\chi_{j_{tot}-2M+1}(\nu)\chi_{j_{tot}-2M+1}(-\nu)}{(2 M)! \Gamma (\Delta_{\f}-1) \Gamma (\Delta_{\f}) \Gamma (\Delta_{\p}-1) \Gamma (\Delta_{\p})}\prod\limits_{i=1}^{2M}\widehat{\gamma}^{(1)}(\nu_i) 
\nonumber\\ &\hspace{3.4cm} \Theta(\nu_1,...,\nu_{2M};\nu)(z\bar{z})^{-\frac{1}{2}(j_{tot}-2M)}\Omega_{i\nu}\left(\frac{1}{2}\log(z/\bar{z})\right),
\label{eq:dDisc_2M_1_Loops}
\end{align}
where we defined the total effective spin,
\begin{align}
j_{tot}=\sum\limits_{i}j(\nu_i).
\end{align}
We can plug (\ref{eq:dDisc_2M_1_Loops}) into the inversion formula to determine the OPE function at $2M-1$ loops. Using the resulting OPE function in the Euclidean partial wave expansion and performing the Sommerfeld-Watson transform yields the full correlator in the Regge limit at this loop order:
\begin{align}
\mathcal{G}^{(2M)\circlearrowleft}(z,\bar{z})\approx \int d\nu_1...d\nu_{2M} d\nu &\frac{e^{i \pi  (j_{tot}-M)}}{1+e^{i \pi  (j_{tot}-2 M)}}\frac{2 \pi ^{2 M} \chi_{j_{tot}-2M+1}(\nu)\chi_{j_{tot}-2M+1}(-\nu)}{(2 M)! \Gamma (\Delta_{\f}-1) \Gamma (\Delta_{\f}) \Gamma (\Delta_{\p}-1) \Gamma (\Delta_{\p}) }
\nonumber \\ &\hspace{0cm}\prod_{k=1}^{2M}\widehat{\gamma}^{(1)}(\nu_k)  \Theta(\nu_1,...,\nu_{2M};\nu)(z\bar{z})^{-\frac{1}{2}(j_{tot}-2M)}\Omega_{i\nu}\left(\frac{1}{2}\log(z/\bar{z})\right).
\end{align}

The analysis is almost identical at $2M$ loops with the double-discontinuity now given by,
\begin{align}
\dDisc_{t}[\mathcal{G}^{(2M+1)}(z,\bar{z})]\approx-\frac{(z\bar{z})^{\Delta_{\f}}}{\left[(1-z)(1-\bar{z})\right]^{\frac{\Delta_\f+\Delta_\p}{2}}}\sum\limits_{h,\bar{h}}\frac{(i \pi )^{2 M}}{(2 M)!}P^{MFT}_{h,\bar{h}}\left(\gamma^{(1)}_{h,\bar{h}}\right)^{2M}\delta P^{(1)}_{h,\bar{h}} g^{a,a}_{h,\bar{h}}(1-z,1-\bar{z}),
\end{align}
so we will just quote the final answer:
\begin{align}
\mathcal{G}^{(2M+1)\circlearrowleft}(z,\bar{z})\approx&\int d\nu_1...d\nu_{2M+1} d\nu\frac{ e^{i \pi  (j_{tot}-M)} }{ \left(e^{i \pi  (j_{tot}-2 M)}-1\right)} \frac{2\pi ^{2 M} \chi_{j_{tot}-2M}(\nu)\chi_{j_{tot}-2M}(-\nu)}{(2 M)! \Gamma (\Delta_{\f}-1) \Gamma (\Delta_{\f}) \Gamma (\Delta_{\p}-1) \Gamma (\Delta_{\p})}
\nonumber \\ & \prod_{k=1}^{2M}\widehat{\gamma}^{(1)}(\nu_k) \widehat{\delta P}^{(1)}(\nu_{2M+1})\Theta(\nu_1,...,\nu_{2M+1};\nu)(z\bar{z})^{\frac{1}{2}(1-j_{tot}+2M)} \Omega_{i\nu}\left(\frac{1}{2}\log(z/\bar{z})\right).
\end{align}

Now we have to compare both expressions with the $t$-channel expansion. For compactness, we define the following double-trace sums:
\begin{align}
\mathcal{S}^{(k)}_{\gamma}(z,\bar{z})=&\frac{(z\bar{z})^{\Delta_{\f}}}{\left[(1-z)(1-\bar{z})\right]^{\frac{\Delta_\f+\Delta_\p}{2}}}\sum\limits_{h,\bar{h}}P^{MFT}_{h,\bar{h}}\frac{1}{k!}(i\pi)^{k}\left(\gamma^{(1)}_{h,\bar{h}}\right)^{k}g^{a,a}_{h,\bar{h}}(1-z,1-\bar{z}) ,\label{eq:SsumGamma}
\\
\mathcal{S}^{(k)}_{\delta P\times \gamma}(z,\bar{z})=&\frac{(z\bar{z})^{\Delta_{\f}}}{\left[(1-z)(1-\bar{z})\right]^{\frac{\Delta_\f+\Delta_\p}{2}}}\sum\limits_{h,\bar{h}}P^{MFT}_{h,\bar{h}}\frac{1}{(k-1)!}(i\pi)^{k-1}\left(\gamma^{(1)}_{h,\bar{h}}\right)^{k-1}\delta P^{(1)}_{h,\bar{h}}g^{a,a}_{h,\bar{h}}(1-z,1-\bar{z}). \label{eq:SsumdpGamma}
\end{align}

To compute both sums we again use the spectral representation for $\gamma^{(1)}_{h,\bar{h}}$ and $\delta P^{(1)}_{h,\bar{h}}$. As in the one-loop case, we should symmetrize the spectral integrals for $\mathcal{S}^{(k)}_{\delta P\times \gamma}$. We will make the replacement:
\begin{align}
\delta P^{(1)}_{h,\bar{h}}\left(\gamma^{(1)}_{h,\bar{h}}\right)^{k-1}=\frac{1}{k}\int\limits_{-\infty}^{\infty} d\nu_1...d\nu_{k}d\nu\sum\limits_{i=1}^{k} \ \widehat{\delta P}^{(1)}(\nu_i)\prod\limits_{\substack{j=1 \\ j\neq i}}^{k}\widehat{\gamma}^{(1)}(\nu_j)\Theta(\nu_1,...,\nu_{k};\nu)\Omega_{i\nu}\left(\log(h/\bar{h})\right).
\end{align}
Then at $2M-1$ loops we find:
\begin{align}
Re\left[\mathcal{G}^{(2M)\circlearrowleft}(z,\bar{z})\right]-\mathcal{S}^{(2M)}_{\gamma}(z,\bar{z})&=0, \label{eq:AllLoopPt1}
\\
Im\left[\mathcal{G}^{(2M)\circlearrowleft}(z,\bar{z})\right]-\mathcal{S}^{(2M)}_{\delta P\times \gamma}(z,\bar{z})&= O(\Delta_{gap}^{-6}), \label{eq:AllLoopPt2}
\end{align}
while at $2M$ loops we have:
\begin{align}
Re\left[\mathcal{G}^{(2M+1)\circlearrowleft}(z,\bar{z})\right]-\mathcal{S}^{(2M+1)}_{\delta P\times \gamma}&= O(\Delta_{gap}^{-6}), \label{eq:AllLoopPt3}
\\
Im\left[\mathcal{G}^{(2M+1)\circlearrowleft}(z,\bar{z})\right]-\mathcal{S}^{(2M+1)}_{\gamma}&=0.  \label{eq:AllLoopPt4}
\end{align}
Therefore at each loop order and for large $\Delta_{gap}$, the tree-level corrections determine the leading Regge behavior. At odd loops the imaginary piece of $\mathcal{G}^{\circlearrowleft}(z,\bar{z})$ starts at $\Delta_{gap}^{-2}$, so the loop effects are suppressed by $\Delta_{gap}^{-4}$ in comparison. Similarly, at even loops the real piece also starts at $\Delta_{gap}^{-2}$, so the same statement carries over.
\subsection{Exponentiation in Impact Parameter Space}
\label{ssec:Eikonalization}
We can now return to the problem of exponentiating the tree-level phase shift. For simplicity, we start by studying the correlator at one-loop. Recall that the eikonal approximation for the correlator in impact parameter space is:
\begin{align}
\mathcal{B}_{eik}(S,L)\ &= \ \mathcal{N}e^{i\delta(S,L)}, \label{eq:EikB}
\\
Re \ \delta(S,L)  \ &= \ \frac{1}{N^{2}} \pi \gamma^{(1)}_{h,\bar{h}},  \label{eq:EikRe}
\\
Im \ \delta(S,L) \ &= \ -\frac{1}{N^{2}}\delta P^{(1)}_{h,\bar{h}},  \label{eq:EikIm}
\end{align}
where $S=h\bar{h}$, $L= \log(\bar{h}/h)$ and the normalization $\mathcal{N}$ is given in (\ref{eq:NConst}).

At one-loop this implies
\begin{align}
\mathcal{B}^{(2)}_{eik}(S,L)=&\frac{\mathcal{N}}{2}\left[\left(\delta P^{(1)}_{h,\bar{h}}\right)^{2}-\pi^{2}\left(\gamma^{(1)}_{h,\bar{h}}\right)^{2}+2\pi i\gamma^{(1)}_{h,\bar{h}}\delta P^{(1)}_{h,\bar{h}}\right]. \label{eq:Eik1lpAnsatz}
\end{align}

If we now take our one-loop expression (\ref{eq:1lpReggeFinal}) and go to impact parameter space we find:
\begin{align}
\mathcal{B}^{(2)}(S,L)=-\int\limits_{-\infty}^{\infty} d\nu_1d\nu_2d\nu\mathcal{N}\frac{\pi^2}{2} \widehat{\gamma}^{(1)}(\nu_1) \widehat{\gamma}^{(1)}(\nu_2) &\left(1+i\tan \left(\frac{1}{2} \pi  (j(\nu_1)+j(\nu_2))\right)\right) 
\nonumber \\ & \quad B(\nu_1,\nu_2;\nu)S^{j(\nu_1)+j(\nu_2)-2}\Omega_{i\nu}(L).
\end{align}
The real piece of our result is simply given by:
\begin{align}
Re[\mathcal{B}^{(2)}(S,L)]= -\mathcal{N}\frac{\pi^{2}}{2}\left(\gamma_{h,\bar{h}}^{(1)}\right)^{2},
\end{align}
where we used the definition of the bubble function, (\ref{eq:HBubble}), to do the $\nu$ integral. In comparison to (\ref{eq:Eik1lpAnsatz}) we are missing the term $\left(\delta P^{(1)}_{h,\bar{h}}\right)^{2}$. However, at large $\Delta_{gap}$ we have $\delta P^{(1)}_{h,\bar{h}}\sim \Delta_{gap}^{-2}$ while  $\gamma^{(1)}_{h,\bar{h}}\sim 1$. Therefore
\begin{align}
Re[\mathcal{B}^{(2)}(S,L)-\mathcal{B}^{(2)}_{eik}(S,L)]= O(\Delta_{gap}^{-4}).
\end{align} 
For the imaginary piece we need to use the explicit form of $j(\nu)$, \eqref{eq:jlargeN}, and expand the integrand at large $\Delta_{gap}$:
\begin{align}
Im[\mathcal{B}^{(2)}(S,L)]=  \mathcal{N}\int\limits_{-\infty}^{\infty} d\nu_1d\nu_2d\nu&\frac{\pi^{3}}{2}\left(\frac{8+\nu_1^2+\nu_2^2}{\Delta_{gap}^{2}}\right)\widehat{\gamma}^{(1)}(\nu_1)\widehat{\gamma}^{(1)}(\nu_2)
\nonumber \\ &B(\nu_1,\nu_2;\nu)S^{j(\nu_1)+j(\nu_2)-2}\Omega_{i\nu}(L)+O(\Delta_{gap}^{-6}).
\end{align}
The eikonal ansatz gives the same result at this order in $\Delta_{gap}$:
\begin{align}
Im[\mathcal{B}^{(2)}_{eik}(S,L)]=  \mathcal{N}\int\limits_{-\infty}^{\infty} d\nu_1d\nu_2d\nu&\frac{\pi^{3}}{2}\left(\frac{8+\nu_1^2+\nu_2^2}{\Delta_{gap}^{2}}\right)\widehat{\gamma}^{(1)}(\nu_1)\widehat{\gamma}^{(1)}(\nu_2)
\nonumber \\ &B(\nu_1,\nu_2;\nu)S^{j(\nu_1)+j(\nu_2)-2}\Omega_{i\nu}(L)+O(\Delta_{gap}^{-6}),
\end{align}
or in other words
\begin{align}
Im[\mathcal{B}^{(2)}(S,L)-\mathcal{B}^{(2)}_{eik}(S,L)]= O(\Delta_{gap}^{-6}).
\end{align}
Recall that $Im[\mathcal{B}^{(2)}(S,L)]\sim \Delta_{gap}^{-2}$, which implies the deviations between our result and the eikonal ansatz are suppressed by $\Delta_{gap}^{-4}$ for both the real and imaginary piece of the phase shift.

At higher orders in $1/N$, we should then only expect the eikonal approximation to be valid up to terms suppressed by $\Delta_{gap}^{-4}$. At order $N^{-2m}$ the eikonal ansatz is:
\begin{align}
\mathcal{B}^{(m)}_{eik}(S,L)&=\frac{(i\pi)^m}{m!}\left(\left(\gamma^{(1)}_{h,\bar{h}}\right)^{m}-\frac{i}{\pi}m\delta P^{(1)}_{h,\bar{h}}\left(\gamma^{(1)}_{h,\bar{h}}\right)^{m-1}+...\right), \label{eq:BeikNth}
\end{align}
where the ellipses stand for terms which are suppressed by $\Delta_{gap}^{-4}$ with respect to the terms we have written. Performing the Fourier transform to go back to position space, we recover the sums \eqref{eq:SsumGamma} and \eqref{eq:SsumdpGamma}:
\begin{align}
\mathcal{G}^{(m)}_{eik}(S,L)=\mathcal{S}^{(m)}_{\gamma}(z,\bar{z})+\mathcal{S}^{(m)}_{\delta P\times\gamma}(z,\bar{z})+... \label{eq:GallOrdersRegge}
\end{align}

We have already verified in (\ref{eq:AllLoopPt1})-(\ref{eq:AllLoopPt4}) that these sums reproduce the correlator at each loop order in the Regge limit, up to terms which are suppressed by $\Delta_{gap}^{-4}$. This implies that at each order in $1/N$, the eikonal ansatz reproduces our result up to terms suppressed at large $\Delta_{gap}$. This completes our proof using AdS/CFT unitarity that the tree-level phase shift exponentiates when $S\gg \Delta_{gap}^{2}\gg 1$ and is therefore in agreement with the prediction from string theory \cite{Brower:2006ea,Brower:2007xg}.

\section{Inelastic Effects and Exact Exponentiation}
\label{sec:MoreStringyCorrections}
In this section we will revisit one of our fundamental assumptions, that at one-loop the Regge limit of $\dDisc_{t}[\mathcal{G}(z,\bar{z})]$ is determined by the sum over the double-trace operators $[\f\p]_{n,J}$. At one-loop there are two ways to correct this assumption: either by allowing for the exchange of single-trace operators or of more general double-trace operators. In string theory these two effects are dual to long string creation and tidal excitations, respectively. Physically, long string creation corresponds to new states being created by the scattering particles, dual to $\f$ and $\p$, due to the presence of extended objects in AdS. Tidal excitations, as we discuss in Section \ref{ssec:tidal}, correspond to an incoming state being excited to a massive state in the scattering process \cite{Amati:1987wq,Amati:1987uf,Camanho:2014apa,Shenker:2014cwa}.\footnote{In string theory this is due to oscillator modes of the original state becoming excited \cite{Camanho:2014apa,Shenker:2014cwa}.}

As we review in the next section, by requiring that our tree-level results are consistent with the Euclidean OPE, we can show single-trace operators must contribute to $\dDisc_{t}[\mathcal{G}(z,\bar{z})]$ at tree-level. By inverting this sum over single-trace operators in the Regge limit, we also reproduce tree-level Regge behavior, or Pomeron exchange. Next, we show that by allowing for similar effects at one-loop, we can find results consistent with exponentiation of the tree-level phase shift, without any assumption on $\Delta_{gap}$. Finally, we discuss tidal excitations, or the exchange of more general double-trace operators, and argue this effect is subleading at large $\Delta_{gap}$.

\subsection{Long String Creation}
\label{sec:MassiveSTOps}
\subsubsection{Tree-Level}
\label{sec:MassiveSTtree}

In general, it is difficult to study sums of massive, single-trace operators in a large $N$ CFT. Unlike the double-trace operators, there is less regularity in their spectrum and OPE coefficients. However, we can extract some information about these operators by requiring that crossing symmetry in the Regge and Euclidean limits are consistent. This idea was used in \cite{Cornalba:2006xm,Cornalba:2007zb} for graviton exchange and \cite{Li:2017lmh} for Pomeron exchange. We will review the arguments here.

We will be studying the tree-level correlator in the limit $z,\bar{z}\rightarrow 0$, with $z/\bar{z}$ held fixed, both in the Euclidean and Regge limits. The crossing equation in the Euclidean regime is:
\begin{align}
\hspace{-.5cm}
(z\bar{z})^{-\Delta_{\f}}\mathcal{G}^{(1)}(z,\bar{z}) \ \approx \ \ &\sum_{h,\bar{h}}P^{MFT}_{h,\bar{h}}\left[ \frac{1}{2}\gamma^{(1)}_{h,\bar{h}}\left(\partial_{h}+\partial_{\bar{h}}\right)+ \delta P^{(1)}_{h,\bar{h}}\right]g^{a,a}_{h,\bar{h}}(1-z,1-\bar{z}) 
\nonumber \\ +&\sum_{\mathcal{X}}P^{(1)}_{\mathcal{X}}g^{a,a}_{\mathcal{X}}(1-z,1-\bar{z}),
\label{eq:EucCrossing}
\end{align}
where as a reminder, $\mathcal{X}$ are single-trace operators in the $\f\times \p $ OPE. The Regge crossing equation is:
\begin{align}
\hspace{-.5cm}
&(z\bar{z})^{-\Delta_{\f}}\mathcal{G}^{(1)\circlearrowleft}(z,\bar{z})
\approx\sum_{h,\bar{h}}P^{MFT}_{h,\bar{h}}\left[ i\pi \gamma^{(1)}_{h,\bar{h}}+ \delta P^{(1)}_{h,\bar{h}}\right]g^{a,a}_{h,\bar{h}}(1-z,1-\bar{z}), 
\label{eq:treeCrossingNum2}
\end{align}
where we dropped the single-trace operators and derivatives with respect to $h$ and $\bar{h}$ since they are both subleading in the Regge limit. 

The sum over the double-traces in \eqref{eq:treeCrossingNum2} is responsible for producing the Regge growth $\mathcal{G}^{(1)\circlearrowleft}(z,\bar{z})\sim (z\bar{z})^{\frac{1}{2}(1-j(0))}$, with the explicit results for the OPE data given in \eqref{eq:treegamma} and \eqref{eq:treedP}. On the other hand, we also know $\mathcal{G}^{(1)}(z,\bar{z})$ is bounded in the limit $z,\bar{z}\rightarrow0$. This is the Euclidean OPE limit, in which case $\mathcal{G}^{(1)}(z,\bar{z})\approx (z\bar{z})^{\frac{1}{2}\Delta_{min}}$ where $\Delta_{min}$ is the lightest, non-identity operator.

The problem is both \eqref{eq:EucCrossing} and \eqref{eq:treeCrossingNum2} contain the same dependence on $\delta P^{(1)}_{h,\bar{h}}$ and from \eqref{eq:treeCrossingNum2} we know this sum produces a divergence in the limit $z,\bar{z}\rightarrow 0$. Each power of the anomalous dimension in \eqref{eq:EucCrossing} comes with the derivative $(\partial_{h}+\partial_{\bar{h}})$, so these terms give a comparatively softer divergence in the same limit and cannot cancel the divergence from $\delta P^{(1)}_{h,\bar{h}}$.
 
To remedy this problem, it was proposed in \cite{Li:2017lmh} that the following sum rule hold:
\begin{align}
\sum\limits_{h,\bar{h}}P^{MFT}_{h,\bar{h}}\delta P^{(1)}_{h,\bar{h}}g^{a,a}_{h,\bar{h}}(1-z,1-\bar{z})\approx -\sum\limits_{\mathcal{X}}P^{(1)}_{\mathcal{X}}g^{a,a}_{\mathcal{X}}(1-z,1-\bar{z}), \label{eqn:sumruleST}
\end{align}
in the limit $z,\bar{z}\rightarrow 0$ with $z/\bar{z}$ fixed. This sum rule is consistent with unitarity, or $P^{(1)}_{\mathcal{X}}>0$, because for $1\leq j(0)<2$ we have $\delta P^{(1)}_{h,\bar{h}}<0$. This relation is a restatement of the optical theorem: the imaginary part of the tree-level phase shift, which is proportional to $\delta P^{(1)}_{h,\bar{h}}$, is related to the production of new states, or the single-trace operators $\mathcal{X}$. We know the left hand side of \eqref{eqn:sumruleST} contains a divergence not present in any individual block, so the same divergence must also appear on the right hand side and come from the tail of the sum. 

Next, we observe the single-trace operators, $\mathcal{X}$, will also contribute to $\dDisc_{t}[\mathcal{G}(z,\bar{z})]$. To ensure our results are consistent, we should plug this single-trace sum into the inversion formula and be able to reproduce the $s$-channel Regge behavior. We then have to consider the sum:
\begin{align}
\dDisc_{t}[\mathcal{G}^{(1)}(z,\bar{z})]=\frac{(z\bar{z})^{\Delta_{\f}}}{\left[(1-z)(1-\bar{z})\right]^{\frac{\Delta_{\f}+\Delta_{\p}}{2}}}\sum\limits_{\mathcal{X}}2P^{(1)}_{\mathcal{X}}\sin^{2}\left(\frac{\pi}{2}(2h_{\mathcal{X}}-\Delta_\f-\Delta_\p)\right)g^{a,a}_{\mathcal{X}}(1-z,1-\bar{z}), \label{eq:dDiscSTSum}
\end{align}
where we are only interested in the leading divergence of \eqref{eq:dDiscSTSum} when $z,\bar{z}\rightarrow0$. We will assume that the spectrum is sufficiently irregular that when we isolate the divergent piece of \eqref{eq:dDiscSTSum}, which can only come from the tail of the sum, that the $\sin^{2}$ term simply averages.\footnote{This is equivalent to our earlier assumption that the single-trace operators add out of phase and cannot reproduce the $s$-channel Regge growth, see the discussion below \eqref{eq:treeCrossing}.} If we make the replacement $2\sin^{2}\left(\frac{\pi}{2}(2h_{\mathcal{X}}-\Delta_\f-\Delta_\p)\right)\rightarrow 1$ when taking the Regge limit of \eqref{eq:dDiscSTSum}, we find:
\begin{align}
\dDisc_{t}[\mathcal{G}^{(1)}(z,\bar{z})]&=\frac{(z\bar{z})^{\Delta_{\f}}}{\left[(1-z)(1-\bar{z})\right]^{\frac{\Delta_{\f}+\Delta_{\p}}{2}}}\sum\limits_{\mathcal{X}}2P^{(1)}_{\mathcal{X}}\sin^{2}\left(\frac{\pi}{2}(2h_{\mathcal{X}}-\Delta_\f-\Delta_\p)\right)g^{a,a}_{\mathcal{X}}(1-z,1-\bar{z})
\nonumber \\[7pt]& \approx \frac{(z\bar{z})^{\Delta_{\f}}}{\left[(1-z)(1-\bar{z})\right]^{\frac{\Delta_{\f}+\Delta_{\p}}{2}}}\sum\limits_{\mathcal{X}}P^{(1)}_{\mathcal{X}}g^{a,a}_{\mathcal{X}}(1-z,1-\bar{z})
\nonumber \\[7pt] &\approx - \frac{(z\bar{z})^{\Delta_{\f}}}{\left[(1-z)(1-\bar{z})\right]^{\frac{\Delta_{\f}+\Delta_{\p}}{2}}}\sum\limits_{h,\bar{h}}P^{MFT}_{h,\bar{h}}\delta P^{(1)}_{h,\bar{h}}g^{a,a}_{h,\bar{h}}(1-z,1-\bar{z}).
\end{align}
In the second line we assumed the $\sin^{2}$ terms average and in the third line we used the sum rule \eqref{eq:dDiscSTSum} to replace the single-trace sum with a double-trace sum.

Now the problem has reduced to the same one as in Section \ref{sec:Regge_One_Lp}. We can determine the double-discontinuity in the Regge limit by approximating the double-trace blocks with the Bessel function approximation \eqref{eq:g4dBessel}, replace the sums over $h$ and $\bar{h}$ with an integral, and then use \eqref{eq:KInt} to evaluate these integrals. Plugging the result for $\dDisc_{t}[\mathcal{G}(z,\bar{z})]$ into the inversion formula determines the poles in spin of $c(\Delta,J)$ at tree-level. Finally, by using the Sommerfeld-Watson transform we can calculate the tree-level correlator in the Regge limit:
\begin{align}
\mathcal{G}^{(1)\circlearrowleft}(z,\bar{z})\approx 2\pi i\int\limits_{-\infty}^{\infty}d\nu \alpha(\nu)(z\bar{z})^{\frac{1}{2}(1-j(\nu))}\Omega_{i\nu}\left(\frac{1}{2}\log(z/\bar{z})\right).
\end{align}
This is of course the contribution of the Pomeron to the Regge limit. Therefore, the single-trace sum which was required to make the Regge and Euclidean crossing equations consistent is also responsible for producing the Pomeron from the inversion formula.\footnote{The problem of reproducing the Pomeron or graviton pole from the inversion formula, and its implications, has also been studied in \cite{Caron-Huot:2017vep,Kraus:2018pax,Kologlu:2019bco}.} 

Note that we are not double counting in this problem, at tree-level $\dDisc_{t}[\mathcal{G}(z,\bar{z})]$ is solely determined by the single-trace operators. The double-trace operators first contribute at one-loop with a term proportional to the anomalous dimensions squared. The corrections to the OPE coefficients of double-trace operators only appeared at tree-level here because we used the sum rule \eqref{eqn:sumruleST} to evaluate the single-trace sum.

\subsubsection{One-Loop}
\label{sec:MassiveSTLoop}

At one-loop we can ask a similar question: does requiring consistency between the Euclidean and Regge limits imply there are new single-trace sums we have to consider? More importantly, can these single-trace sums also contribute at leading order in the Regge limit? It is easy to see from our results in Section \ref{sec:Regge_One_Lp} that this is not the case and our earlier conclusions are self-consistent. To show this, let us compare the crossing equations in the OPE and Regge limits at one-loop. The Euclidean crossing equation is:
\begin{align}
(z\bar{z})^{-\Delta_{\f}}\mathcal{G}^{(2)}(z,\bar{z}) \ \approx \ \ & \sum_{h,\bar{h}}
P^{MFT}_{h,\bar{h}}\bigg[\frac{1}{2}\gamma^{(2)}_{h,\bar{h}}(\partial_{h}+\partial_{\bar{h}})+\delta P^{(2)}_{h,\bar{h}}+\frac{1}{2}\gamma^{(1)}_{h,\bar{h}}\delta P^{(1)}_{h,\bar{h}}(\partial_{h}+\partial_{\bar{h}})
\nonumber \\ &\hspace{.75in}+\frac{1}{8}\left(\gamma^{(1)}_{h,\bar{h}}\right)^{2}\left(\partial_{h}+\partial_{\bar{h}}\right)^{2}\bigg]g^{a,a}_{h,\bar{h}}(1-z,1-\bar{z})
\nonumber \\ +&\sum\limits_{\mathcal{X}}\left(P^{(2)}_{\mathcal{X}}+P^{(1)}_{\mathcal{X}}\gamma^{(1)}_{\mathcal{X}}\partial_{\Delta_{\mathcal{X}}}\right)g^{a,a}_{\mathcal{X}}(1-z,1-\bar{z}).
\label{eq:OPEloopCrossingV3}
\end{align}
In the last line of \eqref{eq:OPEloopCrossingV3} we allowed for one-loop corrections to the OPE coefficients and dimensions of the single-trace operators. For simplicity, we assume there are no new single-traces which first appear at one-loop, but it is straightforward to include them.

The one-loop Regge crossing equation is:
\begin{align}
(z\bar{z})^{-\Delta_{\f}}\mathcal{G}^{(2)\circlearrowleft}(z,\bar{z})\approx \sum_{h,\bar{h}}
P^{MFT}_{h,\bar{h}}\bigg[&i\pi \gamma^{(2)}_{h,\bar{h}}+\delta P^{(2)}_{h,\bar{h}}+i\pi\gamma^{(1)}_{h,\bar{h}}\delta P^{(1)}_{h,\bar{h}}-\frac{\pi^{2}}{2}\left(\gamma^{(1)}_{h,\bar{h}}\right)^{2}\bigg]g^{a,a}_{h,\bar{h}}(1-z,1-\bar{z}),
\label{eq:RegloopCrossingV3}
\end{align}
where again we only keep the leading divergences in the Regge limit.

Comparing \eqref{eq:OPEloopCrossingV3} and \eqref{eq:RegloopCrossingV3}, we see the term which can give the strongest divergence in \eqref{eq:OPEloopCrossingV3} is $\delta P^{(2)}_{h,\bar{h}}$. However, we already showed in \eqref{eq:dpOneLoop} it does not grow like $(h\bar{h})^{2(j(0)-1)}$. Since all the other terms in \eqref{eq:OPEloopCrossingV3} come with factors of $(\partial_{h}+\partial_{\bar{h}})$, their growth in the limit $z,\bar{z}\rightarrow 0$ is softened in comparison to the divergences found in \eqref{eq:RegloopCrossingV3}.

Of course, as each power of $(\partial_{h}+\partial_{\bar{h}})$ only reduces the Regge growth by $(z\bar{z})^{\frac{1}{4}}$, the sums involving anomalous dimensions in \eqref{eq:OPEloopCrossingV3} can yield divergences incompatible with the Euclidean OPE. These divergences will generically be cancelled by the single-trace operators. However, the resulting sums over single-trace operators, when plugged into the Lorentzian inversion formula, will give a subleading contribution to the Regge limit in comparison to the sums over the double-trace operators $[\f\p]_{n,J}$ considered in Section \ref{sec:Regge_One_Lp}. Our previous results, which concerned the leading Regge growth at each loop order, are therefore not affected.

\subsubsection{Exact Exponentiation}

Finally, we will ask a different question: if we tune the one-loop single-trace contribution to the double-discontinuity, is it possible to recover exact exponentiation of the tree-level phase shift at arbitrary $\Delta_{gap}$? This is motivated by the question of how to reproduce the results of \cite{Brower:2007xg}, where they demonstrated that summing the Pomeron ladder diagrams in impact parameter space is equivalent to exponentiating the tree-level phase shift. We will show that in order to reproduce their one-loop result, we need to impose the one-loop sum rule:
\begin{align}
\sum\limits_{\mathcal{X}}\left(P^{(2)}_{\mathcal{X}}+P^{(1)}_{\mathcal{X}}\gamma^{(1)}_{\mathcal{X}}\partial_{\Delta_{\mathcal{X}}}\right)g^{a,a}_{\mathcal{X}}(1-z,1-\bar{z})\approx-\sum\limits_{h,\bar{h}}\frac{1}{2}\left(\delta P^{(1)}_{h,\bar{h}}\right)^{2}g^{a,a}_{h,\bar{h}}(1-z,1-\bar{z}), \label{eq:SumRuleOneLoop}
\end{align}
in the limit $z,\bar{z}\rightarrow 0$. 

To prove this, we first need to include the single-trace operators and their one-loop corrections in the Lorentzian inversion formula. However, to use the sum rule (\ref{eq:SumRuleOneLoop}) in the inversion formula we will also need to assume: 
\begin{align}
\sum\limits_{\mathcal{X}}\left(P^{(2)}_{\mathcal{X}}+P^{(1)}_{\mathcal{X}}\gamma^{(1)}_{\mathcal{X}}\partial_{\Delta_{\mathcal{X}}}\right)&2\sin^{2}\left(\frac{\pi}{2}(\Delta_{\mathcal{X}}-\Delta_{\f}-\Delta_{\p})\right)g^{a,a}_{\mathcal{X}}(1-z,1-\bar{z})
\nonumber \\ &\hspace{-.15cm}\approx\sum\limits_{\mathcal{X}}\left(P^{(2)}_{\mathcal{X}}+P^{(1)}_{\mathcal{X}}\gamma^{(1)}_{\mathcal{X}}\partial_{\Delta_{\mathcal{X}}}\right)g^{a,a}_{\mathcal{X}}(1-z,1-\bar{z}),
\end{align}
to leading order when $z,\bar{z}\rightarrow 0$.

As before, we will argue this is true by recalling that the leading divergences come from the tail of the sum. First, we can note when $\partial_{\Delta_{\mathcal{X}}}$ acts on the $\sin^{2}$ term, we get factors which oscillate rapidly in $\Delta_{\mathcal{X}}$. We will assume this piece of the sum, which adds out of phase, does not produce a divergence in the limit $z,\bar{z}\rightarrow 0$. For the other terms we need to make the same assumption as before: that the spectrum is sufficiently irregular that when we compute the divergent term from the tail of the sum, the $2\sin^{2}$ term averages to one.

If these conditions hold, we can replace the single-trace contribution to $\dDisc_{t}[\mathcal{G}^{(2)}(z,\bar{z})]$ by the double-trace sum on the right hand side of \eqref{eq:SumRuleOneLoop}:
\begin{align}
\dDisc_{t}[\mathcal{G}^{(2)}(z,\bar{z})]\supset &\sum\limits_{\mathcal{X}}\left(P^{(2)}_{\mathcal{X}}+P^{(1)}_{\mathcal{X}}\gamma^{(1)}_{\mathcal{X}}\partial_{\Delta_{\mathcal{X}}}\right)2\sin^{2}\left(\frac{\pi}{2}(\Delta_{\mathcal{X}}-\Delta_{\f}-\Delta_{\p})\right)g^{a,a}_{\mathcal{X}}(1-z,1-\bar{z})
\nonumber \\ \supset &-\sum\limits_{h,\bar{h}}\frac{1}{2}\left(\delta P^{(1)}_{h,\bar{h}}\right)^{2}g^{a,a}_{h,\bar{h}}(1-z,1-\bar{z}).\label{eq:dDisc1lpSum}
\end{align}
In this case, we can follow the same procedure as in the previous section. We approximate the double-trace sum in \eqref{eq:dDisc1lpSum} by an integral and use the Bessel function approximation to calculate $\dDisc_{t}[\mathcal{G}(z,\bar{z})]$ in the Regge limit. Once again, the problem reduces to the one considered in Section \ref{sec:Regge_One_Lp}.

With this new ansatz for the single-trace operators, we now find solving crossing in the Regge limit gives:
\begin{align}
\gamma^{(2)}_{h,\bar{h}}&\approx0,
\\
\delta P^{(2)}_{h,\bar{h}}&\approx\frac{1}{2}\left(\delta P^{(1)}_{h,\bar{h}}\right)^{2},
\end{align}
at leading order for large $h$ and $\bar{h}$. The sum rule \eqref{eq:SumRuleOneLoop} guarantees that this result for $\delta P^{(2)}_{h,\bar{h}}$ is compatible with the standard OPE when plugged into the one-loop, Euclidean crossing equation \eqref{eq:OPEloopCrossingV3}. Any unphysical divergence in \eqref{eq:OPEloopCrossingV3} from the $\delta P^{(2)}_{h,\bar{h}}$ term will be cancelled by the sum over single-trace operators.

Finally, we find the full, one-loop correlator in the Regge limit is:
\begin{align}
(z\bar{z})^{-\Delta_{\f}}\mathcal{G}^{(2)\circlearrowleft}(z,\bar{z})\approx \sum_{h,\bar{h}}
P^{MFT}_{h,\bar{h}}\frac{1}{2}\bigg[\left(\delta P^{(1)}_{h,\bar{h}}\right)^{2}+2i\pi\gamma^{(1)}_{h,\bar{h}}\delta P^{(1)}_{h,\bar{h}}-\pi^{2}\left(\gamma^{(1)}_{h,\bar{h}}\right)^{2}\bigg]g^{a,a}_{h,\bar{h}}(1-z,1-\bar{z}).
\label{eq:RegLoopFinalV2}
\end{align}
In this form the connection to exponentiation is manifest. The eikonal ansatz, (\ref{eq:EikB})-(\ref{eq:EikIm}), says that at one-loop we should have:
\begin{align}
\mathcal{B}^{(2)}_{eik}(S,L)=\frac{\mathcal{N}}{2}\left[\left(\delta P^{(1)}_{h,\bar{h}}\right)^{2}+2\pi i\gamma^{(1)}_{h,\bar{h}}\delta P^{(1)}_{h,\bar{h}}-\pi^{2}\left(\gamma^{(1)}_{h,\bar{h}}\right)^{2}\right], \label{eq:eikAnsatzFinal}
\end{align}
with the standard identifications $S=h\bar{h}$ and $L= \log(\bar{h}/h)$. To compare these two results, we perform the sums in \eqref{eq:RegLoopFinalV2} in the Regge limit and then transform to impact parameter space. In the end, we find exact agreement with the eikonal ansatz \eqref{eq:eikAnsatzFinal}. 

It is important to emphasize that we are not claiming that if the sum rule \eqref{eq:dDisc1lpSum} holds, then the eikonal ansatz \eqref{eq:EikForm} gives a good approximation for the Regge limit at arbitrary $\Delta_{gap}$. As we will discuss in the next section, there are other inelastic effects which appear at $\Delta_{gap}^{-4}$ we have not yet included which will correct the eikonal ansatz. Rather, we are making the following more modest claim: if the one-loop sum rule holds, then the correlator at one-loop contains a term which is consistent with exponentiating the tree-level answer, and therefore the bulk Pomeron loop calculation \cite{Brower:2007xg}, but there will be further corrections we have to consider.

\subsection{Tidal Excitations}
\label{ssec:tidal}
Finally, let us study corrections to the Regge limit from the exchange of more general double-trace operators in the $t$-channel. This problem is simple enough that we can focus directly on the correlator in the Regge limit, $\mathcal{G}^{\circlearrowleft}(z,\bar{z})$. In general, it has the expansion:
\begin{align}
\hspace{-.5cm}
e^{i \pi (\Delta_{\f}+\Delta_{\p})}&(z\bar{z})^{-\Delta_{\f}}\mathcal{G}^{\circlearrowleft}(z,\bar{z})
=\sum_{\mathcal{O}}e^{2i\pi h_{\mathcal{O}}}P_{\mathcal{O}}g^{a,a}_{\mathcal{O}}(1-z,1-\bar{z}),
\label{eq:t_Expansion_Regge}
\end{align}
where the sum runs over all operators in the spectrum, single and multi-trace.

One of the arguments we made in Section \ref{ssec:Tree_Crossing}, and specifically when solving crossing at tree-level in the Regge limit, is that only the double-trace operators $[\f\p]_{n,J}$ add in phase and can lead to Regge behavior. For these double-trace operators:
\begin{align}
h_{[\f\p]_{n,J}}=h_{\f}+h_{\p}+n+\frac{\gamma_{h,\bar{h}}}{2},
\end{align}
so the phase factor in (\ref{eq:t_Expansion_Regge}) is constant when we expand $\gamma_{h,\bar{h}}$ in $1/N$. One can also note that this feature of adding in phase holds for all multi-trace operators. To take these into account we need to recall the standard large $N$ counting:
\begin{align}
\<\f\p[\O_1\O_2]_{n,J}\>\sim\frac{1}{N^{2}}, \qquad \<\f\p[\O_1...\O_m]_{\Delta,J}\>\sim\frac{1}{N^{m}},
\end{align}
where $\O_i$ are some new single-trace operators.

We are interested in one-loop effects, so we can focus on the double-trace operators. The physical picture for how new double-traces can affect the Regge limit is best explained using Witten diagrams. As far as this work is concerned, diagrams involving the Pomeron should be understood as a useful visualization tool for the boundary calculations.\footnote{The AdS Feynman rules associated to the Pomeron were studied in \cite{Brower:2007xg}.}

\begin{figure}
\centering
\includegraphics[scale=.3]{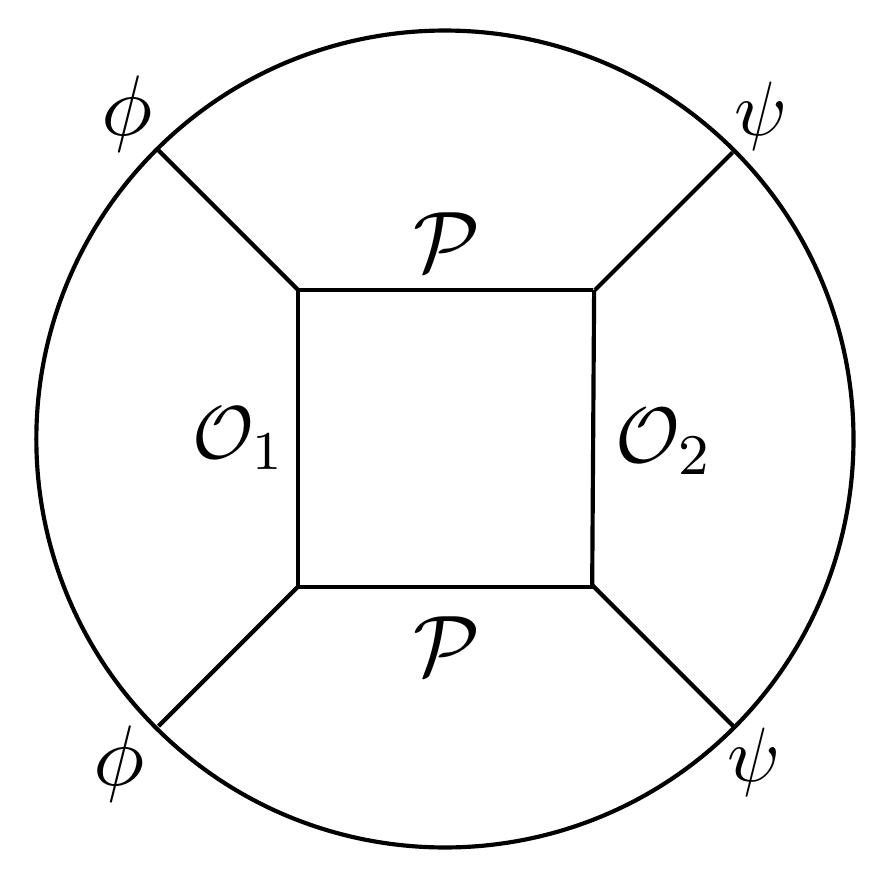}
\caption{Pomeron box graph with a generic, $t$-channel, double-trace cut.}
\label{fig:PomBoxGeneric}
\end{figure}

This caveat aside, tidal excitations can be understood using the box graph shown in figure \ref{fig:PomBoxGeneric}, where we denote the Pomeron by $\mathcal{P}$. The physical picture is that both $\f$ and $\p$ can be excited into new states, $\O_1$ and $\O_2$, after interacting with the Pomeron, or reggeized graviton. This box graph is then a way to glue these two tree-level processes together to obtain a one-loop correction to $\<\f\f\p\p\>$. The process of an incoming state becoming excited to a new state after interacting with a (reggeized) graviton is called a tidal excitation \cite{Amati:1987wq,Amati:1987uf,Amati:1988tn,Giddings:2006vu,Shenker:2014cwa}.
We also see there is a two-Pomeron cut in the $s$-channel and a $[\O_1\O_2]_{n,J}$ cut in the $t$-channel. The $t$-channel cut corresponds to the double-trace exchange:
\begin{align}
\f\p \ \rightarrow \ [\O_1\O_2]_{n,J} \ \rightarrow \ \p\f.
\end{align}
By performing the sum over the $[\O_1\O_2]_{n,J}$ family in the $s$-channel Regge limit, we will reproduce the two-Pomeron cut in the $s$-channel. In other words, the sum over this double-trace family produces the Regge growth $(z\bar{z})^{1-j(0)}$, which we also saw in Section \ref{sec:Regge_One_Lp} from summing over the $[\f\p]_{n,J}$ double-trace family.

On the other hand, if we interpret figure \ref{fig:PomBoxGeneric} as a genuine Witten diagram, it will be proportional to the couplings $\left(c_{\f\O_1 j(0)}c_{\p\O_2j(0)}\right)^{2}$, where we use $j(\nu)$ as a shorthand for the leading Regge trajectory. We are implicitly using the saddle-point approximation to set $\nu=0$. However, the coupling of two distinct operators to the Pomeron is known to be suppressed by $\Delta_{gap}^{-2}$ \cite{Meltzer:2017rtf,Afkhami-Jeddi:2018own}. So although figure \ref{fig:PomBoxGeneric} will contribute at leading order in the Regge limit, it should also be suppressed if we take $\Delta_{gap}\gg1$.

In the rest of this section, we will make the above intuition more concrete. First, to determine how $\f$ and $\p$ couple to $[\O_1\O_2]_{n,J}$, we need to study the four-point function
\begin{align}
\<\f\p\O_1\O_2\>,
\end{align}
in the $t$-channel Regge limit.\footnote{The $t$-channel Regge limit is defined by analytically continuing $z$ counterclockwise around $z=0$ and sending $z,\bar{z}\rightarrow 1$ at the same rate. In general we should also study the $u$-channel, but the main results will be unchanged.} The problem is essentially the same as reviewed in Section \ref{sec:AdSUnitarityRegge}, except now we have four distinct operators. Assuming the Pomeron can be exchanged in the $t$-channel, we find the following scaling at large $h$ and $\bar{h}$ for the $s$-channel OPE coefficients:
\begin{align}
\frac{1}{2^{J}}c_{\f\p[\O_1\O_2]_{n,J}}^{2}\sim \frac{1}{N^{-4}}2^{-2h-2\bar{h}}(h\bar{h})^{2j(0)+\Delta_{\f}+\Delta_{\p}-9/2},
\end{align}
where we made the usual identification $h\approx n$ and $\bar{h}\approx n+J$. Plugging this into the $t$-channel expansion of $\<\f\f\p\p\>$ we find the one-loop Regge behavior:
\begin{align}
\mathcal{G}^{(2)\circlearrowleft}(z,\bar{z})\sim (z\bar{z})^{1-j(0)}.
\end{align}
This implies the $t$-channel exchange, $\f\p\ \rightarrow \ [\O_1\O_2]_{n,J} \ \rightarrow \ \p\f$, will lead to the same Regge growth at one-loop as we saw in Section \ref{sec:Regge_One_Lp}.

Next, we will use unitarity arguments and the stress-tensor Ward identity to show this contribution is suppressed at large $\Delta_{gap}$, see \cite{Costa:2017twz,Meltzer:2017rtf} for more details. To show this, we note the size of the Pomeron contribution to the tree-level, $t$-channel Regge limit of $\<\f\p\O_1\O_2\>$ is proportional to:
\begin{align}
r_{\O_1\O_2}(\nu)=c_{\f\O_2 j(\nu)}c_{\O_1\p j(\nu)}K_{2+i\nu,j(\nu)}.
\end{align}
Suppressing the dependence on $n$ and $J$, crossing symmetry for $\<\f\p\O_1\O_2\>$ in the $t$-channel Regge limit tells us:
\begin{align}
c_{\f\p[\O_1\O_2]_{n,J}}\sim r_{\O_1\O_2}(0),
\end{align}
when $n,J \gg1$ or equivalently when $h,\bar{h}\gg1$.

Finally, we will use the existence of a flat space limit to constrain the size of $r_{\O_1\O_2}(0)$. In general it has the following expansion at large $\Delta_{gap}$ \cite{Costa:2017twz}:
\begin{align}
r_{\O_1\O_2}(\nu)=r_{\O_1\O_2}(0)+\sum\limits_{n=1}^{\infty}\frac{P_{n}(\nu^2)}{\Delta_{gap}^{2n}}, \label{eq:rExpRegge}
\end{align}
where $P_{n}(\nu^{2})$ are polynomials of degree $n$ such that $P_{n}(0)=0$. This form is fixed by requiring that $r(\nu)$ is finite in the limit $\nu^{2}\rightarrow\infty$ with $\nu/\Delta_{gap}$ held fixed. This expansion implies that in theories with a large gap, the value of $r_{\O_1\O_2}(\nu)$ at the stress-tensor point, $\nu=-2i$ in $d=4$, can differ from its value at the intercept point, $\nu=0$, at most by a $\Delta_{gap}^{-2}$ amount. 

This constraint on the OPE coefficients has been used \cite{Costa:2017twz,Meltzer:2017rtf} to bound couplings to the stress-tensor by first bounding the couplings to the Pomeron. Here we are interested in a different problem, we want to directly bound how two local operators couple to the Pomeron. The simplest case to start with is when $\O_1$ and $\O_2$ are two, new, scalar operators. In that case the Ward identity implies $\<\f \O_1 T\>=\<\p\O_2T\>=0$ or:
\begin{align}
r_{\O_1\O_2}\left(-2i\right)=0.
\end{align}
Then, at the $\nu=0$ point we must have $r_{\O_1\O_2}(0)\lesssim\Delta_{gap}^{-2}$.\footnote{We expect the suppression should be $\Delta_{gap}^{-4}$ if $\O_{1,2}$ are new scalar operators and $\Delta_{gap}^{-2}$ if for example $\O_1=\p$ and $\O_2$ is a new scalar operator.} This gives us the overall scaling:
\begin{align}
c_{\f\p[\O_1\O_2]_{n,J}}^{2}\lesssim \Delta_{gap}^{-4}.
\end{align}
Therefore, the Regge growth due to $[\O_1\O_2]_{n,J}$ exchange in the $t$-channel of $\<\f\f\p\p\>$ is necessarily suppressed at large $\Delta_{gap}$.

Next, let us consider the problem when $\O_1=\p$, but $\O_2$ is a new spinning operator. Now we are able to bound the coupling $c_{\O_2\f j(0)}$ directly. This was proven explicitly in \cite{Meltzer:2017rtf} when $\O_2$ is a spin-one operator, $V$,\footnote{The coupling $\<TV\f\>$ is only non-zero when $\Delta_{\O}=\Delta_{V}\pm1$. If this condition is not satisfied then $r(-id/2)=0$ and we can use the same argument as when $\O_2$ was a scalar.} or the stress-tensor itself, $T$. For the latter case, it was shown the OPE coefficients vanish at the $\nu=0$ point:
\begin{align}
r_{\p T}(0)=0.
\end{align}
This does not imply the Pomeron gives a vanishing contribution to the $t$-channel Regge limit of $\<\f\p\p T\>$, only that its total contribution is suppressed at large $\Delta_{gap}$. From the general form (\ref{eq:rExpRegge}) and by imposing crossing symmetry of $\<\f\p\p T\>$ we learn:
\begin{align}
c^{2}_{\f\p[\p T]_{n,J}}\lesssim \Delta_{gap}^{-4}.
\end{align}
Once again, we have found that although new double-trace operators can affect the leading one-loop Regge behavior, unitarity and Ward identities imply this is an effect which is suppressed when $\Delta_{gap}$ is large.

We can also consider more general cases, e.g. when $\O_2$ is a non-conserved, spin-two operator, $M$, or a general, spin-$J$, single-trace operator $X_{\Delta,J}$. Bounds on higher-spin operators have been studied less\footnote{See \cite{Afkhami-Jeddi:2018apj} for constraints on higher-spin particles which sit below $\Delta_{gap}$ and \cite{Meltzer:2018tnm} for bounds on generic CFTs.}, but we expect the general ideas of \cite{Costa:2017twz,Meltzer:2017rtf} should carry over. From these arguments we expect at least the suppression:
\begin{align}
c_{\f j(0)X_{\Delta,J}}\lesssim \Delta_{gap}^{-J}. \label{eq:pomCouplingScaling}
\end{align}
It would be nice to show this explicitly, either by using the previously mentioned unitarity arguments \cite{Costa:2017twz,Meltzer:2017rtf} or by using light-ray operators \cite{Belin:2019mnx,Kologlu:2019bco,Kologlu:2019mfz}.

\section{Discussion}
\label{sec:Discussion}
In this paper, we have studied the Regge limit at one-loop and higher for large $N$, $4d$ CFTs. We studied CFTs with higher-spin, single-trace operators which are dual to AdS theories with higher-spin particles. In such theories, the Regge limit at tree-level is controlled by the leading, single-trace, Regge trajectory. Solving crossing at tree-level then gives corrections to both the anomalous dimensions and OPE coefficients of double-trace operators. In the AdS theory, this implies the phase shift from two-to-two scattering at high energies and fixed impact parameter contains both a real and imaginary piece.

By applying crossing symmetry, we then used the tree-level data to study the Regge limit to all orders in $1/N$. We have argued that in large $N$ CFTs, when $\Delta_{gap}$ is large, the tree-level phase shift exponentiates. To prove this, we showed that other possible corrections to the Regge limit, such as tidal excitations or higher-loop corrections to the double-trace operators, are suppressed when $\Delta_{gap}$ is large. We have also explained how imposing sum rules on the single-trace operators can yield exponentiation of the tree-level answer at arbitrary $\Delta_{gap}$.

To fully understand the Regge limit at higher loops and finite $\Delta_{gap}$, we need a more detailed understanding of tidal excitations. To do this, it is useful to first consider the Regge limit of flat-space, string scattering amplitudes. This was studied in \cite{Amati:1987wq,Amati:1987uf,Amati:1988tn}\footnote{For further work on string theory in the Regge limit see \cite{Amati:1988ww,Amati:1990xe,Amati:1992zb,Amati:1993tb,Muzinich:1987in,Soldate:1986mk,Sundborg:1988tb,Ademollo:1989ag,Ademollo:1990sd,Bellini:1992eb,Giddings:2007bw}.}, where they showed that it was possible to include tidal excitations in the eikonal approximation by promoting the $c$-number phase shift, $\delta$, to an operator, $\hat{\delta}$, on the string Hilbert space.\footnote{This is also reviewed in \cite{Kologlu:2019bco} where they showed the connection to commutativity of shockwaves.} In this case, the phase shift operator will act on the two initial states and produce two, generic outgoing states. If operator eikonalization could be derived in CFTs with higher-spin, single-trace operators, it would yield more evidence that the bulk theory necessarily contains strings.

To understand exponentiation and Regge behavior for more general observables, it will also help to further study how the bulk worldsheet picture considered in \cite{Brower:2006ea,Brower:2007xg,Shenker:2014cwa} is encoded by the boundary CFT. In \cite{Brower:2006ea} they identified the Pomeron as a worldsheet vertex operator and studied it in a semi-classical approximation. One promising avenue to study this from the CFT perspective would be via light-ray operators and their OPE\cite{Banks:2009bj,Hofman:2008ar,Kravchuk:2018htv,Kologlu:2019bco,Kologlu:2019mfz}. It was shown in \cite{Kravchuk:2018htv} that the Pomeron is a generalized light-ray operator $\mathbb{O}_{\Delta,J}$. As one possible application, in this work we derived the existence of two-Pomeron Regge behavior at one-loop by performing sums over local, double-trace operators. The same behavior was derived in \cite{Brower:2007xg}, but in their work they glued together two Pomeron-particle amplitudes. To understand such amplitudes from the CFT perspective will require studying correlation functions with external, light-ray operators, $\mathbb{O}_{\Delta,J}$, of general spin.

We have also made several technical assumptions which can be relaxed. For example, we focused on $d=4$ because the conformal blocks are known in closed form. However, it should be possible to generalize our analysis to arbitrary spacetime dimensions, either by using the impact parameter blocks \cite{Cornalba:2006xm,Cornalba:2007zb,Kulaxizi:2018dxo,Karlsson:2019qfi} or by using the Lorentzian inversion formula to calculate the double-trace data at large $h$ and $\bar{h}$ \cite{Caron-Huot:2017vep}. We also restricted to external, scalar operators for simplicity, but it is in principle straightforward to study external, spinning operators \cite{Li:2015itl,Hofman:2016awc,Elkhidir:2017iov,Sleight:2018epi,Sleight:2018ryu,Albayrak:2019gnz} using differential operators \cite{Costa:2011mg,Costa:2011dw,Karateev:2017jgd,Costa:2018mcg}. One particularly interesting problem is to consider correlation functions of operators on the leading Regge trajectory itself. We expect causality and unitarity will give an intricate set of constraints on these couplings which will yield new information on the space of weakly coupled, gravitational theories with higher-spin particles.

Another interesting, Lorentzian limit we have not discussed thus far is the bulk-point limit \cite{Gary:2009ae,Okuda:2010ym,Maldacena:2015iua}. By studying this limit, we can also probe bulk, high-energy scattering at fixed angles. It is well-known that string theory is exponentially soft in this limit \cite{Gross:1987ar,Gross:1987kza,Mende:1989wt} and this behavior does not occur in Einstein gravity or QFT. Understanding when this behavior emerges from the boundary CFT, and the constraints it imposes on the OPE, is crucial to understanding which CFTs have string theory duals.\footnote{For recent work on how Gross-Mende behavior bounds the Mellin amplitude see \cite{Dodelson:2019ddi}.} 

Finally, we should mention that in general the Regge limit is less understood in comparison to the Euclidean or lightcone OPE. This is due to its intrinsically Lorentzian nature and the fact it is best understood as an expansion in terms of non-local, light-ray operators \cite{Hofman:2008ar,Banks:2009bj,Kravchuk:2018htv}. In terms of the OPE function, $c(\Delta,J)$, the reason is we need to understand its singularities for general complex $J$, while the Euclidean OPE tells us about singularities in $\Delta$ at fixed, integer $J$. We therefore hope that the work presented here will help in elucidating the Regge limit for general, non-perturbative CFTs.

\section*{Acknowledgements}
We thank Soner Albayrak, Dean Carmi, Petr Kravchuk, Daliang Li, Eric Perlmutter, David Poland, David Simmons-Duffin, and Allic Sivaramakrishnan for discussions. The research of DM is supported by the Walter Burke Institute for Theoretical Physics and the Sherman Fairchild Foundation.  This material is based upon work supported by the U.S. Department of Energy, Office of Science, Office of High Energy Physics, under Award Number DE-SC0011632.

\appendix
\section{Definitions and Integrals}
\label{app:Conv}
In this work we follow the standard conventions for CFT two and three-point functions. If we denote scalar operators by $\f_i$ and spinning operators by $\O_{\Delta,J}$, then the kinematic two and three-point functions are:
\begin{align}
&\<\O_{\Delta,J}(x_1,z_1)\O_{\Delta,J}(x_2,z_2)\>=\frac{(z_1\cdot I(x_{12})\cdot z_2)^{J}}{x_{12}^{2\Delta_{\O}}}, \hspace{1.15in} I_{\mu\nu}(x)=\delta_{\mu\nu}-2\frac{x_{\mu}x_{\nu}}{x^{2}},
\\
&\<\f_1(x_1)\f_2(x_2)\O_{\Delta_3,J_3}(x_3,z_3)\>=\frac{(X_3\cdot z_3)^{J_3}}{x_{12}^{\Delta_{12,3}+J_3}x_{13}^{\Delta_{13,2}-J_3}x_{23}^{\Delta_{23,1}-J_3}}, \qquad X_{3}^{\mu}=\frac{x_{13}^{\mu}}{x_{13}^{2}}-\frac{x_{23}^{\mu}}{x_{23}^{2}}.
\end{align}
where $\Delta_{ij,k}=\Delta_i+\Delta_j-\Delta_k$ and $\O_{\Delta,J}(x,z)=\O_{\Delta,J}^{\mu_1...\mu_J}z_{\mu_1}...z_{\mu_J}$. 

If we consider a physical three-point function, which we denote by $\<\f_1\f_2\O_{\Delta_3,J_3}\>_{\Omega}$ then:
\begin{align}
\<\f_1\f_2\O_{\Delta_3,J_3}\>_{\Omega}=c_{123}\<\f_1\f_2\O_{\Delta_3,J_3}\>,
\end{align}
where $c_{123}$ are the OPE coefficients.

The shadow coefficients which relate the conformal partial waves, $F_{\Delta,J}(z,\bar{z})$, to the conformal blocks, $g_{\Delta,J}(z,\bar{z})$, in \eqref{eq:CPW_Def} are defined by:
\begin{align}
S^{\f_1\f_2}_{\Delta,J}=\frac{\pi^{\frac{d}{2}}\Gamma(\Delta-\frac{d}{2})\Gamma(\Delta+J-1)\Gamma(\frac{\widetilde{\Delta}+\Delta_1-\Delta_2+J}{2}) \Gamma(\frac{\widetilde{\Delta}+\Delta_2-\Delta_1+J}{2})}{\Gamma(\Delta-1)\Gamma(\widetilde{\Delta}+J)\Gamma(\frac{\Delta+\Delta_1-\Delta_2+J}{2}) \Gamma(\frac{\Delta+\Delta_2-\Delta_1+J}{2}) }. \label{eq:ShadowDef}
\end{align}
where $\widetilde{\Delta}=d-\Delta$.

Furthermore, our conventions for the $2d$ and $4d$ conformal blocks are:
\begin{align}
g^{(2d),ab}_{h,\bar{h}}&=\frac{1}{1+\delta_{h,\bar{h}}}k_{h}(z)k_{\bar{h}}(\bar{z})+(z\leftrightarrow\bar{z}),
\\
g^{(4d),ab}_{h,\bar{h}}&=\frac{z\bar{z}}{z-\bar{z}}k_{h}(z)k_{\bar{h}}(\bar{z})+(z\leftrightarrow\bar{z}), \label{eq:4dBlocks}
\end{align}
where we also defined the $SL(2,\mathbb{R})$ blocks
\begin{align}
k_{h}(z)=z^{h}{}_2F_1(h+a,h+b,2h;z).
\end{align}
In the body of the paper we only used the $4d$ blocks, so we suppressed this label, but in Appendix \ref{sec:Regge2d} we give results which require the $2d$ blocks.

In \eqref{eq:alpha_def_Regge} we also introduced the factor $K$ defined by:
\begin{align}
K_{\Delta,J}\ = \ &\frac{\Gamma(\Delta+J)\Gamma\left(\Delta+1-\frac{d}{2}\right)(\Delta-1)_{J}}{4^{J-1}\Gamma(\frac{\Delta+J+\Delta_{12}}{2})\Gamma(\frac{\Delta+J-\Delta_{12}}{2})\Gamma(\frac{\Delta+J+\Delta_{34}}{2})\Gamma(\frac{\Delta+J-\Delta_{34}}{2})}
\nonumber \\ &\frac{1}{\Gamma\left(\frac{\Delta_1+\Delta_2-\Delta+J}{2}\right)\Gamma\left(\frac{\Delta_3+\Delta_4-\Delta+J}{2}\right)\Gamma\big(\frac{\Delta_1+\Delta_2-\widetilde{\Delta}+J}{2}\big)\Gamma\big(\frac{\Delta_3+\Delta_4-\widetilde{\Delta}+J}{2}\big)}.
\label{eq:KDef}
\end{align}

Finally the harmonic function $\Omega_{i\nu}(L)$ for $H_{d-1}$ is:
\begin{align}
\Omega_{i\nu}(L)=\frac{\nu sinh(\pi \nu)\Gamma(\frac{d}{2}-1+i\nu)\Gamma(\frac{d}{2}-1-i\nu)}{2^{d-1}\pi^{\frac{d+1}{2}}\Gamma(\frac{d-1}{2})}{}_2F_1\left(\frac{d}{2}-1+i\nu,\frac{d}{2}-1-i\nu,\frac{d-1}{2},-sinh^{2}\left(\frac{L}{2}\right)\right) \label{eq:HarmDefd}.
\end{align}


\section{CFTs in $d=2$}
\label{sec:Regge2d}

In this paper we have focused on large $N$ CFTs in $d=4$, but everything can be repeated with very minor changes for $d=2$. We expect a similar generalization will hold for general, even spacetime dimensions where we also know the blocks in closed form. In this appendix we will focus on the $2d$ global, conformal group and restrict to light, external operators. We will not discuss Virasoro symmetry, but it would be interesting to extend our analysis to this larger symmetry.

\noindent In $2d$ the Bessel function approximation for the blocks in the limit $z^{-\frac{1}{2}}\sim\bar{z}^{-\frac{1}{2}}\sim h \sim \bar{h}\gg1$ is:
\begin{align}
g^{(2d),a,b}_{h,\bar{h}}(1-z,1-\bar{z})\approx \frac{\sqrt{h\bar{h}}}{\pi}2^{2(h+\bar{h})}K_{a+b}(2h\sqrt{z})K_{a+b}(2\bar{h}\sqrt{\bar{z}})+(z\leftrightarrow\bar{z}).
\end{align}
Other objects also simplify, such as the harmonic function,
\begin{align}
\Omega_{i\nu}(L)=\frac{cos(\nu L)}{2\pi},
\end{align}
and the MFT OPE coefficients,
\begin{align}
P^{(2d),MFT}_{h,\bar{h}}\approx\frac{\pi  2^{-2 (h+\bar{h}-2)} h^{\Delta_{\f}+\Delta_{\p}-\frac{3}{2}} \bar{h}^{\Delta_{\f}+\Delta_{\p}-\frac{3}{2}}}{\Gamma (\Delta_{\f})^2 \Gamma (\Delta_{\p})^2}.
\end{align}
One nice feature of working in $d=2$ is all the $h$ and $\bar{h}$ integrals are trivial. We will not need to introduce the bubble function $B(\nu_1,\nu_2;\nu)$ and the associated, extra spectral integral for $\nu$. Since the $2d$ analysis, when we restrict to the global conformal group, is identical in form to the $4d$ results presented in the body of the paper, we will simply state the final results here.

We again consider the correlator of pairwise identical scalars, $\<\f\f\p\p\>$, at tree-level and require that the $t$-channel expansion reproduce the leading Regge trajectory in the $s$-channel. We use the spectral representation and will leave it implicit that $d=2$:
\begin{align}
\gamma^{(1)}_{h,\bar{h}}&\approx\int\limits_{-\infty}^{\infty} d\nu \widehat{\gamma}^{(1)}(\nu)(h\bar{h})^{j(\nu)-1}\Omega_{i\nu}\left(\log(h/\bar{h})\right),
\\
\delta P^{(1)}_{h,\bar{h}}&\approx\int\limits_{-\infty}^{\infty} d\nu \widehat{\delta P}^{(1)}(\nu)(h\bar{h})^{j(\nu)-1}\Omega_{i\nu}\left(\log(h/\bar{h})\right),
\end{align}
and crossing symmetry implies:
\begin{align}
\frac{1}{N^{2}}\widehat{\gamma}^{(1)}(\nu)&=\frac{2\Re[\alpha(\nu)]\Gamma(\Delta_\f)^{2}\Gamma(\Delta_\p)^{2}}{\chi_{j(\nu)}(\nu)\chi_{j(\nu)}(-\nu)},
\\
\frac{1}{N^{2}}\widehat{\delta P}^{(1)}(\nu)&=\frac{-2\pi\Im[\alpha(\nu)]\Gamma(\Delta_\f)^{2}\Gamma(\Delta_\p)^{2}}{\chi_{j(\nu)}(\nu)\chi_{j(\nu)}(-\nu)}.
\end{align}

Performing the inversion formula in the $z,\bar{z}\rightarrow0$ limit then yields the one-loop correlator in the Regge limit:
\begin{align}
\mathcal{G}^{(2)\circlearrowleft}(z,\bar{z})\approx-\int\limits_{-\infty}^{\infty} d\nu_1d\nu_2\frac{\pi}{2}\frac{e^{i \pi  (j(\nu_1)+j(\nu_2))}}{1+e^{i \pi  (j(\nu_1)+j(\nu_2))}}\frac{\widehat{\gamma}^{(1)}(\nu_1)\widehat{\gamma}^{(1)}(\nu_2)}{\Gamma(\Delta_\f)^{2}\Gamma(\Delta_\p)^{2}}(z\bar{z})^{\frac{1}{2}(2-j(\nu_1)-j(\nu_2))}\Omega_{i(\nu_1+\nu_2)}\left(-\frac{1}{2}\log(z/\bar{z})\right)
\end{align}

We can then solve the crossing equation in the Regge limit, \eqref{eq:loopCrossingV2}, to calculate one-loop corrections to the double-trace operators:
\begin{align}
\delta P^{(2)}_{h,\bar{h}}&\approx0
\\
\gamma^{(2)}_{h,\bar{h}}&\approx\int\limits_{-\infty}^{\infty} d\nu_1d\nu_2 \widehat{\gamma}^{(2)}(\nu_1,\nu_2)(h\bar{h})^{j(\nu_1)+j(\nu_2)-2}\Omega_{i(\nu_1+\nu_2)}\left(-\frac{1}{2}\log(h/\bar{h})\right)
\\
\widehat{\gamma}^{(2)}(\nu_1,\nu_2)&=-\frac{\pi}{2} \widehat{\gamma}^{(1)}(\nu_1) \widehat{\gamma}^{(1)}(\nu_2) \tan \left(\frac{1}{2} \pi  j(\nu_1)\right) \tan \left(\frac{1}{2} \pi  j(\nu_2)\right) \tan \left(\frac{1}{2} \pi  (j(\nu_1)+j(\nu_2))\right).
\end{align}
As in the $4d$ case, we find that in the limit $\Delta_{gap}\rightarrow \infty$ or $j(\nu)\rightarrow 2$, that the leading correction to the one-loop anomalous dimension vanishes in the Regge limit. This is consistent with pure Einstein gravity in the bulk, where tree-level graviton exchange exponentiates in the Regge limit. We also find the one-loop corrections to $\gamma^{(2)}_{h,\bar{h}}$ are suppressed by $\Delta_{gap}^{-6}$ and therefore give a subleading effect at one-loop when $\Delta_{gap}$ is large.

\bibliographystyle{utphys}
\bibliography{biblio}

\providecommand{\href}[2]{#2}\begingroup\raggedright\begin{thebibliography}{100}

\bibitem{Polyakov:1974gs}
A.~M. Polyakov, ``{Nonhamiltonian approach to conformal quantum field
  theory},''
{\em Zh. Eksp. Teor. Fiz.} {\bfseries 66} (1974) 23--42.

\bibitem{Rattazzi:2008pe}
R.~Rattazzi, V.~S. Rychkov, E.~Tonni, and A.~Vichi, ``{Bounding scalar operator
  dimensions in 4D CFT},''
  \href{http://dx.doi.org/10.1088/1126-6708/2008/12/031}{{\em JHEP} {\bfseries
  12} (2008) 031},
\href{http://arxiv.org/abs/0807.0004}{{\ttfamily arXiv:0807.0004 [hep-th]}}.

\bibitem{Poland:2018epd}
D.~Poland, S.~Rychkov, and A.~Vichi, ``{The Conformal Bootstrap: Theory,
  Numerical Techniques, and Applications},''
\href{http://arxiv.org/abs/1805.04405}{{\ttfamily arXiv:1805.04405 [hep-th]}}.

\bibitem{Maldacena:1997re}
J.~M. Maldacena, ``{The Large N limit of superconformal field theories and
  supergravity},'' \href{http://dx.doi.org/10.1023/A:1026654312961,
  10.4310/ATMP.1998.v2.n2.a1}{{\em Int. J. Theor. Phys.} {\bfseries 38} (1999)
  1113--1133}, \href{http://arxiv.org/abs/hep-th/9711200}{{\ttfamily
  arXiv:hep-th/9711200 [hep-th]}}.
[Adv. Theor. Math. Phys.2,231(1998)].

\bibitem{Witten:1998qj}
E.~Witten, ``{Anti-de Sitter space and holography},''
  \href{http://dx.doi.org/10.4310/ATMP.1998.v2.n2.a2}{{\em Adv. Theor. Math.
  Phys.} {\bfseries 2} (1998) 253--291},
\href{http://arxiv.org/abs/hep-th/9802150}{{\ttfamily arXiv:hep-th/9802150
  [hep-th]}}.

\bibitem{Gubser:1998bc}
S.~S. Gubser, I.~R. Klebanov, and A.~M. Polyakov, ``{Gauge theory correlators
  from noncritical string theory},''
  \href{http://dx.doi.org/10.1016/S0370-2693(98)00377-3}{{\em Phys. Lett.}
  {\bfseries B428} (1998) 105--114},
\href{http://arxiv.org/abs/hep-th/9802109}{{\ttfamily arXiv:hep-th/9802109
  [hep-th]}}.

\bibitem{Fitzpatrick:2012yx}
A.~L. Fitzpatrick, J.~Kaplan, D.~Poland, and D.~Simmons-Duffin, ``{The Analytic
  Bootstrap and AdS Superhorizon Locality},''
  \href{http://dx.doi.org/10.1007/JHEP12(2013)004}{{\em JHEP} {\bfseries 1312}
  (2013) 004},
\href{http://arxiv.org/abs/1212.3616}{{\ttfamily arXiv:1212.3616 [hep-th]}}.

\bibitem{Komargodski:2012ek}
Z.~Komargodski and A.~Zhiboedov, ``{Convexity and Liberation at Large Spin},''
  \href{http://dx.doi.org/10.1007/JHEP11(2013)140}{{\em JHEP} {\bfseries 1311}
  (2013) 140},
\href{http://arxiv.org/abs/1212.4103}{{\ttfamily arXiv:1212.4103 [hep-th]}}.

\bibitem{Caron-Huot:2017vep}
S.~Caron-Huot, ``{Analyticity in Spin in Conformal Theories},''
\href{http://arxiv.org/abs/1703.00278}{{\ttfamily arXiv:1703.00278 [hep-th]}}.

\bibitem{Heemskerk:2009pn}
I.~Heemskerk, J.~Penedones, J.~Polchinski, and J.~Sully, ``{Holography from
  Conformal Field Theory},''
  \href{http://dx.doi.org/10.1088/1126-6708/2009/10/079}{{\em JHEP} {\bfseries
  0910} (2009) 079},
\href{http://arxiv.org/abs/0907.0151}{{\ttfamily arXiv:0907.0151 [hep-th]}}.

\bibitem{Maldacena:2012sf}
J.~Maldacena and A.~Zhiboedov, ``{Constraining conformal field theories with a
  slightly broken higher spin symmetry},''
  \href{http://dx.doi.org/10.1088/0264-9381/30/10/104003}{{\em
  Class.Quant.Grav.} {\bfseries 30} (2013) 104003},
\href{http://arxiv.org/abs/1204.3882}{{\ttfamily arXiv:1204.3882 [hep-th]}}.

\bibitem{Camanho:2014apa}
X.~O. Camanho, J.~D. Edelstein, J.~Maldacena, and A.~Zhiboedov, ``{Causality
  Constraints on Corrections to the Graviton Three-Point Coupling},''
  \href{http://dx.doi.org/10.1007/JHEP02(2016)020}{{\em JHEP} {\bfseries 02}
  (2016) 020},
\href{http://arxiv.org/abs/1407.5597}{{\ttfamily arXiv:1407.5597 [hep-th]}}.

\bibitem{Caron-Huot:2016icg}
S.~Caron-Huot, Z.~Komargodski, A.~Sever, and A.~Zhiboedov, ``{Strings from
  Massive Higher Spins: The Asymptotic Uniqueness of the Veneziano
  Amplitude},''
\href{http://arxiv.org/abs/1607.04253}{{\ttfamily arXiv:1607.04253 [hep-th]}}.

\bibitem{Brower:2006ea}
R.~C. Brower, J.~Polchinski, M.~J. Strassler, and C.-I. Tan, ``{The Pomeron and
  gauge/string duality},''
  \href{http://dx.doi.org/10.1088/1126-6708/2007/12/005}{{\em JHEP} {\bfseries
  12} (2007) 005},
\href{http://arxiv.org/abs/hep-th/0603115}{{\ttfamily arXiv:hep-th/0603115
  [hep-th]}}.

\bibitem{Brower:2007xg}
R.~C. Brower, M.~J. Strassler, and C.-I. Tan, ``{On The Pomeron at Large 't
  Hooft Coupling},''
  \href{http://dx.doi.org/10.1088/1126-6708/2009/03/092}{{\em JHEP} {\bfseries
  03} (2009) 092},
\href{http://arxiv.org/abs/0710.4378}{{\ttfamily arXiv:0710.4378 [hep-th]}}.

\bibitem{Cornalba:2006xk}
L.~Cornalba, M.~S. Costa, J.~Penedones, and R.~Schiappa, ``{Eikonal
  Approximation in AdS/CFT: From Shock Waves to Four-Point Functions},''
  \href{http://dx.doi.org/10.1088/1126-6708/2007/08/019}{{\em JHEP} {\bfseries
  08} (2007) 019},
\href{http://arxiv.org/abs/hep-th/0611122}{{\ttfamily arXiv:hep-th/0611122
  [hep-th]}}.

\bibitem{Cornalba:2006xm}
L.~Cornalba, M.~S. Costa, J.~Penedones, and R.~Schiappa, ``{Eikonal
  Approximation in AdS/CFT: Conformal Partial Waves and Finite N Four-Point
  Functions},'' \href{http://dx.doi.org/10.1016/j.nuclphysb.2007.01.007}{{\em
  Nucl. Phys.} {\bfseries B767} (2007) 327--351},
\href{http://arxiv.org/abs/hep-th/0611123}{{\ttfamily arXiv:hep-th/0611123
  [hep-th]}}.

\bibitem{Cornalba:2007zb}
L.~Cornalba, M.~S. Costa, and J.~Penedones, ``{Eikonal approximation in
  AdS/CFT: Resumming the gravitational loop expansion},''
  \href{http://dx.doi.org/10.1088/1126-6708/2007/09/037}{{\em JHEP} {\bfseries
  09} (2007) 037},
\href{http://arxiv.org/abs/0707.0120}{{\ttfamily arXiv:0707.0120 [hep-th]}}.

\bibitem{Cornalba:2007fs}
L.~Cornalba, ``{Eikonal methods in AdS/CFT: Regge theory and multi-reggeon
  exchange},''
\href{http://arxiv.org/abs/0710.5480}{{\ttfamily arXiv:0710.5480 [hep-th]}}.

\bibitem{Costa:2012cb}
M.~S. Costa, V.~Goncalves, and J.~Penedones, ``{Conformal Regge theory},''
  \href{http://dx.doi.org/10.1007/JHEP12(2012)091}{{\em JHEP} {\bfseries 1212}
  (2012) 091},
\href{http://arxiv.org/abs/1209.4355}{{\ttfamily arXiv:1209.4355 [hep-th]}}.

\bibitem{Afkhami-Jeddi:2016ntf}
N.~Afkhami-Jeddi, T.~Hartman, S.~Kundu, and A.~Tajdini, ``{Einstein gravity
  3-point functions from conformal field theory},''
\href{http://arxiv.org/abs/1610.09378}{{\ttfamily arXiv:1610.09378 [hep-th]}}.

\bibitem{Afkhami-Jeddi:2017rmx}
N.~Afkhami-Jeddi, T.~Hartman, S.~Kundu, and A.~Tajdini, ``{Shockwaves from the
  Operator Product Expansion},''
\href{http://arxiv.org/abs/1709.03597}{{\ttfamily arXiv:1709.03597 [hep-th]}}.

\bibitem{Afkhami-Jeddi:2018own}
N.~Afkhami-Jeddi, S.~Kundu, and A.~Tajdini, ``{A Conformal Collider for
  Holographic CFTs},''
\href{http://arxiv.org/abs/1805.07393}{{\ttfamily arXiv:1805.07393 [hep-th]}}.

\bibitem{KPZ2017}
M.~Kulaxizi, A.~Parnachev, and A.~Zhiboedov, ``{Bulk Phase Shift, CFT Regge
  Limit and Einstein Gravity},''
\href{http://arxiv.org/abs/1705.02934}{{\ttfamily arXiv:1705.02934 [hep-th]}}.

\bibitem{Costa:2017twz}
M.~S. Costa, T.~Hansen, and J.~Penedones, ``{Bounds for OPE coefficients on the
  Regge trajectory},''
\href{http://arxiv.org/abs/1707.07689}{{\ttfamily arXiv:1707.07689 [hep-th]}}.

\bibitem{Meltzer:2017rtf}
D.~Meltzer and E.~Perlmutter, ``{Beyond $a = c$: gravitational couplings to
  matter and the stress tensor OPE},''
  \href{http://dx.doi.org/10.1007/JHEP07(2018)157}{{\em JHEP} {\bfseries 07}
  (2018) 157},
\href{http://arxiv.org/abs/1712.04861}{{\ttfamily arXiv:1712.04861 [hep-th]}}.

\bibitem{Maldacena:2015waa}
J.~Maldacena, S.~H. Shenker, and D.~Stanford, ``{A bound on chaos},''
  \href{http://dx.doi.org/10.1007/JHEP08(2016)106}{{\em JHEP} {\bfseries 08}
  (2016) 106},
\href{http://arxiv.org/abs/1503.01409}{{\ttfamily arXiv:1503.01409 [hep-th]}}.

\bibitem{Hartman:2015lfa}
T.~Hartman, S.~Jain, and S.~Kundu, ``{Causality Constraints in Conformal Field
  Theory},'' \href{http://dx.doi.org/10.1007/JHEP05(2016)099}{{\em JHEP}
  {\bfseries 05} (2016) 099},
\href{http://arxiv.org/abs/1509.00014}{{\ttfamily arXiv:1509.00014 [hep-th]}}.

\bibitem{DAppollonio:2015fly}
G.~D'Appollonio, P.~Di~Vecchia, R.~Russo, and G.~Veneziano, ``{Regge behavior
  saves String Theory from causality violations},''
  \href{http://dx.doi.org/10.1007/JHEP05(2015)144}{{\em JHEP} {\bfseries 05}
  (2015) 144},
\href{http://arxiv.org/abs/1502.01254}{{\ttfamily arXiv:1502.01254 [hep-th]}}.

\bibitem{Kravchuk:2018htv}
P.~Kravchuk and D.~Simmons-Duffin, ``{Light-ray operators in conformal field
  theory},''
\href{http://arxiv.org/abs/1805.00098}{{\ttfamily arXiv:1805.00098 [hep-th]}}.

\bibitem{Li:2017lmh}
D.~Li, D.~Meltzer, and D.~Poland, ``{Conformal Bootstrap in the Regge Limit},''
\href{http://arxiv.org/abs/1705.03453}{{\ttfamily arXiv:1705.03453 [hep-th]}}.

\bibitem{Aichelburg:1970dh}
P.~C. Aichelburg and R.~U. Sexl, ``{On the Gravitational field of a massless
  particle},''
\href{http://dx.doi.org/10.1007/BF00758149}{{\em Gen. Rel. Grav.} {\bfseries 2}
  (1971) 303--312}.

\bibitem{Dray:1984ha}
T.~Dray and G.~'t~Hooft, ``{The Gravitational Shock Wave of a Massless
  Particle},''
\href{http://dx.doi.org/10.1016/0550-3213(85)90525-5}{{\em Nucl. Phys.}
  {\bfseries B253} (1985) 173--188}.

\bibitem{tHooft:1987vrq}
G.~'t~Hooft, ``{Graviton Dominance in Ultrahigh-Energy Scattering},''
\href{http://dx.doi.org/10.1016/0370-2693(87)90159-6}{{\em Phys. Lett.}
  {\bfseries B198} (1987) 61--63}.

\bibitem{Amati:1987wq}
D.~Amati, M.~Ciafaloni, and G.~Veneziano, ``{Superstring Collisions at
  Planckian Energies},''
\href{http://dx.doi.org/10.1016/0370-2693(87)90346-7}{{\em Phys. Lett.}
  {\bfseries B197} (1987) 81}.

\bibitem{Amati:1987uf}
D.~Amati, M.~Ciafaloni, and G.~Veneziano, ``{Classical and Quantum Gravity
  Effects from Planckian Energy Superstring Collisions},''
\href{http://dx.doi.org/10.1142/S0217751X88000710}{{\em Int. J. Mod. Phys.}
  {\bfseries A3} (1988) 1615--1661}.

\bibitem{Cornalba:2008qf}
L.~Cornalba, M.~S. Costa, and J.~Penedones, ``{Eikonal Methods in AdS/CFT: BFKL
  Pomeron at Weak Coupling},''
  \href{http://dx.doi.org/10.1088/1126-6708/2008/06/048}{{\em JHEP} {\bfseries
  06} (2008) 048},
\href{http://arxiv.org/abs/0801.3002}{{\ttfamily arXiv:0801.3002 [hep-th]}}.

\bibitem{Cornalba:2009ax}
L.~Cornalba, M.~S. Costa, and J.~Penedones, ``{Deep Inelastic Scattering in
  Conformal QCD},'' \href{http://dx.doi.org/10.1007/JHEP03(2010)133}{{\em JHEP}
  {\bfseries 03} (2010) 133},
\href{http://arxiv.org/abs/0911.0043}{{\ttfamily arXiv:0911.0043 [hep-th]}}.

\bibitem{Amati:1988tn}
D.~Amati, M.~Ciafaloni, and G.~Veneziano, ``{Can Space-Time Be Probed Below the
  String Size?},''
\href{http://dx.doi.org/10.1016/0370-2693(89)91366-X}{{\em Phys. Lett.}
  {\bfseries B216} (1989) 41--47}.

\bibitem{Giddings:2006vu}
S.~B. Giddings, ``{Locality in quantum gravity and string theory},''
  \href{http://dx.doi.org/10.1103/PhysRevD.74.106006}{{\em Phys. Rev.}
  {\bfseries D74} (2006) 106006},
\href{http://arxiv.org/abs/hep-th/0604072}{{\ttfamily arXiv:hep-th/0604072
  [hep-th]}}.

\bibitem{Shenker:2014cwa}
S.~H. Shenker and D.~Stanford, ``{Stringy effects in scrambling},''
  \href{http://dx.doi.org/10.1007/JHEP05(2015)132}{{\em JHEP} {\bfseries 05}
  (2015) 132},
\href{http://arxiv.org/abs/1412.6087}{{\ttfamily arXiv:1412.6087 [hep-th]}}.

\bibitem{ssw}
D.~Simmons-Duffin, D.~Stanford, and E.~Witten, ``{A spacetime derivation of the
  Lorentzian OPE inversion formula},''
\href{http://arxiv.org/abs/1711.03816}{{\ttfamily arXiv:1711.03816 [hep-th]}}.

\bibitem{Costa:2013zra}
M.~S. Costa, J.~Drummond, V.~Goncalves, and J.~Penedones, ``{The role of
  leading twist operators in the Regge and Lorentzian OPE limits},''
  \href{http://dx.doi.org/10.1007/JHEP04(2014)094}{{\em JHEP} {\bfseries 04}
  (2014) 094},
\href{http://arxiv.org/abs/1311.4886}{{\ttfamily arXiv:1311.4886 [hep-th]}}.

\bibitem{DO1}
F.~Dolan and H.~Osborn, ``{Conformal four point functions and the operator
  product expansion},''
  \href{http://dx.doi.org/10.1016/S0550-3213(01)00013-X}{{\em Nucl.Phys.}
  {\bfseries B599} (2001) 459--496},
\href{http://arxiv.org/abs/hep-th/0011040}{{\ttfamily arXiv:hep-th/0011040
  [hep-th]}}.

\bibitem{DO2}
F.~Dolan and H.~Osborn, ``{Conformal partial waves and the operator product
  expansion},'' \href{http://dx.doi.org/10.1016/j.nuclphysb.2003.11.016}{{\em
  Nucl.Phys.} {\bfseries B678} (2004) 491--507},
\href{http://arxiv.org/abs/hep-th/0309180}{{\ttfamily arXiv:hep-th/0309180
  [hep-th]}}.

\bibitem{DO3}
F.~Dolan and H.~Osborn, ``{Conformal Partial Waves: Further Mathematical
  Results},''
\href{http://arxiv.org/abs/1108.6194v2}{{\ttfamily arXiv:1108.6194v2
  [hep-th]}}.

\bibitem{Fitzpatrick:2011dm}
A.~L. Fitzpatrick and J.~Kaplan, ``{Unitarity and the Holographic S-Matrix},''
  \href{http://dx.doi.org/10.1007/JHEP10(2012)032}{{\em JHEP} {\bfseries 1210}
  (2012) 032},
\href{http://arxiv.org/abs/1112.4845}{{\ttfamily arXiv:1112.4845 [hep-th]}}.

\bibitem{Aharony:2016dwx}
O.~Aharony, L.~F. Alday, A.~Bissi, and E.~Perlmutter, ``{Loops in AdS from
  Conformal Field Theory},''
\href{http://arxiv.org/abs/1612.03891}{{\ttfamily arXiv:1612.03891 [hep-th]}}.

\bibitem{Giombi:2017hpr}
S.~Giombi, C.~Sleight, and M.~Taronna, ``{Spinning AdS Loop Diagrams: Two Point
  Functions},'' \href{http://dx.doi.org/10.1007/JHEP06(2018)030}{{\em JHEP}
  {\bfseries 06} (2018) 030},
\href{http://arxiv.org/abs/1708.08404}{{\ttfamily arXiv:1708.08404 [hep-th]}}.

\bibitem{Cardona:2017tsw}
C.~Cardona, ``{Mellin-(Schwinger) representation of One-loop Witten diagrams in
  AdS},''
\href{http://arxiv.org/abs/1708.06339}{{\ttfamily arXiv:1708.06339 [hep-th]}}.

\bibitem{Yuan:2017vgp}
E.~Y. Yuan, ``{Loops in the Bulk},''
\href{http://arxiv.org/abs/1710.01361}{{\ttfamily arXiv:1710.01361 [hep-th]}}.

\bibitem{Yuan:2018qva}
E.~Y. Yuan, ``{Simplicity in AdS Perturbative Dynamics},''
\href{http://arxiv.org/abs/1801.07283}{{\ttfamily arXiv:1801.07283 [hep-th]}}.

\bibitem{Liu:2018jhs}
J.~Liu, E.~Perlmutter, V.~Rosenhaus, and D.~Simmons-Duffin, ``{$d$-dimensional
  SYK, AdS Loops, and $6j$ Symbols},''
\href{http://arxiv.org/abs/1808.00612}{{\ttfamily arXiv:1808.00612 [hep-th]}}.

\bibitem{Bertan:2018afl}
I.~Bertan, I.~Sachs, and E.~D. Skvortsov, ``{Quantum $\phi^4$ Theory in
  AdS${}_4$ and its CFT Dual},''
  \href{http://dx.doi.org/10.1007/JHEP02(2019)099}{{\em JHEP} {\bfseries 02}
  (2019) 099},
\href{http://arxiv.org/abs/1810.00907}{{\ttfamily arXiv:1810.00907 [hep-th]}}.

\bibitem{Bertan:2018khc}
I.~Bertan and I.~Sachs, ``{Loops in Anti-de Sitter Space},''
  \href{http://dx.doi.org/10.1103/PhysRevLett.121.101601}{{\em Phys. Rev.
  Lett.} {\bfseries 121} no.~10, (2018) 101601},
\href{http://arxiv.org/abs/1804.01880}{{\ttfamily arXiv:1804.01880 [hep-th]}}.

\bibitem{Ponomarev:2019ofr}
D.~Ponomarev, ``{From bulk loops to boundary large-N expansion},''
\href{http://arxiv.org/abs/1908.03974}{{\ttfamily arXiv:1908.03974 [hep-th]}}.

\bibitem{Carmi:2019ocp}
D.~Carmi, ``{Loops in AdS: From the Spectral Representation to Position
  Space},''
\href{http://arxiv.org/abs/1910.14340}{{\ttfamily arXiv:1910.14340 [hep-th]}}.

\bibitem{Meltzer:2019nbs}
D.~Meltzer, E.~Perlmutter, and A.~Sivaramakrishnan, ``{Unitarity Methods in
  AdS/CFT},''
\href{http://arxiv.org/abs/1912.09521}{{\ttfamily arXiv:1912.09521 [hep-th]}}.

\bibitem{Alday:2017gde}
L.~F. Alday, A.~Bissi, and E.~Perlmutter, ``{Holographic Reconstruction of AdS
  Exchanges from Crossing Symmetry},''
\href{http://arxiv.org/abs/1705.02318}{{\ttfamily arXiv:1705.02318 [hep-th]}}.

\bibitem{Bissi:2019kkx}
A.~Bissi, P.~Dey, and T.~Hansen, ``{Dispersion Relation for CFT Four-Point
  Functions},''
\href{http://arxiv.org/abs/1910.04661}{{\ttfamily arXiv:1910.04661 [hep-th]}}.

\bibitem{Carmi:2019cub}
D.~Carmi and S.~Caron-Huot, ``{A Conformal Dispersion Relation: Correlations
  from Absorption},''
\href{http://arxiv.org/abs/1910.12123}{{\ttfamily arXiv:1910.12123 [hep-th]}}.

\bibitem{Kologlu:2019bco}
M.~Kologlu, P.~Kravchuk, D.~Simmons-Duffin, and A.~Zhiboedov, ``{Shocks,
  Superconvergence, and a Stringy Equivalence Principle},''
\href{http://arxiv.org/abs/1904.05905}{{\ttfamily arXiv:1904.05905 [hep-th]}}.

\bibitem{Kologlu:2019mfz}
M.~Kologlu, P.~Kravchuk, D.~Simmons-Duffin, and A.~Zhiboedov, ``{The light-ray
  OPE and conformal colliders},''
\href{http://arxiv.org/abs/1905.01311}{{\ttfamily arXiv:1905.01311 [hep-th]}}.

\bibitem{Fitzpatrick:2019efk}
A.~L. Fitzpatrick, K.-W. Huang, and D.~Li, ``{Probing Universalities in d$>$2
  CFTs: from Black Holes to Shockwaves},''
\href{http://arxiv.org/abs/1907.10810}{{\ttfamily arXiv:1907.10810 [hep-th]}}.

\bibitem{Penedones:2007ns}
J.~Penedones, {\em {High Energy Scattering in the AdS/CFT Correspondence}}.
\newblock PhD thesis, Porto U., 2007.
\newblock
\href{http://arxiv.org/abs/0712.0802}{{\ttfamily arXiv:0712.0802 [hep-th]}}.
\newblock

\bibitem{Penedones:2010ue}
J.~Penedones, ``{Writing CFT correlation functions as AdS scattering
  amplitudes},'' \href{http://dx.doi.org/10.1007/JHEP03(2011)025}{{\em JHEP}
  {\bfseries 1103} (2011) 025},
\href{http://arxiv.org/abs/1011.1485}{{\ttfamily arXiv:1011.1485 [hep-th]}}.

\bibitem{Fitzpatrick:2011hu}
A.~L. Fitzpatrick and J.~Kaplan, ``{Analyticity and the Holographic
  S-Matrix},'' \href{http://dx.doi.org/10.1007/JHEP10(2012)127}{{\em JHEP}
  {\bfseries 1210} (2012) 127},
\href{http://arxiv.org/abs/1111.6972}{{\ttfamily arXiv:1111.6972 [hep-th]}}.

\bibitem{Carmi:2018qzm}
D.~Carmi, L.~Di~Pietro, and S.~Komatsu, ``{A Study of Quantum Field Theories in
  AdS at Finite Coupling},''
  \href{http://dx.doi.org/10.1007/JHEP01(2019)200}{{\em JHEP} {\bfseries 01}
  (2019) 200},
\href{http://arxiv.org/abs/1810.04185}{{\ttfamily arXiv:1810.04185 [hep-th]}}.

\bibitem{Kraus:2018pax}
P.~Kraus and A.~Sivaramakrishnan, ``{Light-state Dominance from the Conformal
  Bootstrap},'' \href{http://dx.doi.org/10.1007/JHEP08(2019)013}{{\em JHEP}
  {\bfseries 08} (2019) 013},
\href{http://arxiv.org/abs/1812.02226}{{\ttfamily arXiv:1812.02226 [hep-th]}}.

\bibitem{Afkhami-Jeddi:2018apj}
N.~Afkhami-Jeddi, S.~Kundu, and A.~Tajdini, ``{A Bound on Massive Higher Spin
  Particles},'' \href{http://dx.doi.org/10.1007/JHEP04(2019)056}{{\em JHEP}
  {\bfseries 04} (2019) 056},
\href{http://arxiv.org/abs/1811.01952}{{\ttfamily arXiv:1811.01952 [hep-th]}}.

\bibitem{Meltzer:2018tnm}
D.~Meltzer, ``{Higher Spin ANEC and the Space of CFTs},''
  \href{http://dx.doi.org/10.1007/JHEP07(2019)001}{{\em JHEP} {\bfseries 07}
  (2019) 001},
\href{http://arxiv.org/abs/1811.01913}{{\ttfamily arXiv:1811.01913 [hep-th]}}.

\bibitem{Belin:2019mnx}
A.~Belin, D.~M. Hofman, and G.~Mathys, ``{Einstein gravity from ANEC
  correlators},'' \href{http://dx.doi.org/10.1007/JHEP08(2019)032}{{\em JHEP}
  {\bfseries 08} (2019) 032},
\href{http://arxiv.org/abs/1904.05892}{{\ttfamily arXiv:1904.05892 [hep-th]}}.

\bibitem{Amati:1988ww}
D.~Amati and C.~Klimcik, ``{Strings in a Shock Wave Background and Generation
  of Curved Geometry from Flat Space String Theory},''
\href{http://dx.doi.org/10.1016/0370-2693(88)90355-3}{{\em Phys. Lett.}
  {\bfseries B210} (1988) 92--96}.

\bibitem{Amati:1990xe}
D.~Amati, M.~Ciafaloni, and G.~Veneziano, ``{Higher Order Gravitational
  Deflection and Soft Bremsstrahlung in Planckian Energy Superstring
  Collisions},''
\href{http://dx.doi.org/10.1016/0550-3213(90)90375-N}{{\em Nucl. Phys.}
  {\bfseries B347} (1990) 550--580}.

\bibitem{Amati:1992zb}
D.~Amati, M.~Ciafaloni, and G.~Veneziano, ``{Planckian scattering beyond the
  semiclassical approximation},''
\href{http://dx.doi.org/10.1016/0370-2693(92)91366-H}{{\em Phys. Lett.}
  {\bfseries B289} (1992) 87--91}.

\bibitem{Amati:1993tb}
D.~Amati, M.~Ciafaloni, and G.~Veneziano, ``{Effective action and all order
  gravitational eikonal at Planckian energies},''
\href{http://dx.doi.org/10.1016/0550-3213(93)90367-X}{{\em Nucl. Phys.}
  {\bfseries B403} (1993) 707--724}.

\bibitem{Muzinich:1987in}
I.~J. Muzinich and M.~Soldate, ``{High-Energy Unitarity of Gravitation and
  Strings},''
\href{http://dx.doi.org/10.1103/PhysRevD.37.359}{{\em Phys. Rev.} {\bfseries
  D37} (1988) 359}.

\bibitem{Soldate:1986mk}
M.~Soldate, ``{Partial Wave Unitarity and Closed String Amplitudes},''
\href{http://dx.doi.org/10.1016/0370-2693(87)90302-9}{{\em Phys. Lett.}
  {\bfseries B186} (1987) 321--327}.

\bibitem{Sundborg:1988tb}
B.~Sundborg, ``{High-energy Asymptotics: The One Loop String Amplitude and
  Resummation},''
\href{http://dx.doi.org/10.1016/0550-3213(88)90014-4}{{\em Nucl. Phys.}
  {\bfseries B306} (1988) 545--566}.

\bibitem{Ademollo:1989ag}
M.~Ademollo, A.~Bellini, and M.~Ciafaloni, ``{Superstring Regge Amplitudes and
  Emission Vertices},''
\href{http://dx.doi.org/10.1016/0370-2693(89)91609-2}{{\em Phys. Lett.}
  {\bfseries B223} (1989) 318--324}.

\bibitem{Ademollo:1990sd}
M.~Ademollo, A.~Bellini, and M.~Ciafaloni, ``{Superstring Regge Amplitudes and
  Graviton Radiation at Planckian Energies},''
\href{http://dx.doi.org/10.1016/0550-3213(90)90626-O}{{\em Nucl. Phys.}
  {\bfseries B338} (1990) 114--142}.

\bibitem{Bellini:1992eb}
A.~Bellini, M.~Ademollo, and M.~Ciafaloni, ``{Superstring one loop and
  gravitino contributions to Planckian scattering},''
  \href{http://dx.doi.org/10.1016/0550-3213(93)90238-K}{{\em Nucl. Phys.}
  {\bfseries B393} (1993) 79--94},
\href{http://arxiv.org/abs/hep-th/9207113}{{\ttfamily arXiv:hep-th/9207113
  [hep-th]}}.

\bibitem{Giddings:2007bw}
S.~B. Giddings, D.~J. Gross, and A.~Maharana, ``{Gravitational effects in
  ultrahigh-energy string scattering},''
  \href{http://dx.doi.org/10.1103/PhysRevD.77.046001}{{\em Phys. Rev.}
  {\bfseries D77} (2008) 046001},
\href{http://arxiv.org/abs/0705.1816}{{\ttfamily arXiv:0705.1816 [hep-th]}}.

\bibitem{Banks:2009bj}
T.~Banks and G.~Festuccia, ``{The Regge Limit for Green Functions in Conformal
  Field Theory},'' \href{http://dx.doi.org/10.1007/JHEP06(2010)105}{{\em JHEP}
  {\bfseries 06} (2010) 105},
\href{http://arxiv.org/abs/0910.2746}{{\ttfamily arXiv:0910.2746 [hep-th]}}.

\bibitem{Hofman:2008ar}
D.~M. Hofman and J.~Maldacena, ``{Conformal collider physics: Energy and charge
  correlations},'' \href{http://dx.doi.org/10.1088/1126-6708/2008/05/012}{{\em
  JHEP} {\bfseries 05} (2008) 012},
\href{http://arxiv.org/abs/0803.1467}{{\ttfamily arXiv:0803.1467 [hep-th]}}.

\bibitem{Kulaxizi:2018dxo}
M.~Kulaxizi, G.~S. Ng, and A.~Parnachev, ``{Black Holes, Heavy States, Phase
  Shift and Anomalous Dimensions},''
  \href{http://dx.doi.org/10.21468/SciPostPhys.6.6.065}{{\em SciPost Phys.}
  {\bfseries 6} (2019) 065},
\href{http://arxiv.org/abs/1812.03120}{{\ttfamily arXiv:1812.03120 [hep-th]}}.

\bibitem{Karlsson:2019qfi}
R.~Karlsson, M.~Kulaxizi, A.~Parnachev, and P.~Tadic, ``{Black Holes and
  Conformal Regge Bootstrap},''
\href{http://arxiv.org/abs/1904.00060}{{\ttfamily arXiv:1904.00060 [hep-th]}}.

\bibitem{Li:2015itl}
D.~Li, D.~Meltzer, and D.~Poland, ``{Conformal Collider Physics from the
  Lightcone Bootstrap},'' \href{http://dx.doi.org/10.1007/JHEP02(2016)143}{{\em
  JHEP} {\bfseries 02} (2016) 143},
\href{http://arxiv.org/abs/1511.08025}{{\ttfamily arXiv:1511.08025 [hep-th]}}.

\bibitem{Hofman:2016awc}
D.~M. Hofman, D.~Li, D.~Meltzer, D.~Poland, and F.~Rejon-Barrera, ``{A Proof of
  the Conformal Collider Bounds},''
  \href{http://dx.doi.org/10.1007/JHEP06(2016)111}{{\em JHEP} {\bfseries 06}
  (2016) 111},
\href{http://arxiv.org/abs/1603.03771}{{\ttfamily arXiv:1603.03771 [hep-th]}}.

\bibitem{Elkhidir:2017iov}
E.~Elkhidir and D.~Karateev, ``{Scalar-Fermion Analytic Bootstrap in 4D},''
  \href{http://dx.doi.org/10.1007/JHEP06(2019)026}{{\em JHEP} {\bfseries 06}
  (2019) 026},
\href{http://arxiv.org/abs/1712.01554}{{\ttfamily arXiv:1712.01554 [hep-th]}}.

\bibitem{Sleight:2018epi}
C.~Sleight and M.~Taronna, ``{Spinning Mellin Bootstrap: Conformal Partial
  Waves, Crossing Kernels and Applications},''
  \href{http://dx.doi.org/10.1002/prop.201800038}{{\em Fortsch. Phys.}
  {\bfseries 66} (2018) 8},
\href{http://arxiv.org/abs/1804.09334}{{\ttfamily arXiv:1804.09334 [hep-th]}}.

\bibitem{Sleight:2018ryu}
C.~Sleight and M.~Taronna, ``{Anomalous Dimensions from Crossing Kernels},''
  \href{http://dx.doi.org/10.1007/JHEP11(2018)089}{{\em JHEP} {\bfseries 11}
  (2018) 089},
\href{http://arxiv.org/abs/1807.05941}{{\ttfamily arXiv:1807.05941 [hep-th]}}.

\bibitem{Albayrak:2019gnz}
S.~Albayrak, D.~Meltzer, and D.~Poland, ``{More Analytic Bootstrap:
  Nonperturbative Effects and Fermions},''
  \href{http://dx.doi.org/10.1007/JHEP08(2019)040}{{\em JHEP} {\bfseries 08}
  (2019) 040},
\href{http://arxiv.org/abs/1904.00032}{{\ttfamily arXiv:1904.00032 [hep-th]}}.

\bibitem{Costa:2011mg}
M.~S. Costa, J.~Penedones, D.~Poland, and S.~Rychkov, ``{Spinning Conformal
  Correlators},'' \href{http://dx.doi.org/10.1007/JHEP11(2011)071}{{\em JHEP}
  {\bfseries 1111} (2011) 071},
\href{http://arxiv.org/abs/1107.3554}{{\ttfamily arXiv:1107.3554 [hep-th]}}.

\bibitem{Costa:2011dw}
M.~S. Costa, J.~Penedones, D.~Poland, and S.~Rychkov, ``{Spinning Conformal
  Blocks},'' \href{http://dx.doi.org/10.1007/JHEP11(2011)154}{{\em JHEP}
  {\bfseries 1111} (2011) 154},
\href{http://arxiv.org/abs/1109.6321}{{\ttfamily arXiv:1109.6321 [hep-th]}}.

\bibitem{Karateev:2017jgd}
D.~Karateev, P.~Kravchuk, and D.~Simmons-Duffin, ``{Weight Shifting Operators
  and Conformal Blocks},''
\href{http://arxiv.org/abs/1706.07813}{{\ttfamily arXiv:1706.07813 [hep-th]}}.

\bibitem{Costa:2018mcg}
M.~S. Costa and T.~Hansen, ``{AdS Weight Shifting Operators},''
  \href{http://dx.doi.org/10.1007/JHEP09(2018)040}{{\em JHEP} {\bfseries 09}
  (2018) 040},
\href{http://arxiv.org/abs/1805.01492}{{\ttfamily arXiv:1805.01492 [hep-th]}}.

\bibitem{Gary:2009ae}
M.~Gary, S.~B. Giddings, and J.~Penedones, ``{Local bulk S-matrix elements and
  CFT singularities},''
  \href{http://dx.doi.org/10.1103/PhysRevD.80.085005}{{\em Phys. Rev.}
  {\bfseries D80} (2009) 085005},
\href{http://arxiv.org/abs/0903.4437}{{\ttfamily arXiv:0903.4437 [hep-th]}}.

\bibitem{Okuda:2010ym}
T.~Okuda and J.~Penedones, ``{String scattering in flat space and a scaling
  limit of Yang-Mills correlators},''
  \href{http://dx.doi.org/10.1103/PhysRevD.83.086001}{{\em Phys. Rev.}
  {\bfseries D83} (2011) 086001},
\href{http://arxiv.org/abs/1002.2641}{{\ttfamily arXiv:1002.2641 [hep-th]}}.

\bibitem{Maldacena:2015iua}
J.~Maldacena, D.~Simmons-Duffin, and A.~Zhiboedov, ``{Looking for a bulk
  point},''
\href{http://arxiv.org/abs/1509.03612}{{\ttfamily arXiv:1509.03612 [hep-th]}}.

\bibitem{Gross:1987ar}
D.~J. Gross and P.~F. Mende, ``{String Theory Beyond the Planck Scale},''
\href{http://dx.doi.org/10.1016/0550-3213(88)90390-2}{{\em Nucl. Phys.}
  {\bfseries B303} (1988) 407--454}.

\bibitem{Gross:1987kza}
D.~J. Gross and P.~F. Mende, ``{The High-Energy Behavior of String Scattering
  Amplitudes},''
\href{http://dx.doi.org/10.1016/0370-2693(87)90355-8}{{\em Phys. Lett.}
  {\bfseries B197} (1987) 129--134}.

\bibitem{Mende:1989wt}
P.~F. Mende and H.~Ooguri, ``{Borel Summation of String Theory for Planck Scale
  Scattering},''
\href{http://dx.doi.org/10.1016/0550-3213(90)90202-O}{{\em Nucl. Phys.}
  {\bfseries B339} (1990) 641--662}.

\bibitem{Dodelson:2019ddi}
M.~Dodelson and H.~Ooguri, ``{The High Energy Behavior of Mellin Amplitudes},''
\href{http://arxiv.org/abs/1911.05274}{{\ttfamily arXiv:1911.05274 [hep-th]}}.

\end{thebibliography}\endgroup

\end{document}